\documentclass[fleqn,usenatbib]{mnras}

\usepackage{newtxtext,newtxmath}
\usepackage{comment}

\usepackage[T1]{fontenc}

\DeclareRobustCommand{\VAN}[3]{#2}
\let\VANthebibliography\thebibliography
\def\thebibliography{\DeclareRobustCommand{\VAN}[3]{##3}\VANthebibliography}

\usepackage{graphicx}	
\usepackage{amsmath}	
\usepackage{cleveref}
\usepackage{upgreek}
\usepackage{xcolor}
\usepackage{longtable}
\usepackage[nolist]{acronym}
\usepackage{subfigure}
\usepackage{caption}

\usepackage{pdflscape}
\usepackage{ragged2e}

\crefformat{section}{\S#2#1#3} 
\crefformat{subsection}{\S#2#1#3}
\crefformat{subsubsection}{\S#2#1#3}

\definecolor{sof}{rgb}{0.5, 0.0, 0.5}

\definecolor{wf}{rgb}{0.8, 0.5, 0}

\definecolor{wl}{rgb}{0.3, 0.3, 0}

\title[ATA observations of FRB 20220912A]{Characterization of the Repeating FRB 20220912A with the Allen Telescope Array}

\author[S. Z. Sheikh et al.]{\parbox{\textwidth}
  {Sofia Z. Sheikh$^{1,2,3}$\thanks{E-mail: ssheikh@seti.org},
  Wael Farah$^{1,2}$,
  Alexander W. Pollak$^{1,2}$,
  Andrew, P. V., Siemion$^{1,2,4,5}$,
  Mohammed A. Chamma$^{6}$,
  Luigi F. Cruz$^{1,2}$,
  Roy H. Davis$^{1}$,
  David R. DeBoer$^{1,2}$,
  Vishal Gajjar$^{1,2}$,
  Phil Karn$^{1}$,
  Jamar Kittling$^{1,7}$,
  Wenbin Lu$^{8}$,
  Mark Masters$^{1}$,
  Pranav Premnath$^{1,9}$,
  Sarah Schoultz$^{1}$,
  Carol Shumaker$^{1}$,
  Gurmehar Singh$^{1,10}$,
  Michael Snodgrass$^{1}$
  } \\ \\ \\
\parbox{\textwidth}{
  $^{1}$SETI Institute, 339 Bernardo Ave, Suite 200 Mountain View, CA 94043, USA\\
  $^{2}$Berkeley SETI Research Center, University of California, Berkeley, CA 94720, USA\\
  $^{3}$Penn State Extraterrestrial Intelligence Center, 525 Davey Laboratory, The Pennsylvania State University, University Park, PA, 16802, USA\\
  $^{4}$Department of Physics and Astronomy, University of Manchester, UK\\
  $^{5}$University of Malta, Institute of Space Sciences and Astronomy, Msida, MSD2080, Malta\\
  $^{6}$Department of Physics \& Astronomy, McMaster University, 1280 Main Street West, Hamilton, ON, L8S 4M1, Canada\\
  $^{7}$Wesleyan University, Middletown, CT 06459, USA\\
  $^{8}$Departments of Astronomy and Theoretical Astrophysics Center, UC Berkeley, Berkeley, CA 94720, USA\\
  $^{9}$Department of Physics \& Astronomy, The University of California, Irvine, Irvine, CA 92697, USA\\
  $^{10}$Purdue University, West Lafayette, IN 47907, USA
}
}

\date{Accepted 2023 November 20. Received 2023 November 17; in original form 2023 May 19}

\pubyear{2023}

\begin{document}
\label{firstpage}
\pagerange{\pageref{firstpage}--\pageref{lastpage}}
\maketitle

\begin{abstract}
FRB\,20220912A is a repeating \ac{FRB} that was discovered in Fall 2022 and remained highly active for several months. We report the detection of 35 \ac{FRB}s from 541 hours of follow-up observations of this source using the recently refurbished Allen Telescope Array, covering 1344 MHz of bandwidth primarily centered at 1572 MHz. All 35 \ac{FRB}s were detected in the lower half of the band with non-detections in the upper half and covered fluences from 4--431~Jy-ms (median$=$48.27~Jy-ms). We find consistency with previous repeater studies for a range of spectrotemporal features including: bursts with downward frequency drifting over time; a positive correlation between bandwidth and center frequency; and a decrease in sub-burst duration over time. We report an apparent decrease in the center frequency of observed bursts over the 2 months of the observing campaign (corresponding to a drop of $6.21\pm 0.76$ MHz per day). We predict a cut-off fluence for FRB~20220912A of $F_\textrm{max}\lesssim 10^4$~Jy-ms, for this source to be consistent with the all-sky rate, and find that FRB~20220912A significantly contributed to the all-sky \ac{FRB} rate at a level of a few percent for fluences of $\sim$100~Jy-ms. Finally, we investigate characteristic timescales and sub-burst periodicities and find a) a median inter-subburst timescale of 5.82$\pm$1.16~ms in the multi-component bursts and b) no evidence of strict periodicity even in the most evenly-spaced multi-component burst in the sample. Our results demonstrate the importance of wideband observations of \ac{FRB}s, and provide an important set of observational parameters against which to compare \ac{FRB} progenitor and emission mechanism models. \acresetall

\end{abstract}




\section{Introduction}

\acp{FRB}, subsecond-duration coherent flashes in the radio spectrum that originate at cosmological distances, are one of the most intriguing phenomena in the last decade of time-domain astronomy. \ac{FRB}s have been detected at radio frequencies from 120~MHz \citep{pastor-marazuela2021chromatic} all the way up to 8~GHz \citep{Gajjar2018, Michilli2018}, with isotropic-equivalent spectral luminosities of $10^{27}$ to $10^{34}$ erg~s$^{-1}$Hz$^{-1}$ \citep{nimmo2022burst}, and excess \acp{DM} (subtracting the galactic contribution using electron density models) of several tens to a few thousand pc~cm$^{-3}$ \citep{petroff2022fast}. Most \ac{FRB}s have only been observed a single time, but some, known as ``repeaters,'' are seen more than once; the repeater fraction is still evolving with more observation, but, at the moment, only 2.6\% of \ac{CHIME} \ac{FRB} sources have been seen to repeat \citep{andersen2023chime}. The first of these repeaters discovered, FRB20121102A \citep[e.g., ][]{Spitler2016, Scholz2016, Zhang2018_ML, gourdji19_121102_sample, Hessels2019, majid20_121102_high_frequencies, pearlman2020multiwavelength, Agarwal2020_GBT, li21_121102_FAST, andersen2023chime} has now produced thousands of recorded bursts across different instruments and frequency ranges. As of the writing of this paper, 46 FRB sources have been confirmed to repeat, the majority of which do not show any burst-to-burst periodicity \citep[{though two sources, FRB121102 and FRB180916, show periodic activity windows}, e.g., ][]{chime20_periodic_180916, Pilia2020}. Many FRB sources have been precisely localized by interferometric arrays \citep{Bannister2019, Marcote2020_host_galaxy, kirsten2022precise, ravi2023dsalocalization}, which provide important clues to the nature of the emitting objects and their host environments. Repeaters are particularly valuable for interrogating the emission mechanism of \ac{FRB}s, as their bursts can be analyzed as a self-contained sample of observational data arising from a single physical object \citep[e.g., ][]{Hessels2019, li21_121102_FAST,chamma2022broad, jahns2023november}.

The emission mechanism of \ac{FRB}s still remains a mystery \citep[see][for recent reviews]{Cordes2019_review, zhang20_review, petroff2022fast, Bailes2022_frb_review}, though a consensus is emerging that magnetars are likely responsible for at least a sub-population due to an \ac{FRB}-like event from a galactic magnetar \citep{Bochenek2020_stare2, chime2020magnetar}. For magnetar-related classes of models, there is still considerable debate on the detailed radiation mechanism, with various explanations distinguished by, among other things, their distance from the progenitor magnetar \citep{petroff2022fast}. Magnetospheric models operate at a few neutron star radii from the magnetar \citep[e.g., ][]{lu20_unified_model}, while shock models operate much further out, at characteristic scales of $10^{10}$~cm \citep[e.g., ][]{Metzger2019}. Most radiation mechanism models for repeaters sufficiently explain the broad features of repeating \ac{FRB} emission (e.g., coherent emission, an energy budget consistent with repeating bursts, non-cataclysmic sources) and therefore will likely be differentiated via more specific spectrotemporal behaviour.

The observed population of repeating \ac{FRB}s does show consistent spectrotemporal features, such as the tendency of emission from a single burst to decrease in frequency over the duration of the burst, often in discrete steps associated with distinct ``subpulses'' or ``sub-bursts'' \citep{Hessels2019}. This so-called ``sad trombone'' effect, or downward drift, is not an unbreakable rule, however, as others have reported positive drift rates in some bursts \citep[e.g., ][]{kumar2022circularly, zhou2022upward}. For each \ac{FRB}, features such as the duration, central frequency, frequency extent, and drift rate can be combined with flux and polarization information to reveal consistent properties across bursts and sources. Specific spectrotemporal features such as upward-drifting ``happy trombones,'' sub-burst periodicities \citep{andersen2023chime}, or $<$100 ns sub-burst structure \citep{majid21_rapid_variability, nimmo2022burst} might provide the clues necessary to understand the full nature of \ac{FRB}s. For example, \citet{zhou2022upward} notes the similarity of drifting behaviour in \ac{FRB}s to that of certain pulsars, hinting at a potential similarity in emission mechanism or environment. Similarly, the small emission regions (tens to thousands of meters) implied by $<$100 ns sub-burst structure lends itself to a magnetospheric origin \citep{nimmo2022burst}.

On 2022 October 15, a new repeating source, FRB~20220912A, was reported by \citet{mckinven2022nine}; nine bursts at 400~MHz were detected by \ac{CHIME} in a three day period in September 2022. The original detection had inferred J2000 coordinates of RA $= 347.29(4)^\circ$, Dec $= +48.70(3)^\circ$. Other Astronomer's Telegrams (ATels) soon followed, showing detections at L-band \citep{herrmann2022fast} and an improved localization from the \ac{DSA-110} at 23h09m04.9s +48d42m25.4s (J2000) \citep{ravi2022erratum}. This location is coincident with the potential host galaxy PSO J347.2702+48.7066 \citep{mckinven2022nine, ravi2022erratum}. The repeating \ac{FRB} has since been detected at frequencies from 300~MHz \citep{bhusare2022uGMRT}\footnote{There is also a claimed marginal detection at 111~MHz} \citep{fedorova2022detection} to 2.3~GHz \citep{rajwade2022detection, perera2022detection}, with some large single-dish telescopes seeing L-band burst rates of over 100 bursts per hour \citep[e.g., ][]{feng2022extreme}. Recent published work on the source in \citet{zhang2023fast} and \citet{feng2023extreme} characterize this source as the fourth extremely active \ac{FRB}, but the first one in a relatively clean environment as derived from its polarization information and steady \ac{DM}, providing a unique laboratory for understanding which \ac{FRB} properties come from the emission mechanism intrinsically. 

Repeating \ac{FRB}s show varying behaviour across wide bandwidths, for instance, periodicity in activity that is phase-delayed in time over frequency for FRB~20180916B \citep[e.g., ][]{sand_FRB_2023}. They are also limited in bandwidth \citep[$100$s of MHz up to a few GHz; e.g., ][]{gourdji19_121102_sample} around their unpredictable center frequencies. Given these characteristics, wide bandwidth receivers and simultaneous observation at different bands will be critical to characterize them. The \ac{ATA} has both of these features, making it a uniquely suited instrument for \ac{FRB} observation.

In this paper, we report an observing campaign of FRB~20220912A conducted with the \ac{ATA} between October and December of 2022. In Section \ref{sec:observations_and_data}, we summarize the current status of the refurbishment of the \ac{ATA}, describe the observational campaign of FRB~20220912A, and detail the search pipeline used to detect and validate \ac{FRB}s. In Section \ref{sec:data_reduction}, we discuss our methods for data pre-processing and extraction of spectrotemporal properties. In Section \ref{sec:analysis}, we use the data and properties from the previous section to compute the all-sky rate, quantify correlations between spectrotemporal properties, and investigate potential sub-burst periodicity. Finally, we discuss our results and conclude in Section \ref{sec:conclusion}.

\section{Observations and Data}
\label{sec:observations_and_data}

\subsection{The Allen Telescope Array: Instrument Specifications}
\label{ssec:ATA}

The \ac{ATA} is a 42-element interferometer consisting of 6.1m dishes hosted on the Hat Creek Radio Observatory in northern California owned and operated by the SETI Institute, Mtn. View, CA. In late 2019, the instrument began a refurbishment program aimed at improving the sensitivity and robustness of the telescope feeds and revamping the \ac{DSP} system. A full description of the analog and \ac{DSP} upgrades will be presented in Pollak et al. (in prep.) and Farah et al. (in prep.) respectively, but here we will include the essential details to accompany the observations of FRB~20220912A. 

Each \ac{ATA} dish is an offset Gregorian and can slew in both the azimuth and elevation direction. The refurbished ``Antonio'' log-periodic feeds are dual-polarization and sensitive to a large instantaneous frequency range covering the 1 to 11 GHz band \citep{welch2009allen}. Each feed is placed in a cryostat and is cooled to a temperature of $\sim70$K. Analog signals from each antenna are amplified and sent over optical fiber. Each antenna's signal is then split and mixed to produce up to 4 independently tunable signal chains, denoted `a', `b', `c', and `d'. 

Digitization of the antenna signals is performed on 16-channel input Xilinx ZYNQ UltraScale+ \ac{RFSoC} boards, where data get channelized, packetized, and transmitted to the network on a 100 GB ethernet link. A delay engine and fringe rotator are also included as part of the firmware on the \ac{FPGA} boards such that voltage data from all antennas are delayed and phase-centered relative to a user-defined sky coordinate (usually the center of the \ac{ATA} antenna primary beam). Five \ac{RFSoC} boards are currently deployed as part of the prototype \ac{DSP} system which supports the digitisation of 20 dual-polarization antennas over 2 frequency tunings. Upgrades to outfit the entire \ac{ATA} with newly refurbished feeds and deploy more digitizers to cover all the available tunings are planned for the near future.

Channelized voltages are received by a cluster of 8 compute nodes where data are processed depending on the observer-selected \ac{DSP} backend. An xGPU-based, which accelerates cross-correlations (x) using Graphical Processing Units (GPUs) \citep{xgpu}, can be selected to generate interferometric visibilities that can be used for imaging and for delay and phase calibrating the beamformer. Our beamformer, \ac{BLADE}, is a custom-built GPU-based beamformer that was developed in-house, tested, and deployed on the \ac{ATA} cluster (Cruz et al., in prep.) and has since been used for novel science, such as tracking the reverse shock emission of gamma-ray burst 221009A \citep{bright2023precise}. Telescope users can customise the \ac{ATA} backend according to science cases. This includes selecting the time-integration length of the correlator, setting the number of beams to produce with the beamformer, and setting the polarization data output. 

Table \ref{tab:ata} contains a summary of the relevant instrument properties for the \ac{FRB} observations described here.

\begin{table}
    \centering
    \caption{Specifications of the Allen Telescope Array in Fall 2022}.
    \label{tab:ata}
    \begin{tabular}{ll}
        \hline
        \hline
        Parameter & Value \\
        \hline
        Antenna diameter & 6.1 meter \\
        Longest baseline & 300 meter \\
        Frequency coverage & 1-11 GHz \\
        Primary beam FWHM (at 1GHz; at 11GHz) & 3.5$^{\circ}$; 0.3$^{\circ}$\\
        Synthesized beam FWHM (at 1GHz; at 11GHz) & 4.2$'$; 0.38$'$ \\
        Processed bandwidth per tuning & 672 MHz \\
        Number of available simultaneous tunings & 2 \\
        Number of simultaneous polarizations & 2 \\
        Number of beamformed antennas & 20 \\
        \hline
    \end{tabular}
\end{table}

\subsection{Observing Campaign}
\label{ssec:observations} 

Just 56 minutes after the initial announcement of FRB~20220912A by \citet{mckinven2022nine}, the \ac{ATA} began a follow-up campaign of the source. During the time period between 15 October 2022 and 31 December 2022, we observed the source in a series of 70 observations, with a median observation length of 8 hours, divided into 30 minute scans. This led to a total observing time of 541 hours on-source. We ended the campaign when the source's activity had decreased, but not fully ceased, due to observing resource constraints. We recorded data with the 20-element beamformer, placing a single phase-centered synthesized beam on the target at the center of the primary beam; initially, we used the inferred J2000 coordinates from the \ac{CHIME} detection of FRB~20220912A, then switched coordinates to the updated \ac{DSA-110} coordinates once they were available on 25 October 2022. Given the synthesized beamsize of the \ac{ATA} at 2~GHz (approximately 2 arcmin), the original \ac{FRB} coordinates would have placed the true source at the edge of the beamformed beam, which affected the first 31 hours of observation. \citet{sheikh2022bright} described the detection of 8 bursts from FRB~20220912A with the \ac{ATA}; the 8 bursts in that announcement are included in this work.

We exercised the \ac{ATA}'s tuning flexibility to observe the source at two different 672~MHz tunings simultaneously. The center frequencies of the tunings varied at the beginning of the campaign, but settled at their final values of 1236 MHz and 1908 MHz on 4 November 2022. This setting allows for the two tunings to provide continuous frequency coverage from 900--2244~MHz. Before 4 November 2022, 12 sessions had tunings of 1400~MHz and 3000~MHz, 2 sessions had tunings of 1400~MHz and 6000~MHz, and 1 session had tunings of 1150~MHz and 1850~MHz --- this amounts to 76 / 541 hours of data recorded at the non-final central frequencies.

Before every observing session, we observed a phase calibrator (3C48, 3C286, or 3C147) in the correlator mode to assess sensitivity and calibrate the instrument with updated delay and phase solutions prior to science observations. In addition, from 7 November 2022 until the end of the campaign, we added an observation of pulsar J0332+5434 for general system validation and validation of the \texttt{SPANDAK} pulse detection pipeline described in Section \ref{ssec:search}.

\subsection{Searching for FRBs}
\label{ssec:search}

The \ac{ATA} beamformer used for the observations described in the previous section currently produces Stokes I 32-bit \texttt{SIGPROC} ``filterbank''\footnote{\href{https://sigproc.sourceforge.net/sigproc.pdf}{https://sigproc.sourceforge.net/sigproc.pdf}} files at a frequency and time resolution of 0.5~MHz and $64\upmu\textrm{s}$, respectively. These data were not coherently dedispersed to any particular \ac{DM} during data-recording, but were stored for off-line \ac{FRB} searching. The beamformer is capable of producing cross-polarisation data products that can enable polarization studies. However, given the intricate nature of polarisation calibration, we deliberately chose not to record individual polarisations for this work (though we intend to do so in future studies). Moreover, efforts are currently underway to implement a beamformer mode which supports coherent dedispersion.

In this project, we used the \texttt{SPANDAK} pipeline \citep{Gajjar2022}, which is a wrapper for \texttt{HEIMDALL}, a GPU-accelerated search code for dispersed signals in radio astronomical data \citep{Barsdell_PhD}. \texttt{SPANDAK} implements additional filtering and candidate evaluation procedures on the \texttt{HEIMDALL} outputs and produces tables of burst candidates along with diagnostic plots that are later visually reviewed by an observer. 

We first decimate the filterbank files in bit depth from 32-bit to 8-bit so they are compatible with the existing format requirements for \texttt{HEIMDALL}. We then splice the seven nodes of data for each of the two tunings together, creating one 8-bit filterbank file covering the entirety of the tuning centered at \ac{LO} b and another covering the entirety of the tuning centered at \ac{LO}c. Due to the data volume and processing time, we opted not to perform \ac{RFI} removal before searching. 

We assign a \ac{DM} search range of $\pm 10\%$ from the nominal dispersion measure of 219.46 pc~cm${^{-3}}$ as reported by the \ac{CHIME} repeater catalog\footnote{\url{https://www.chime-frb.ca/repeaters}}, with a \ac{DM} step size calculated such that the \ac{SNR} loss between \ac{DM} trials is no more than 0.1\%. We then run \texttt{SPANDAK} with a \ac{SNR} threshold of 10, a boxcar maximum width of $2^{16}$ samples (4.194 seconds), and a maximum-candidates-per-second (maxCsec) value of 15, which is slightly higher than the \texttt{SPANDAK} default. The maxCsec parameter is observatory and band dependent as it is strongly correlated with the \ac{RFI} environment. For these frequencies at the \ac{ATA}, we determined that 15 was an appropriate value via test observations of J0332+5434. We execute the search code over each spliced tuning independently, corresponding to a fixed search bandwidth of 672~MHz. The incoherent dedispersion and searching code is then executed and logged. More information about how the \texttt{SPANDAK} candidate plots are constructed can be found in Figure 9 of \citet{gajjar2021galactic}.

The median number of top candidates per observing session was 9, where a top candidate is a \texttt{HEIMDALL} candidate given at least a B--C rank by \texttt{SPANDAK} due to its broadband characteristics and trial DM-vs-SNR response. The low number of candidates made additional filtering for false-positive \ac{RFI} unnecessary in most cases.

In 541 hours of observation, we detect 35 bursts from FRB~20220912A. All 35 bursts were detected in \ac{LO}b, the lower 672~MHz tuning, with none detected (or visible simultaneously in dedispersed archives) in \ac{LO}c. This is consistent with low detection rates at S-band and non-detections in C-band \citep[e.g., ][]{kirsten2022precise}. Dynamic spectra for all the bursts detected with the \ac{ATA} are shown in Figure~\ref{fig:dynamic_spectra_no_sp}. As suggested by the single-tuning detections, the bursts from FRB~20220912A are highly spectrally limited, which was similarly observed in FRB~20180916B \citep{sand_FRB_2023} and FRB~20121102A \citep{Law2017} --- a deeper investigation of spectral extent is performed in Section \cref{sec:analysis}. 

\begin{figure*}
    \centering
    \subfigure{\includegraphics[width=0.24\textwidth]{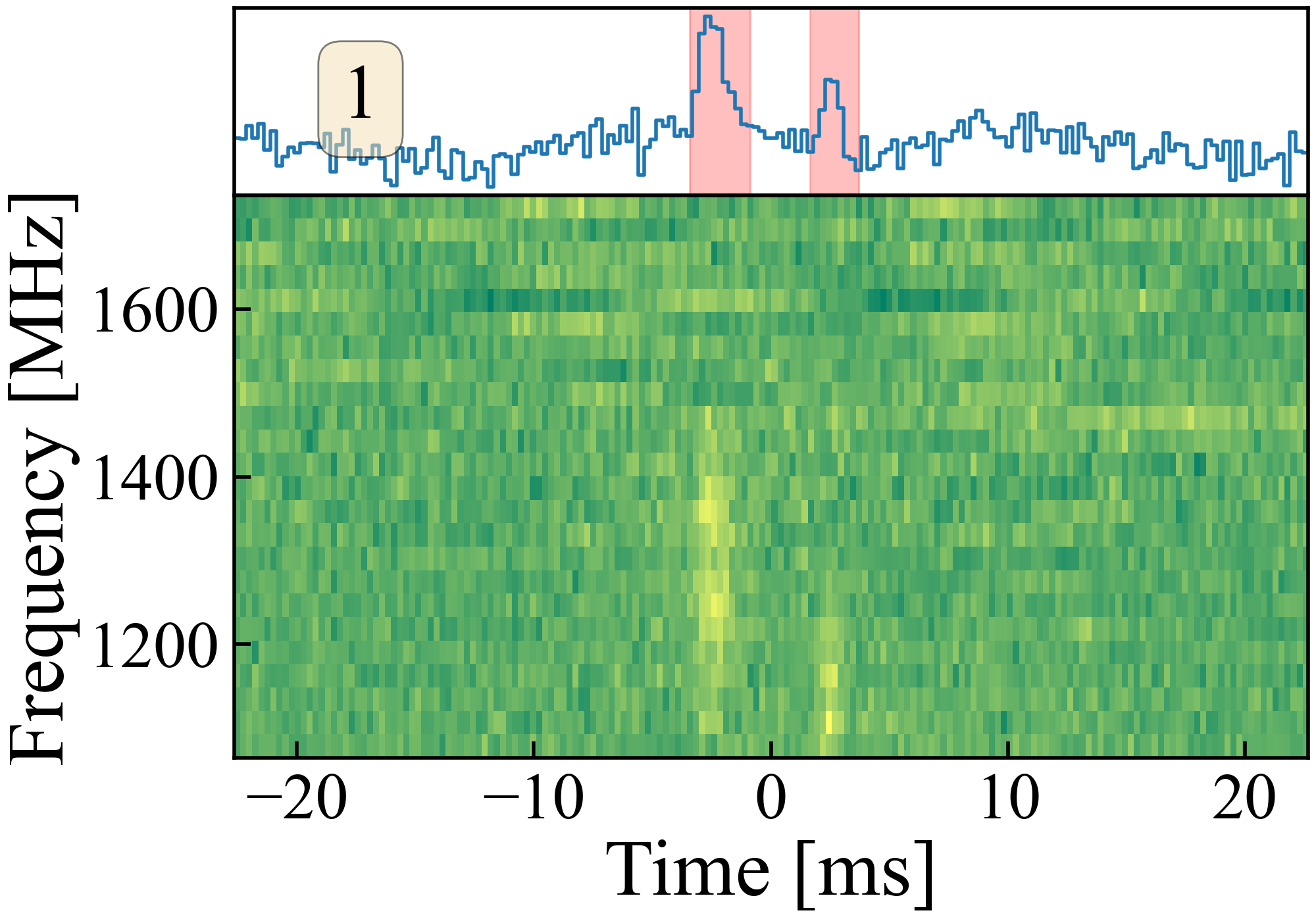}}
    \subfigure{\includegraphics[width=0.24\textwidth]{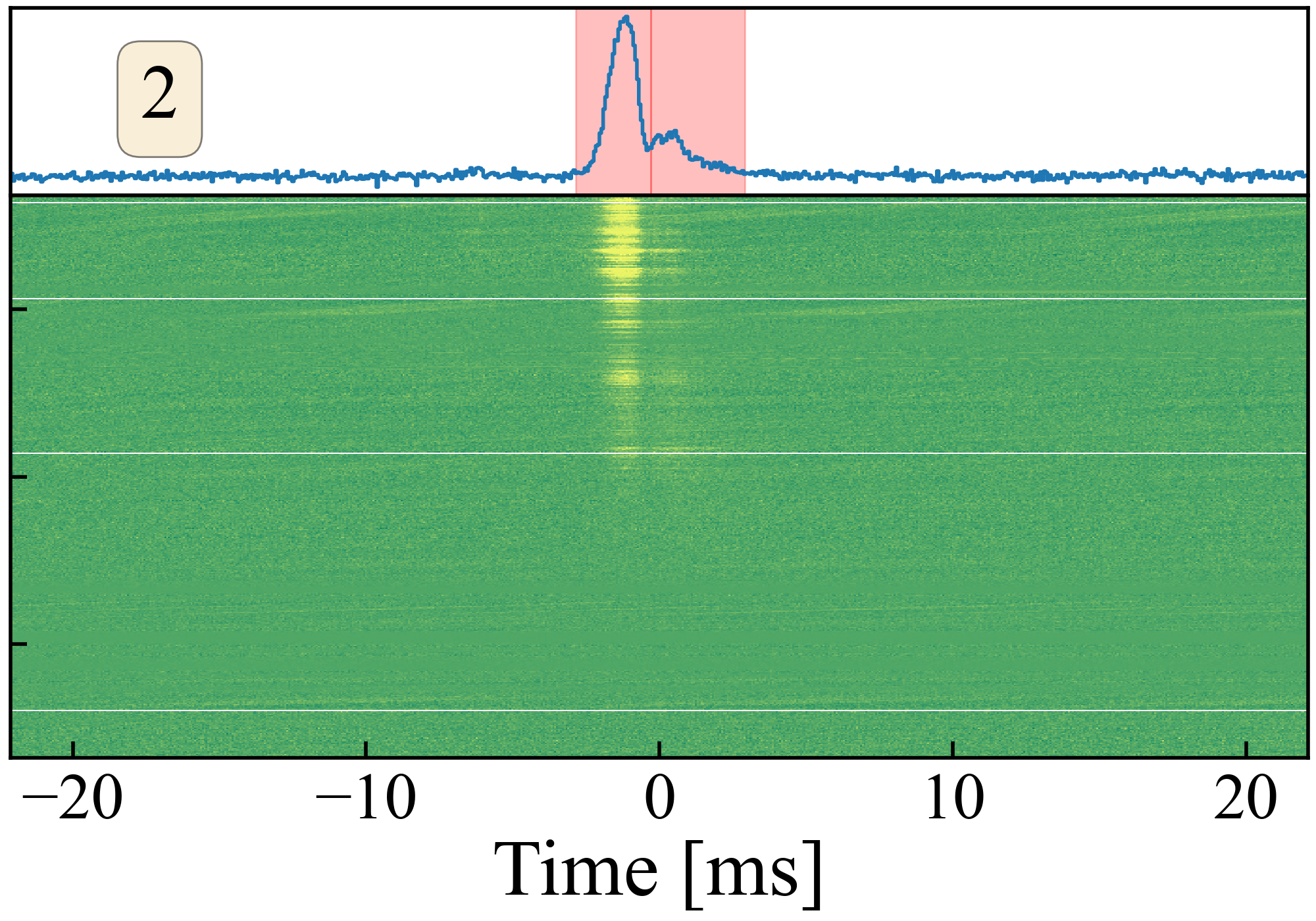}}
    \subfigure{\includegraphics[width=0.24\textwidth]{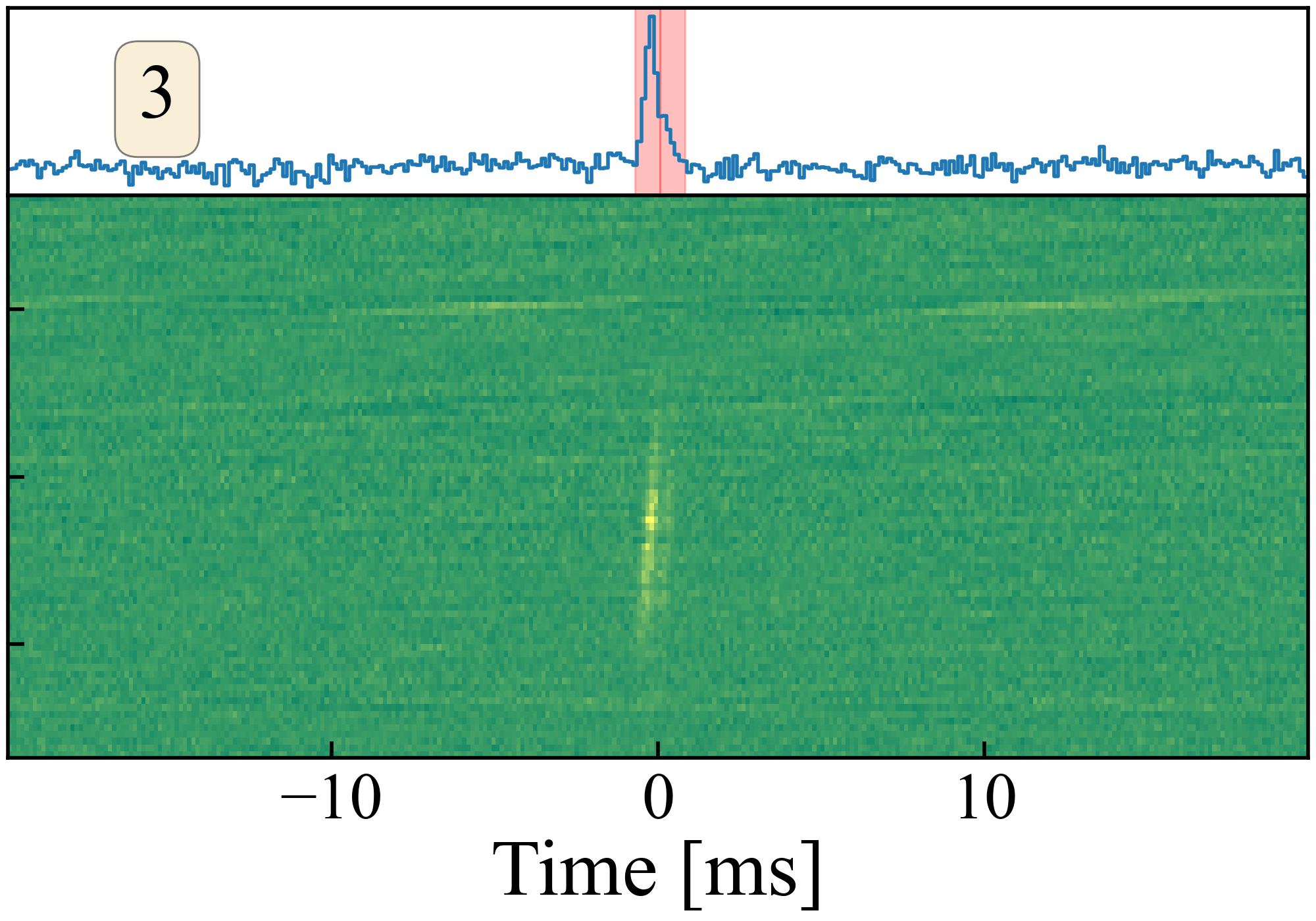}}
    \subfigure{\includegraphics[width=0.24\textwidth]{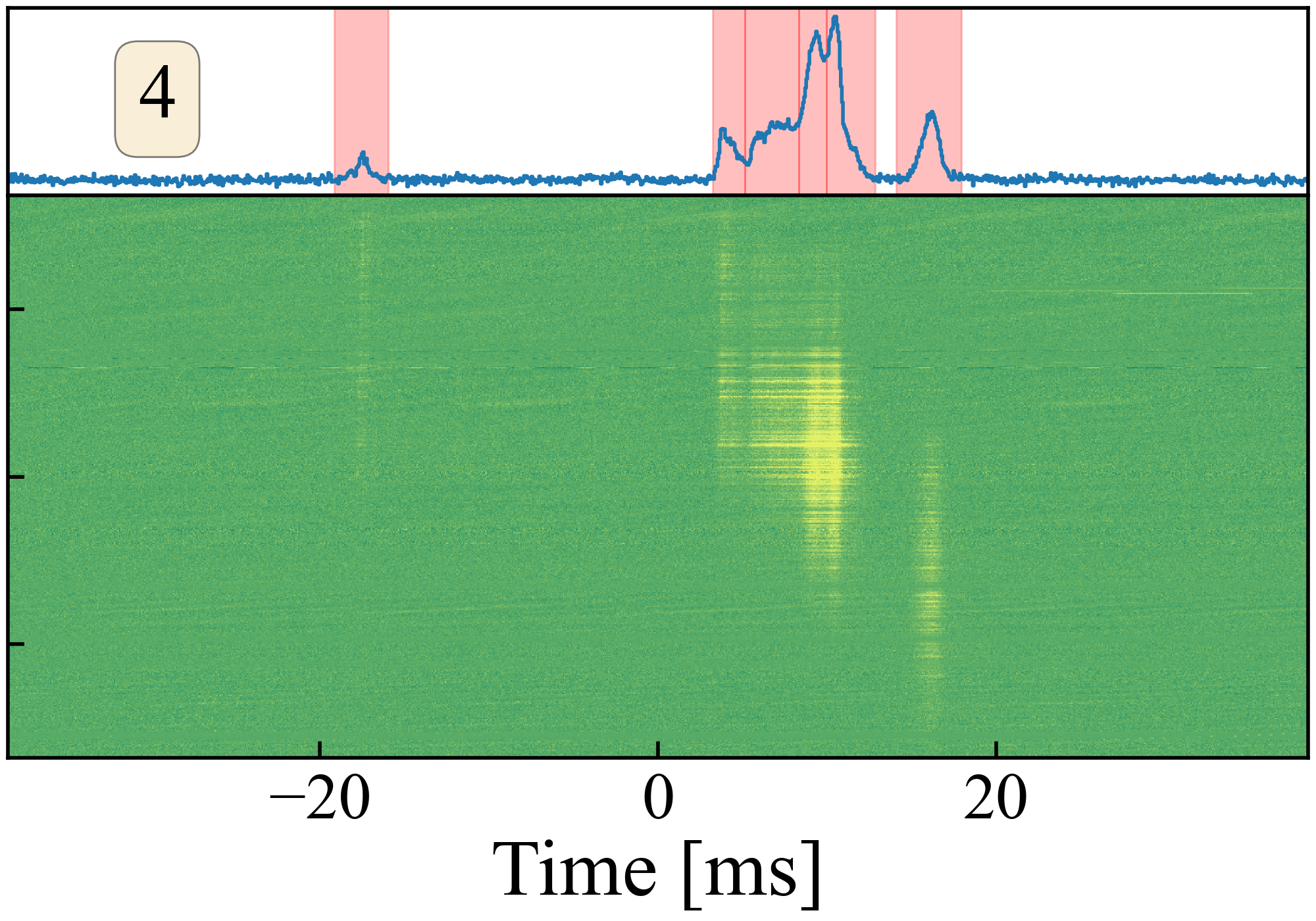}}
    
    \subfigure{\includegraphics[width=0.24\textwidth]{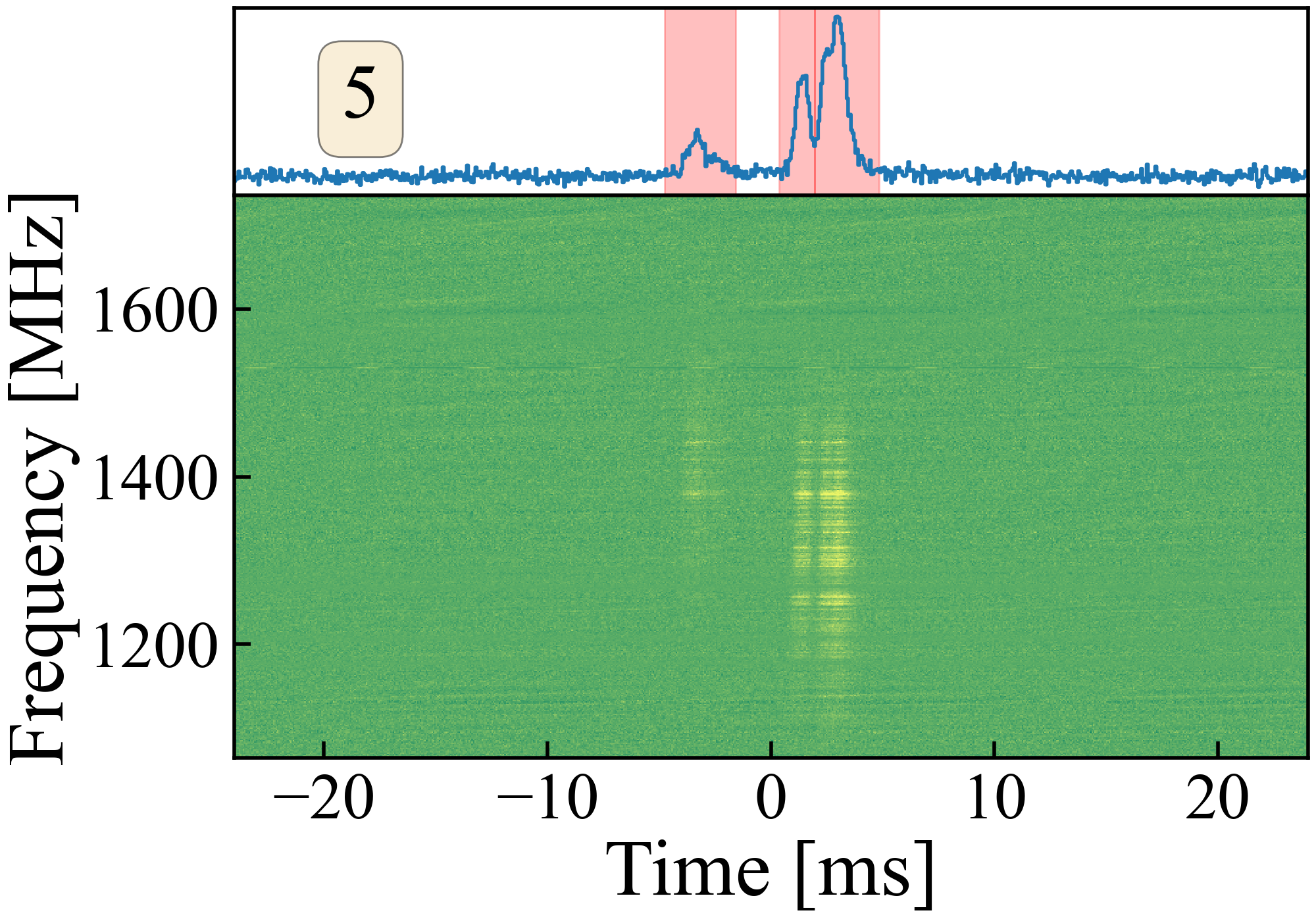}}
    \subfigure{\includegraphics[width=0.24\textwidth]{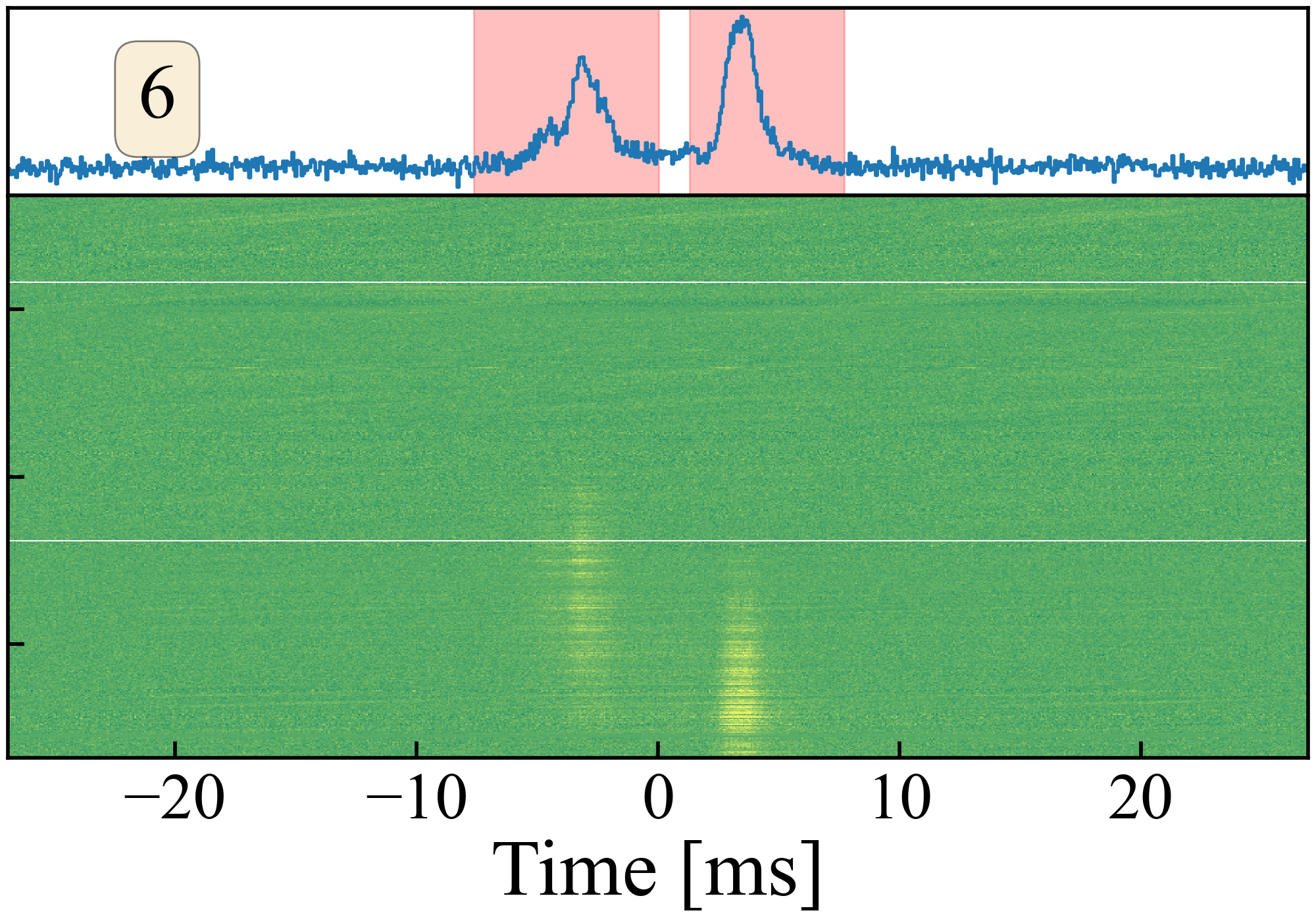}}
    \subfigure{\includegraphics[width=0.24\textwidth]{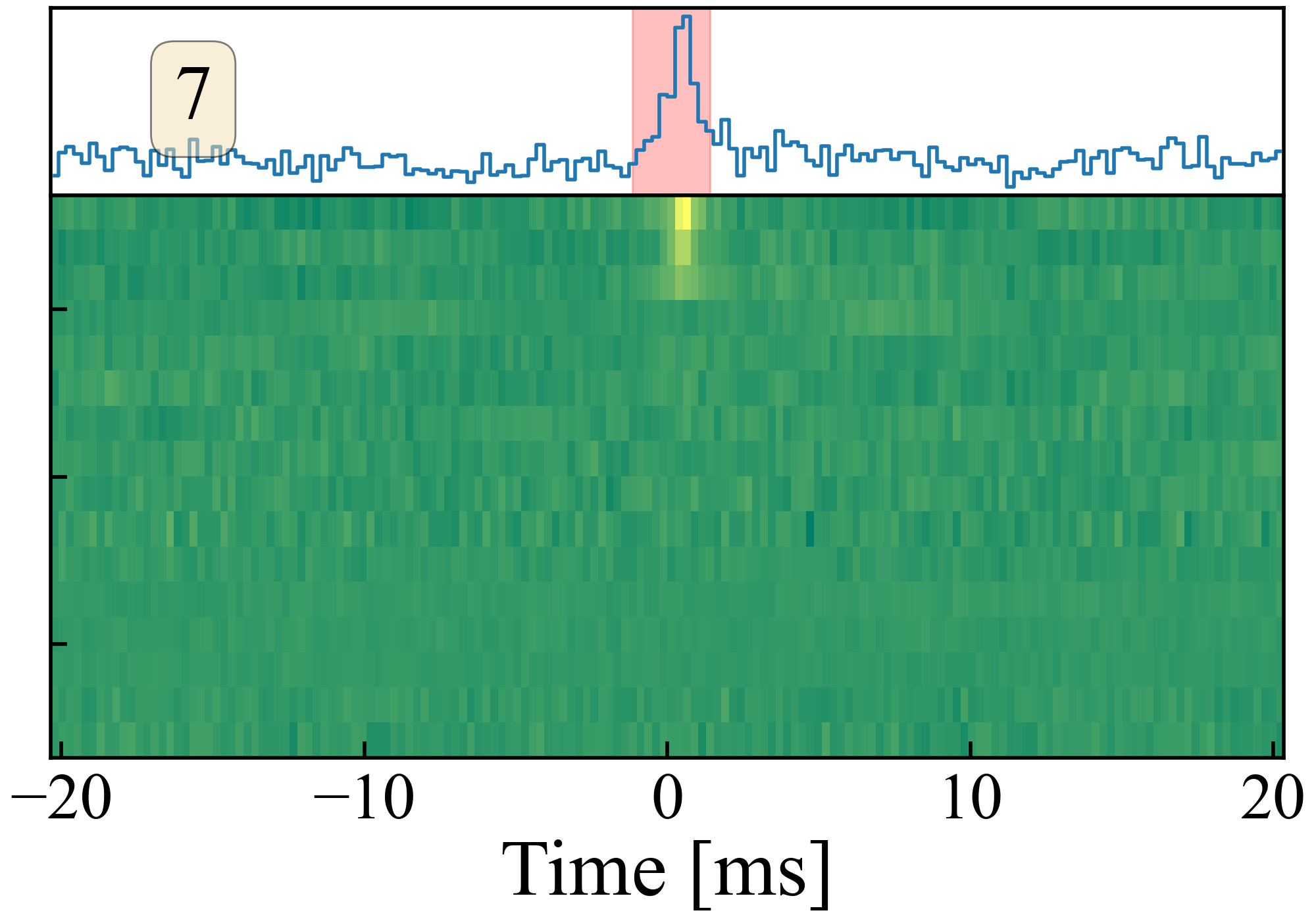}}
    \subfigure{\includegraphics[width=0.24\textwidth]{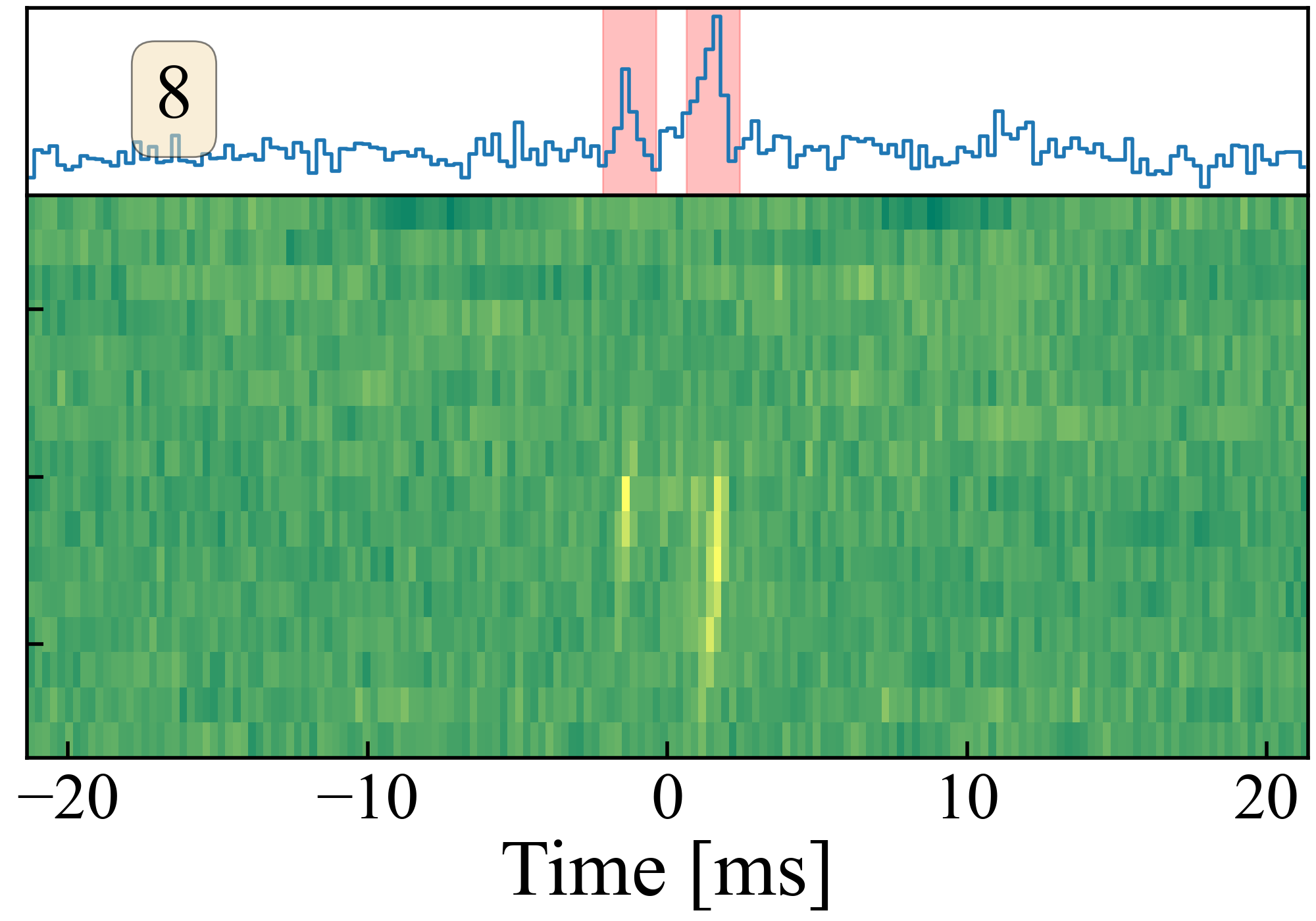}}
    
    \subfigure{\includegraphics[width=0.24\textwidth]{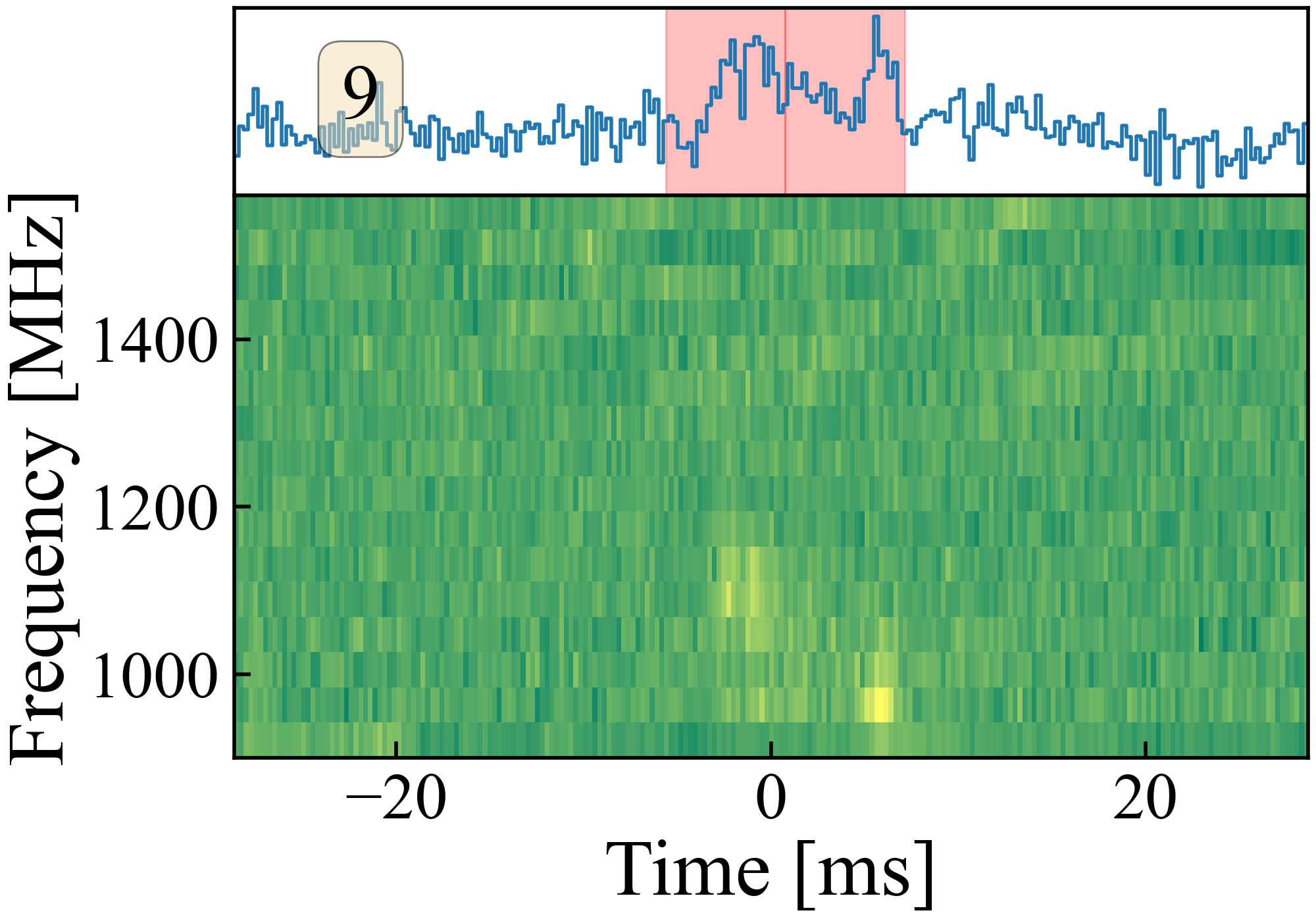}}
    \subfigure{\includegraphics[width=0.24\textwidth]{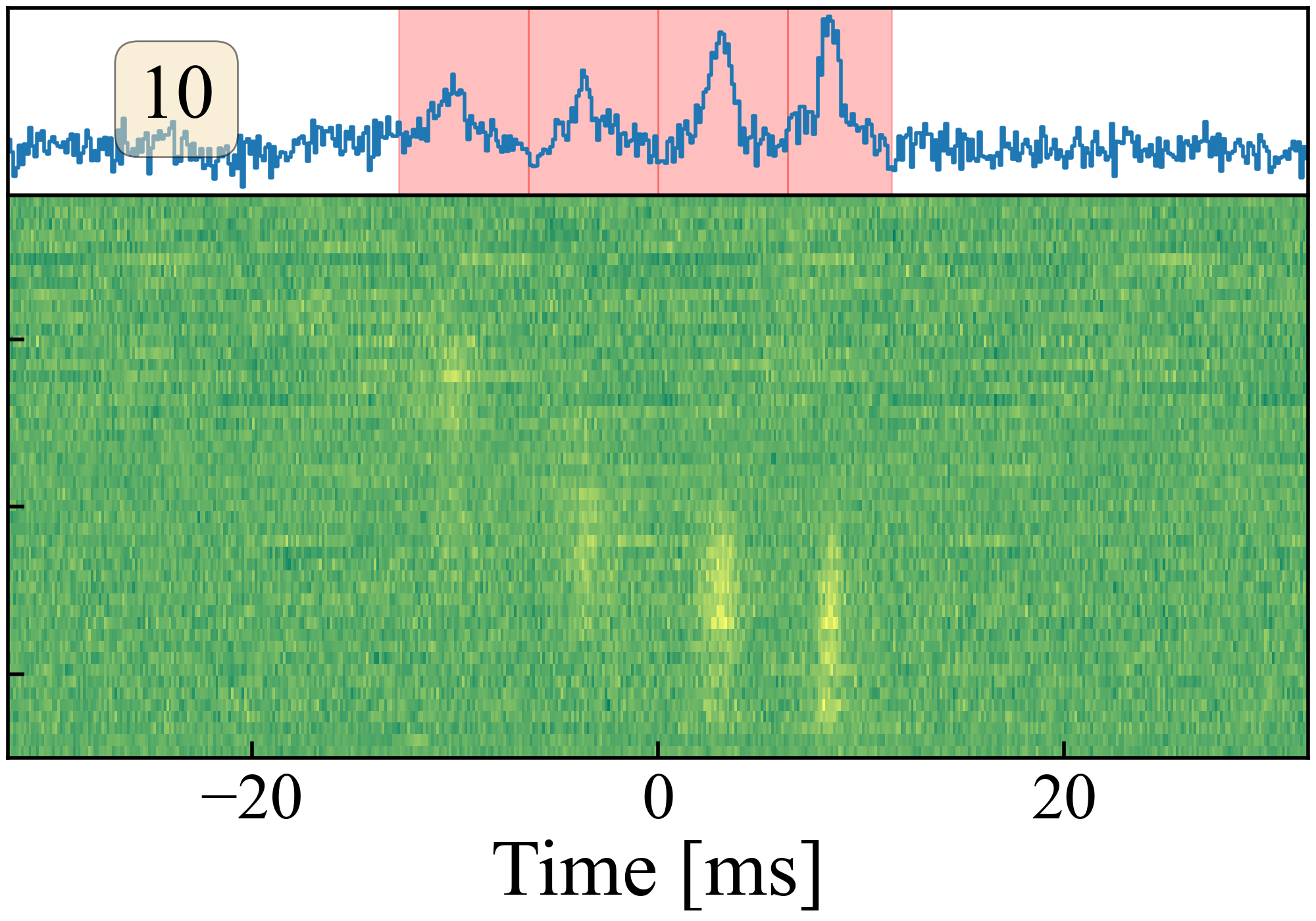}}
    \subfigure{\includegraphics[width=0.24\textwidth]{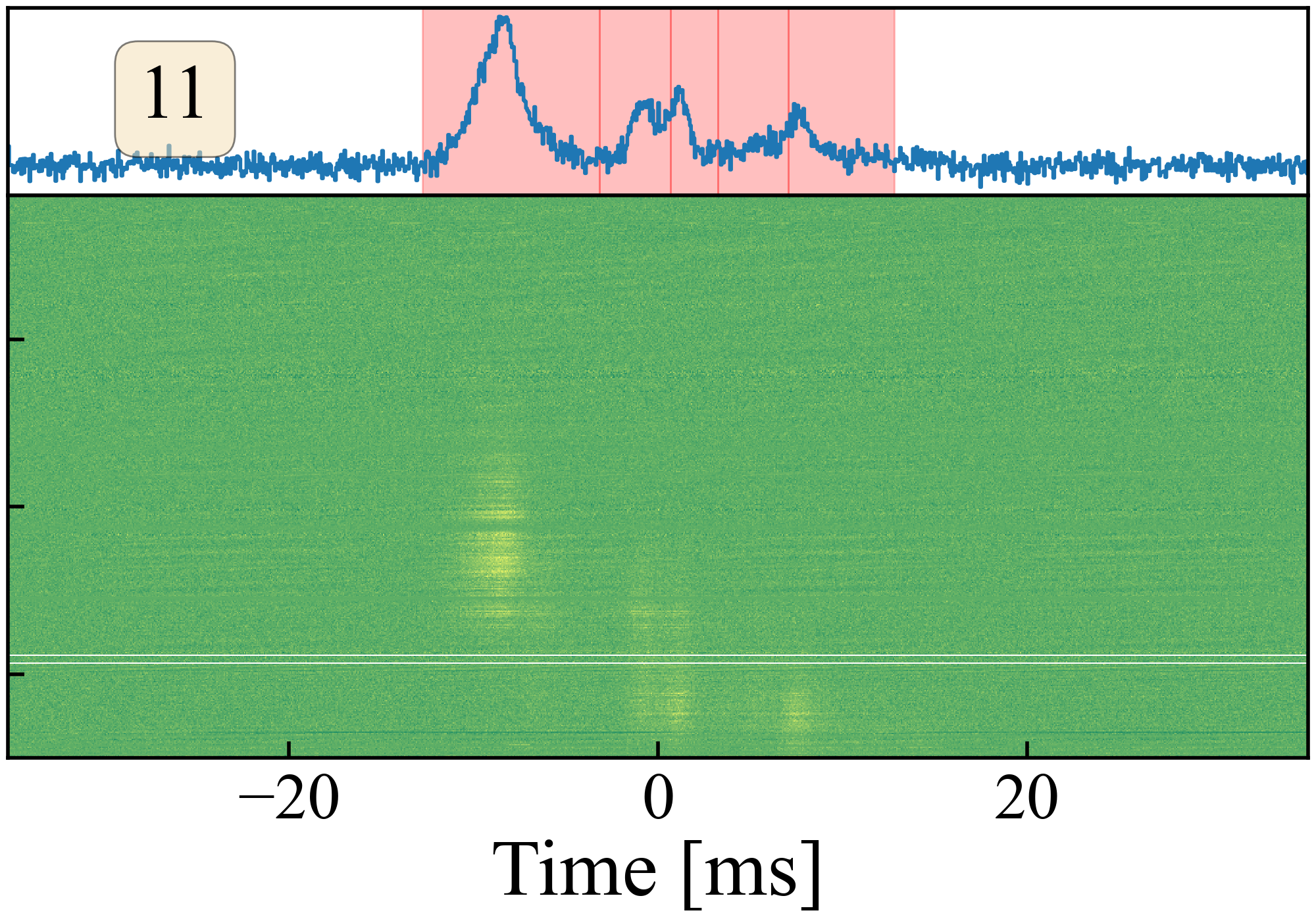}}
    \subfigure{\includegraphics[width=0.24\textwidth]{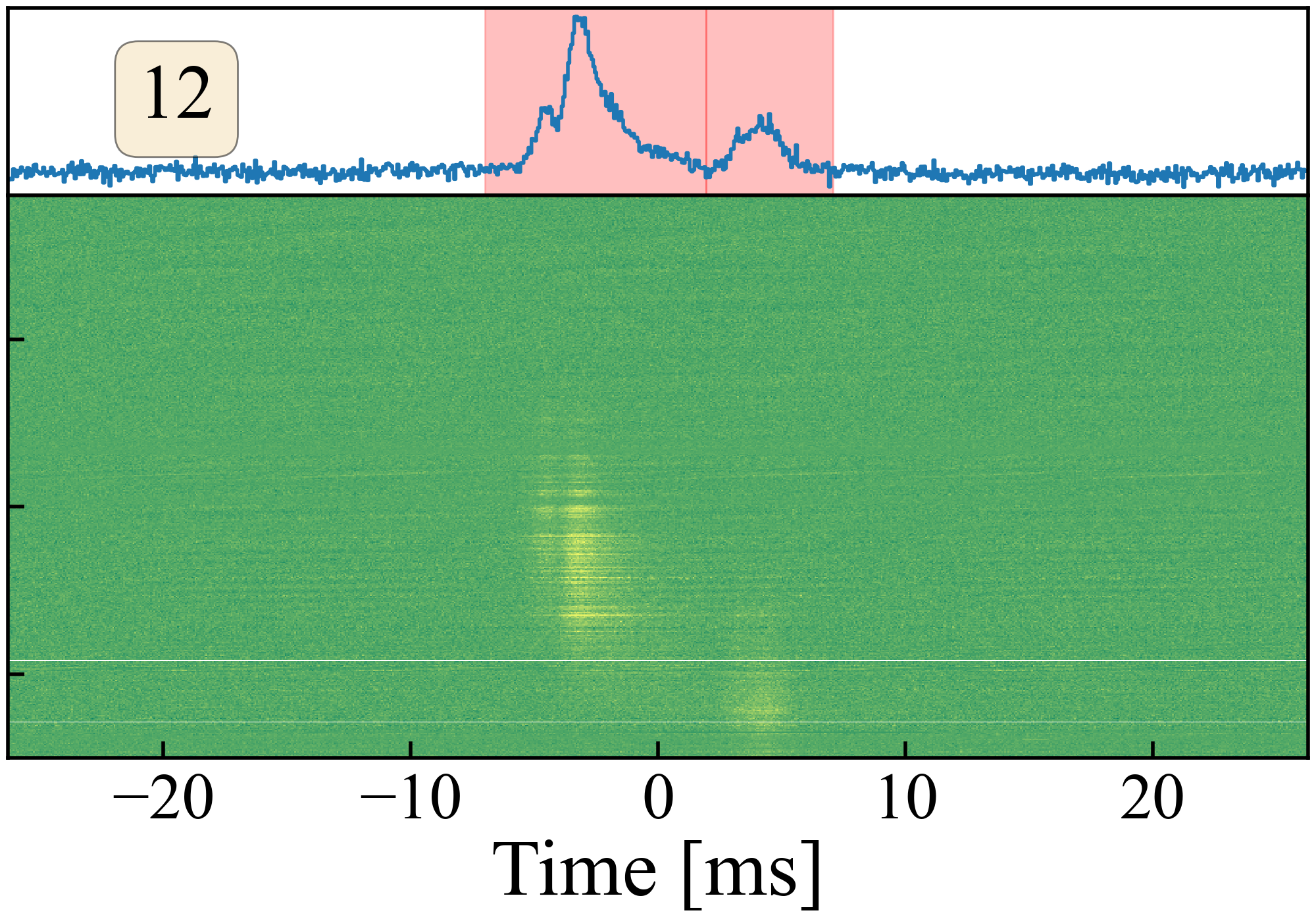}}
    
    \subfigure{\includegraphics[width=0.24\textwidth]{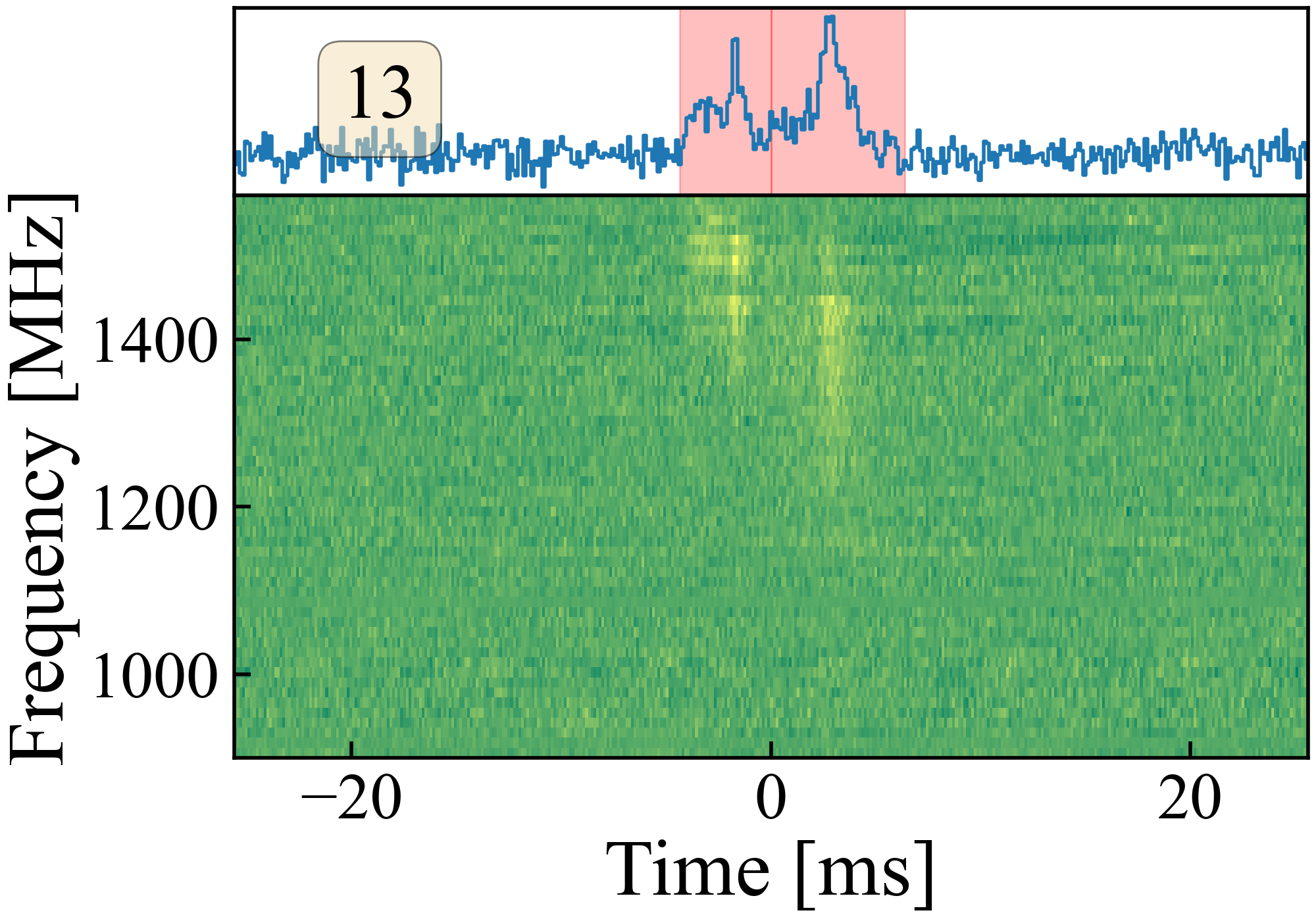}}
    \subfigure{\includegraphics[width=0.24\textwidth]{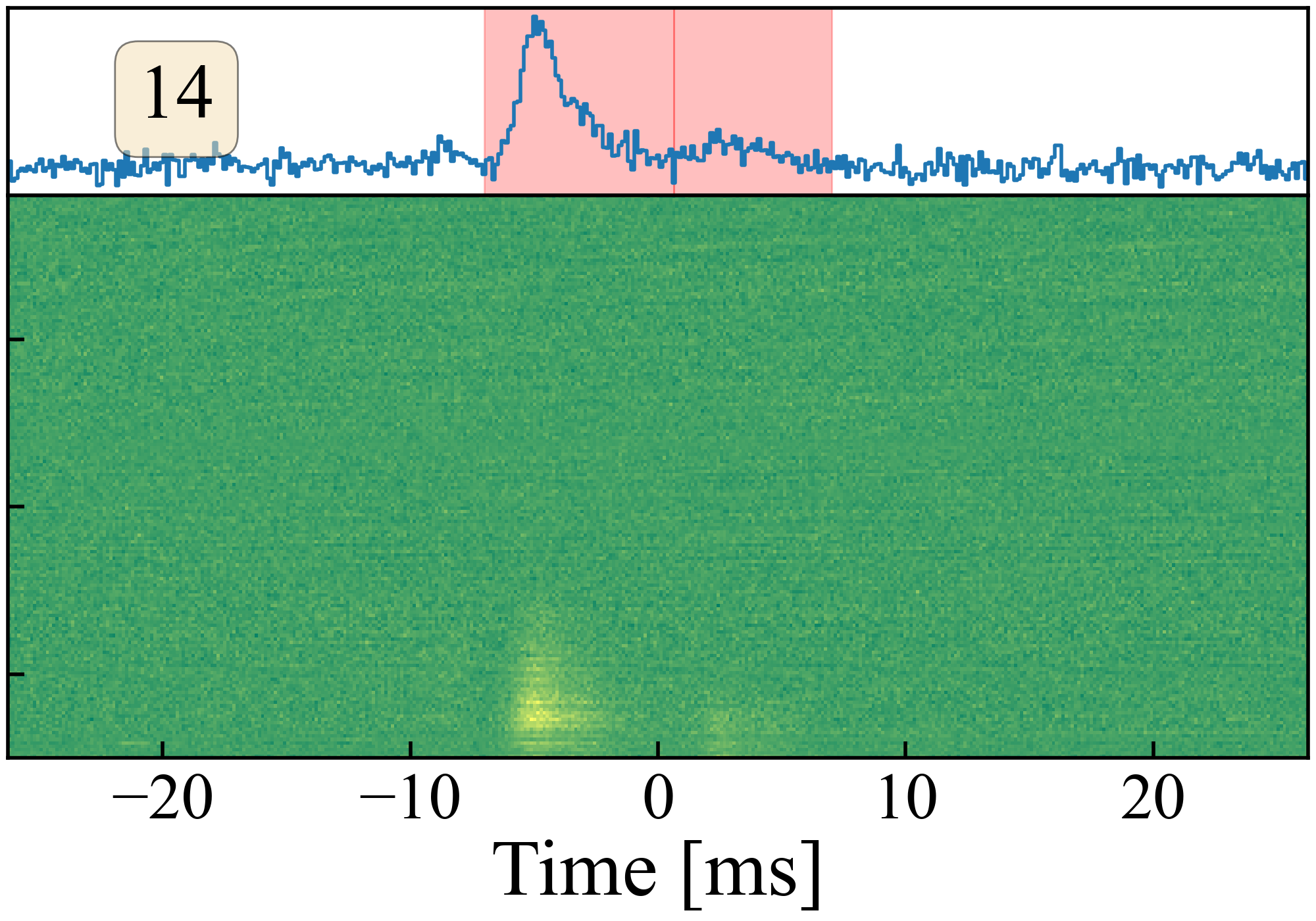}}
    \subfigure{\includegraphics[width=0.24\textwidth]{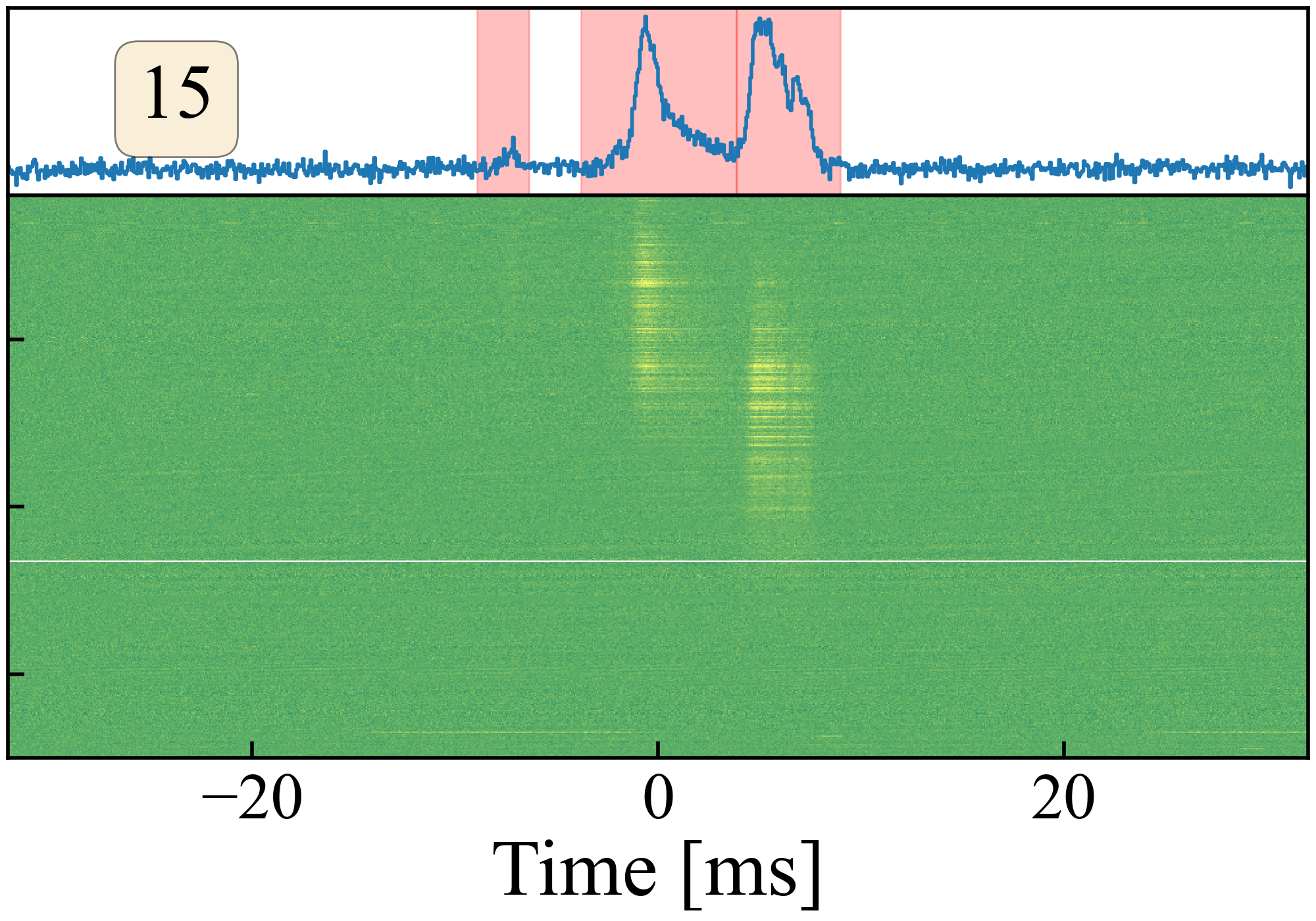}}
    \subfigure{\includegraphics[width=0.24\textwidth]{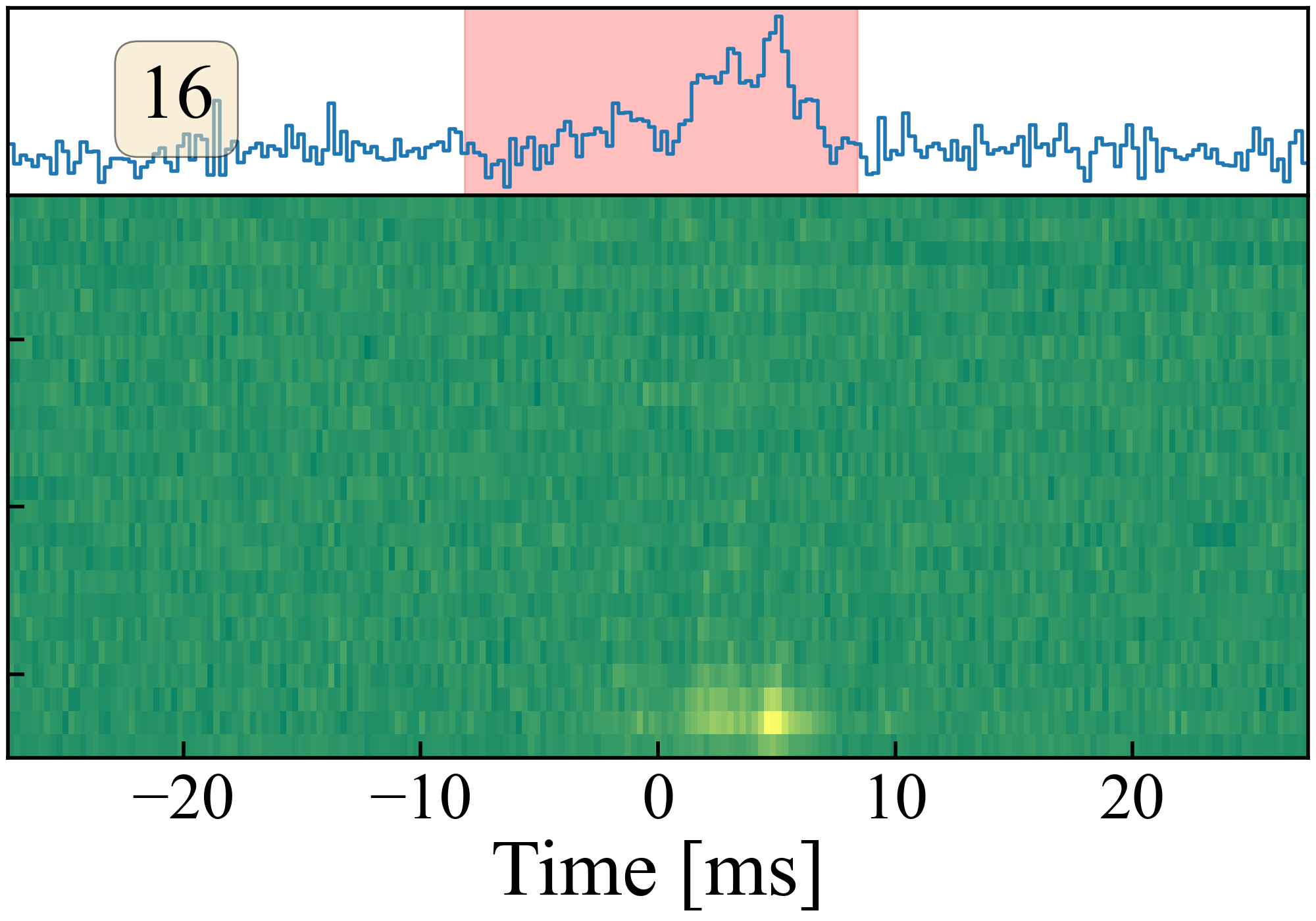}}

    \subfigure{\includegraphics[width=0.24\textwidth]{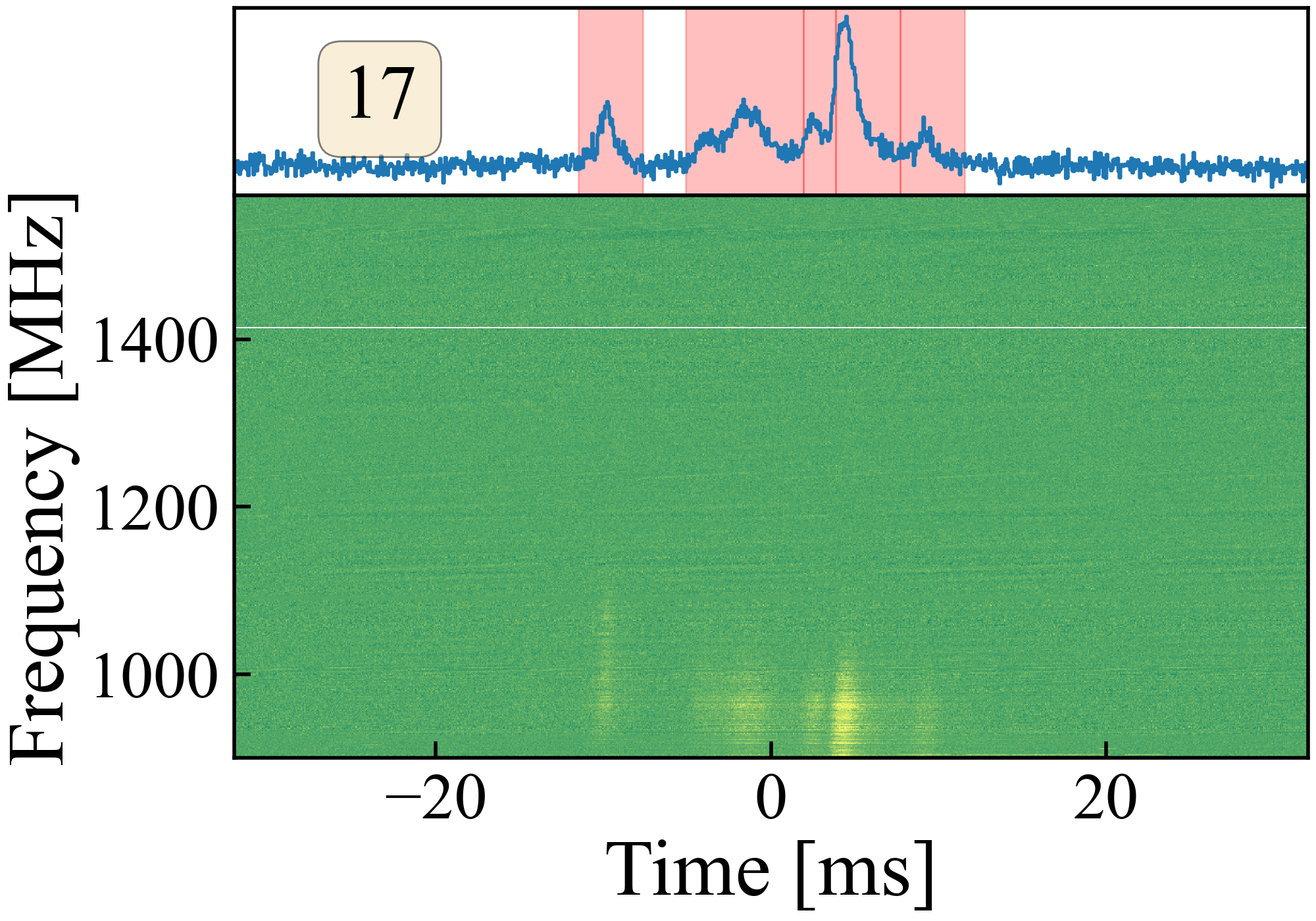}}
    \subfigure{\includegraphics[width=0.24\textwidth]{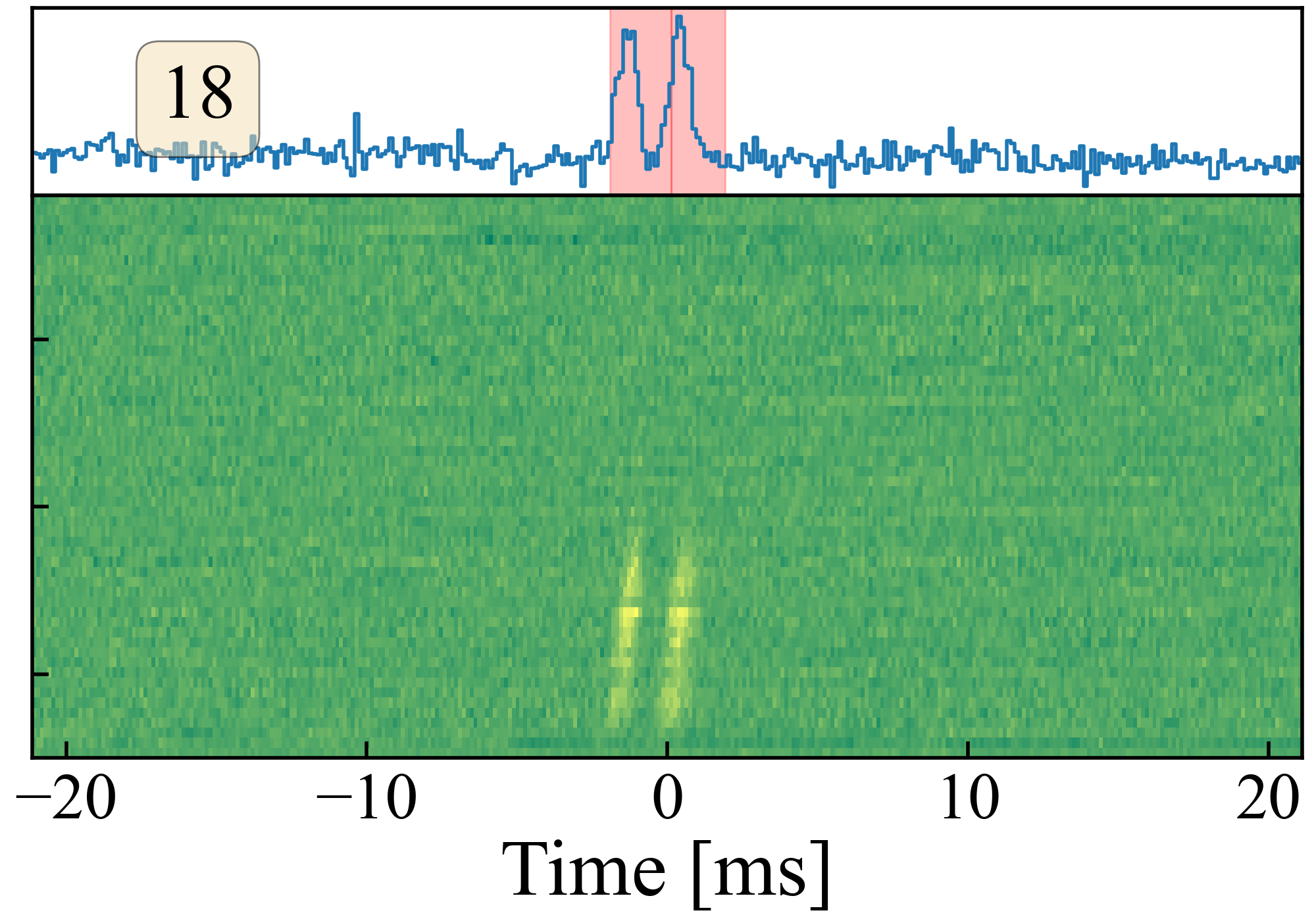}}
    \subfigure{\includegraphics[width=0.24\textwidth]{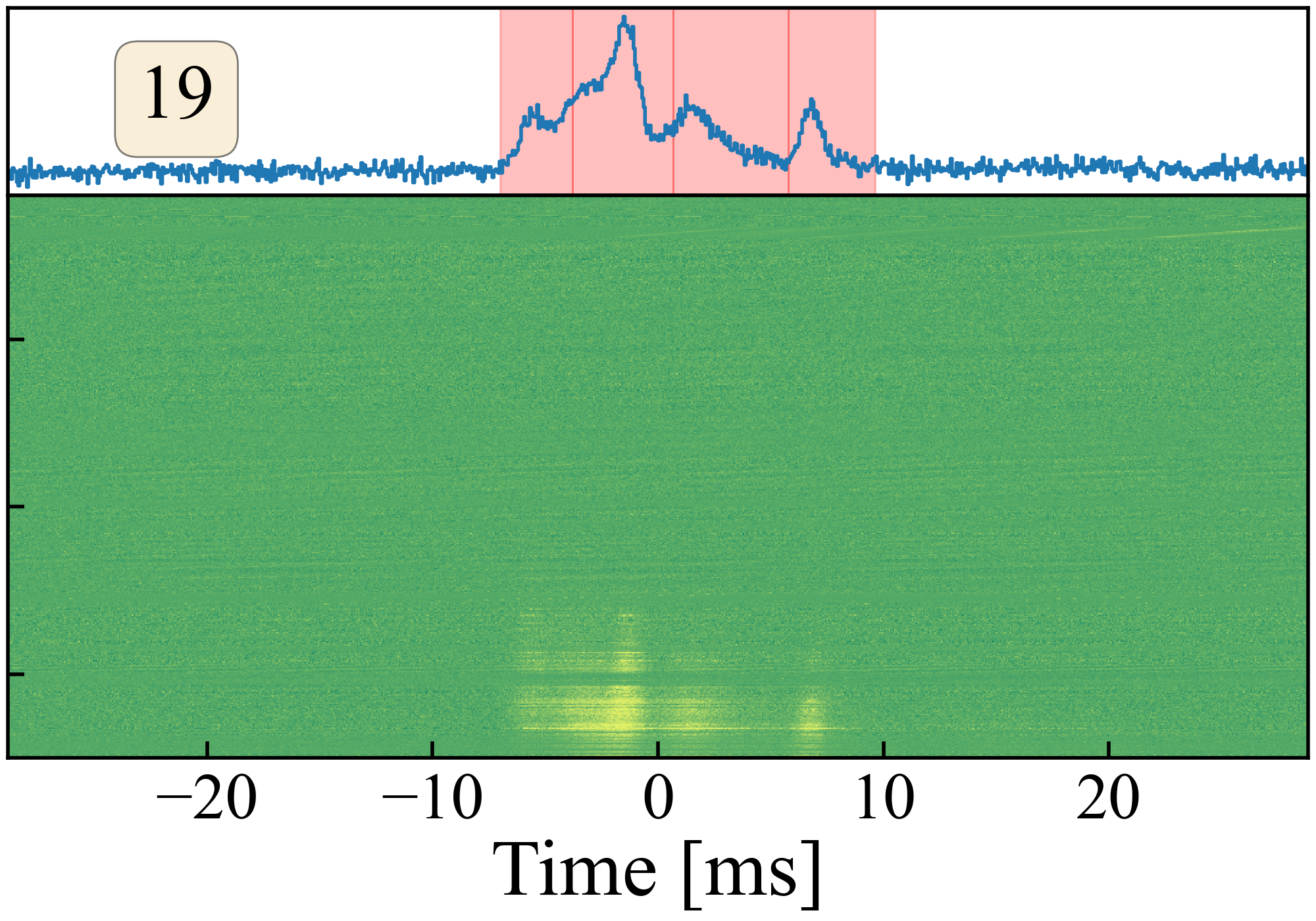}}
    \subfigure{\includegraphics[width=0.24\textwidth]{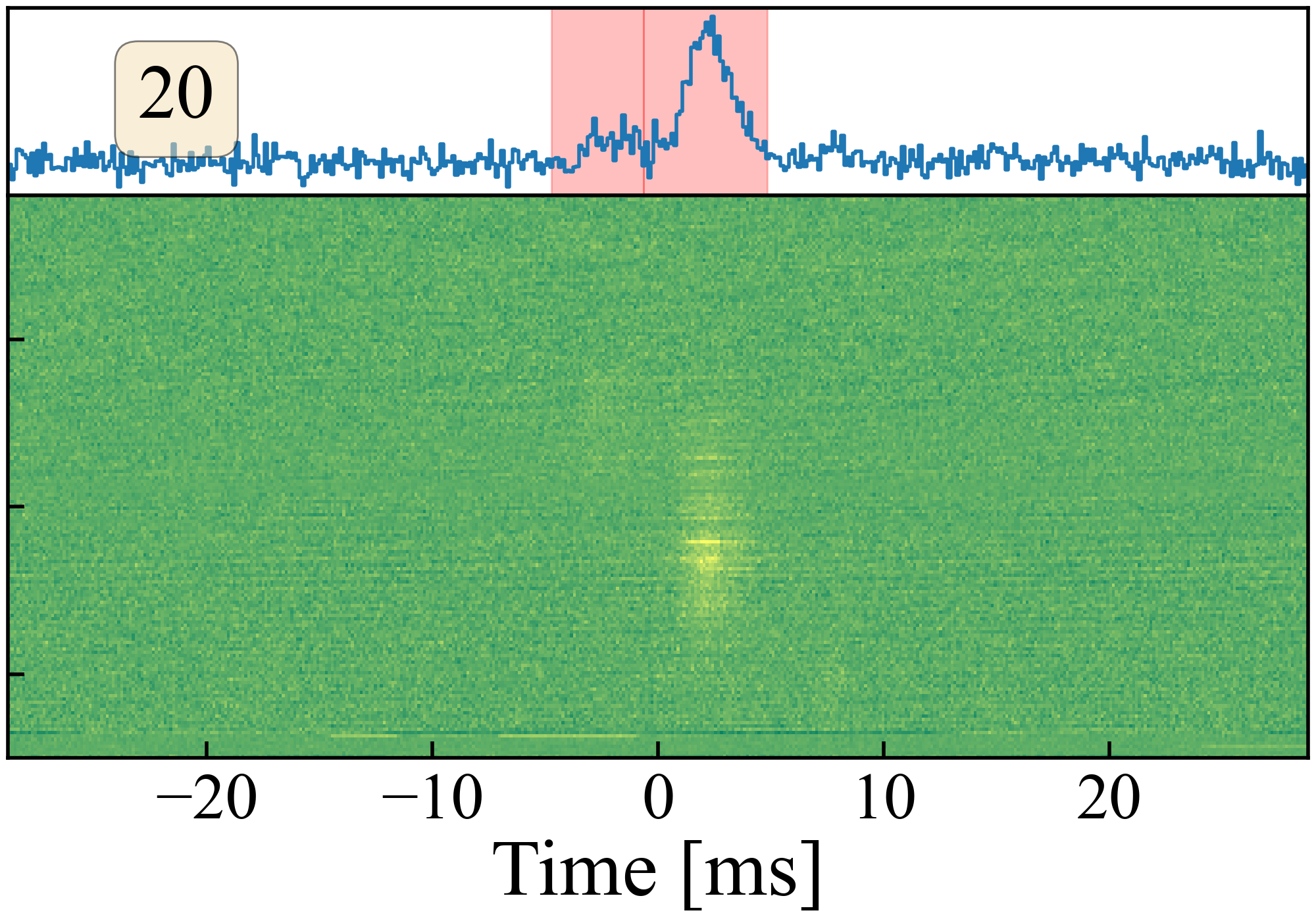}}
    
    \subfigure{\includegraphics[width=0.24\textwidth]{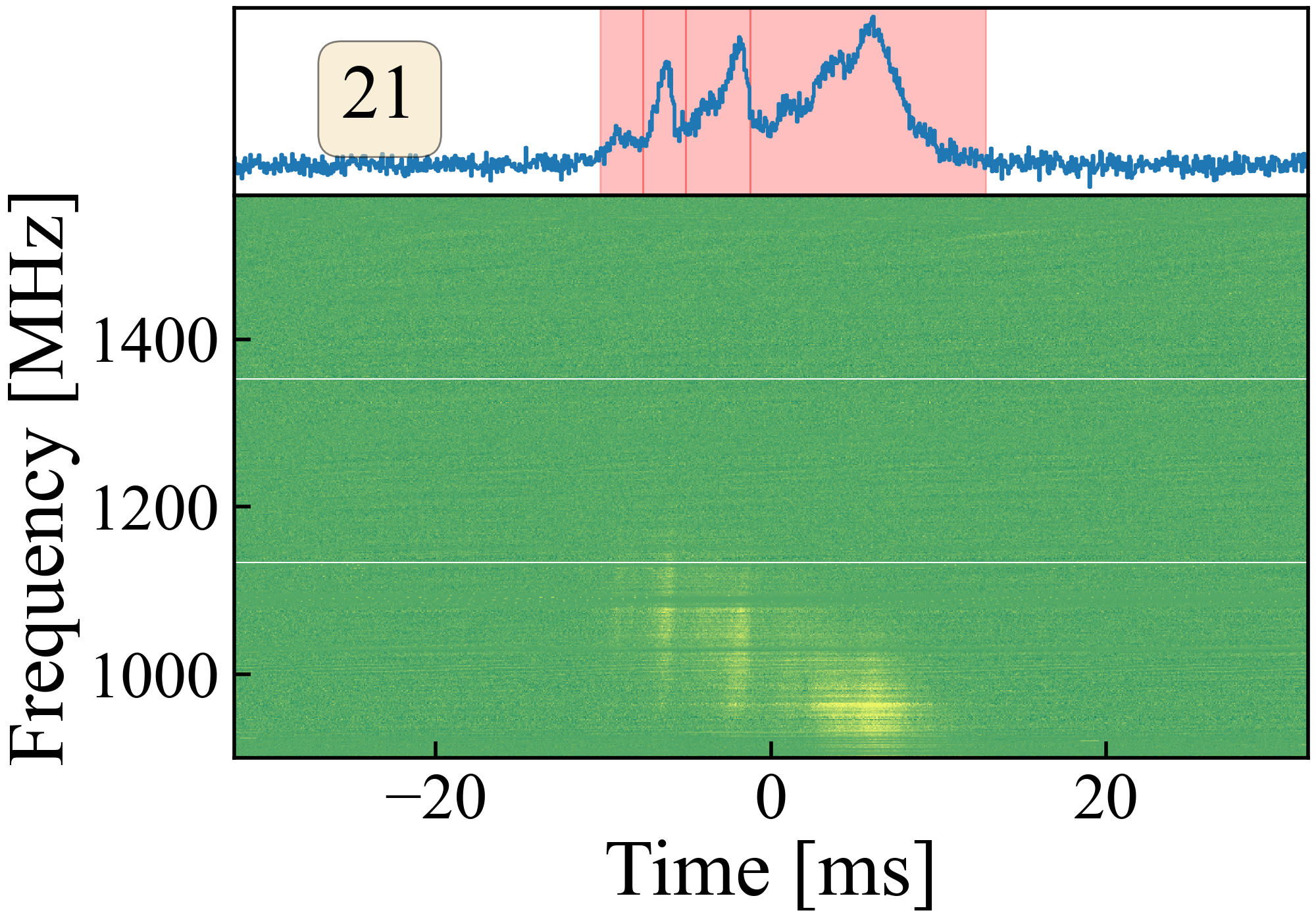}}
    \subfigure{\includegraphics[width=0.24\textwidth]{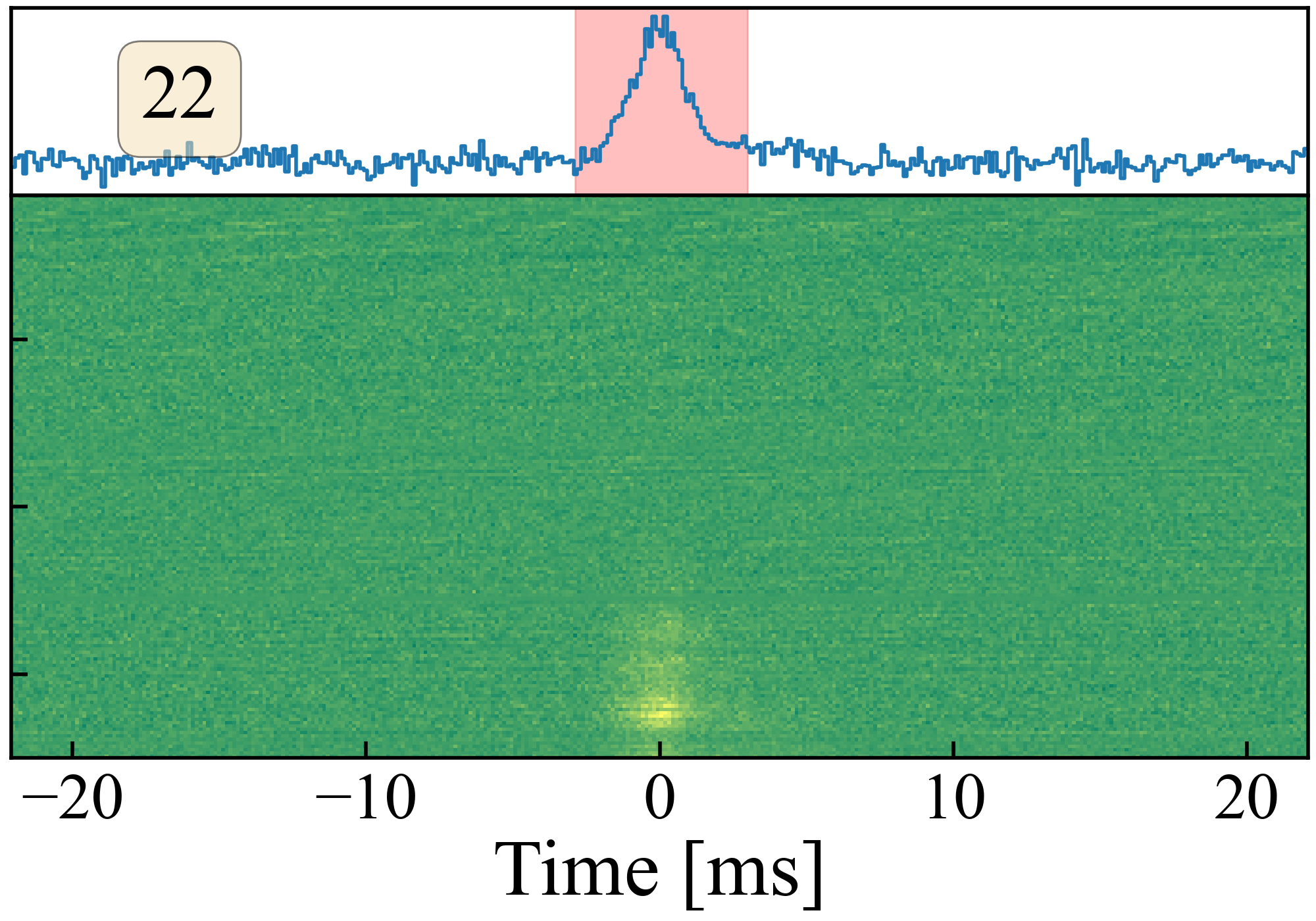}}
    \subfigure{\includegraphics[width=0.24\textwidth]{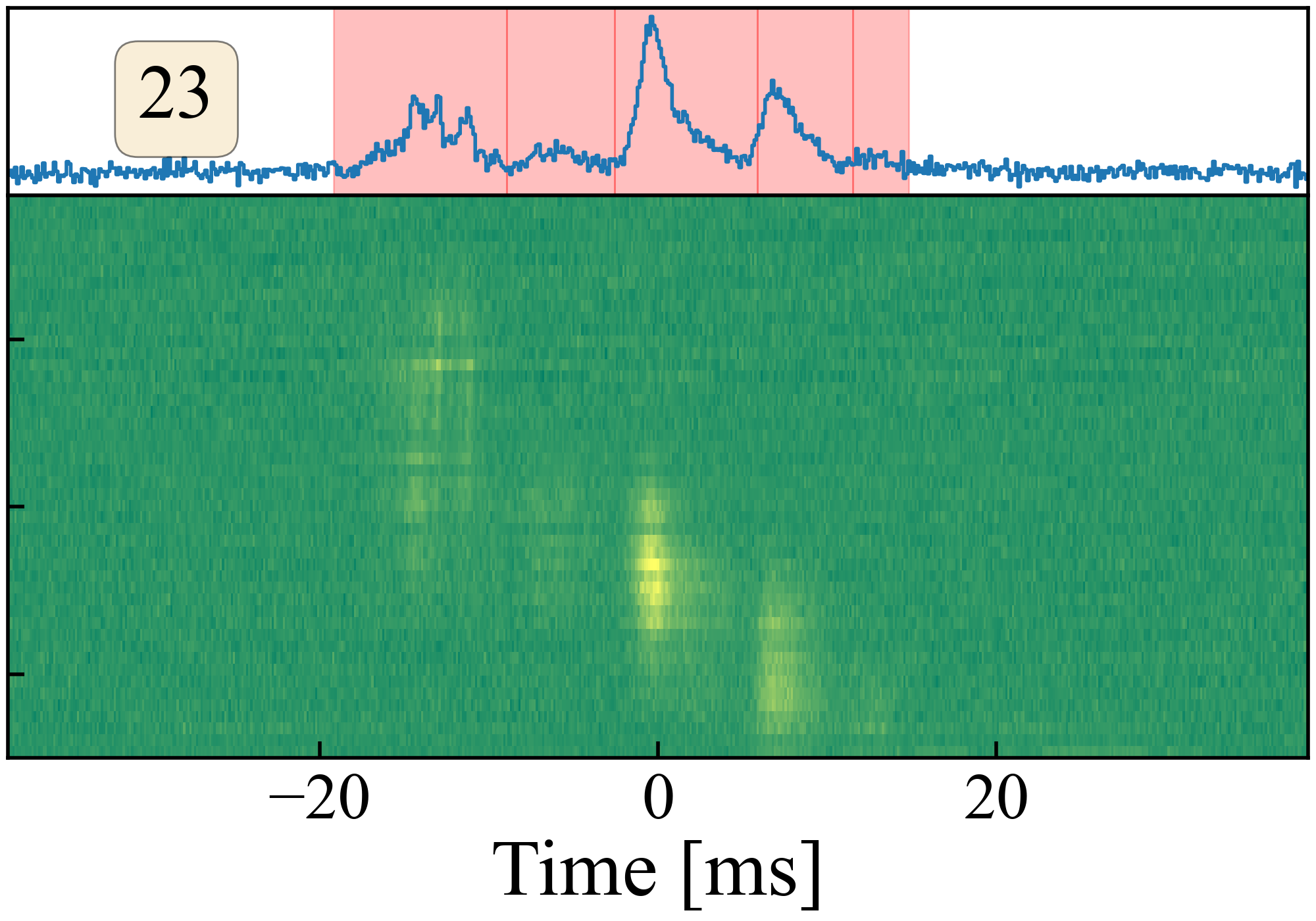}}
    \subfigure{\includegraphics[width=0.24\textwidth]{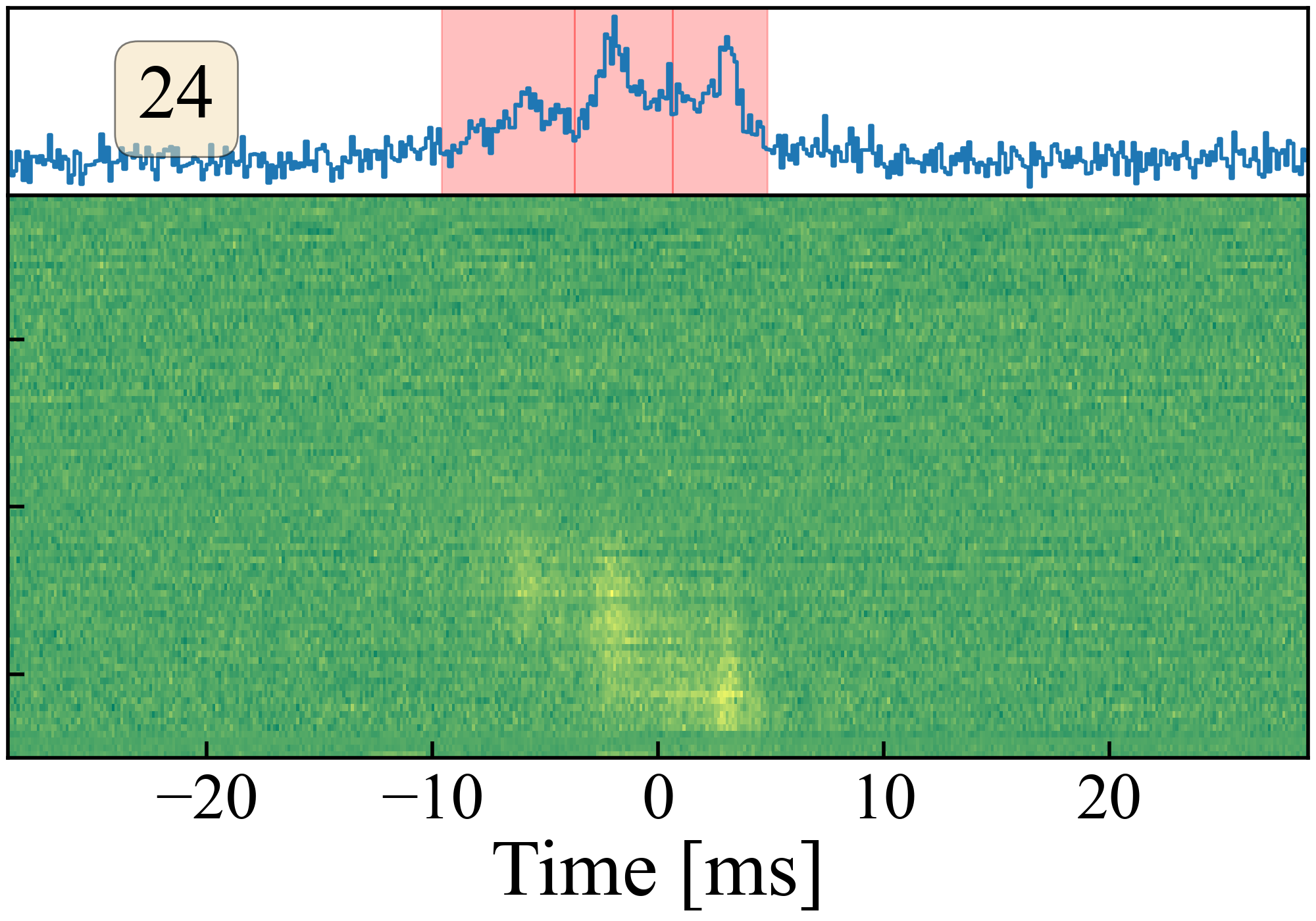}}

    \caption{Dynamic spectra (or ``waterfall'' plots) for all the bursts from FRB\,20220912A detected using the Allen Telescope Array, the frequency-averaged pulse profiles, and the time-averaged spectra. The red-shaded regions in the time series plots denote the time span of the defined sub-bursts, with red vertical lines demarcating adjacent sub-bursts.}
\end{figure*}

\begin{figure*}
    \ContinuedFloat
    \subfigure{\includegraphics[width=0.24\textwidth]{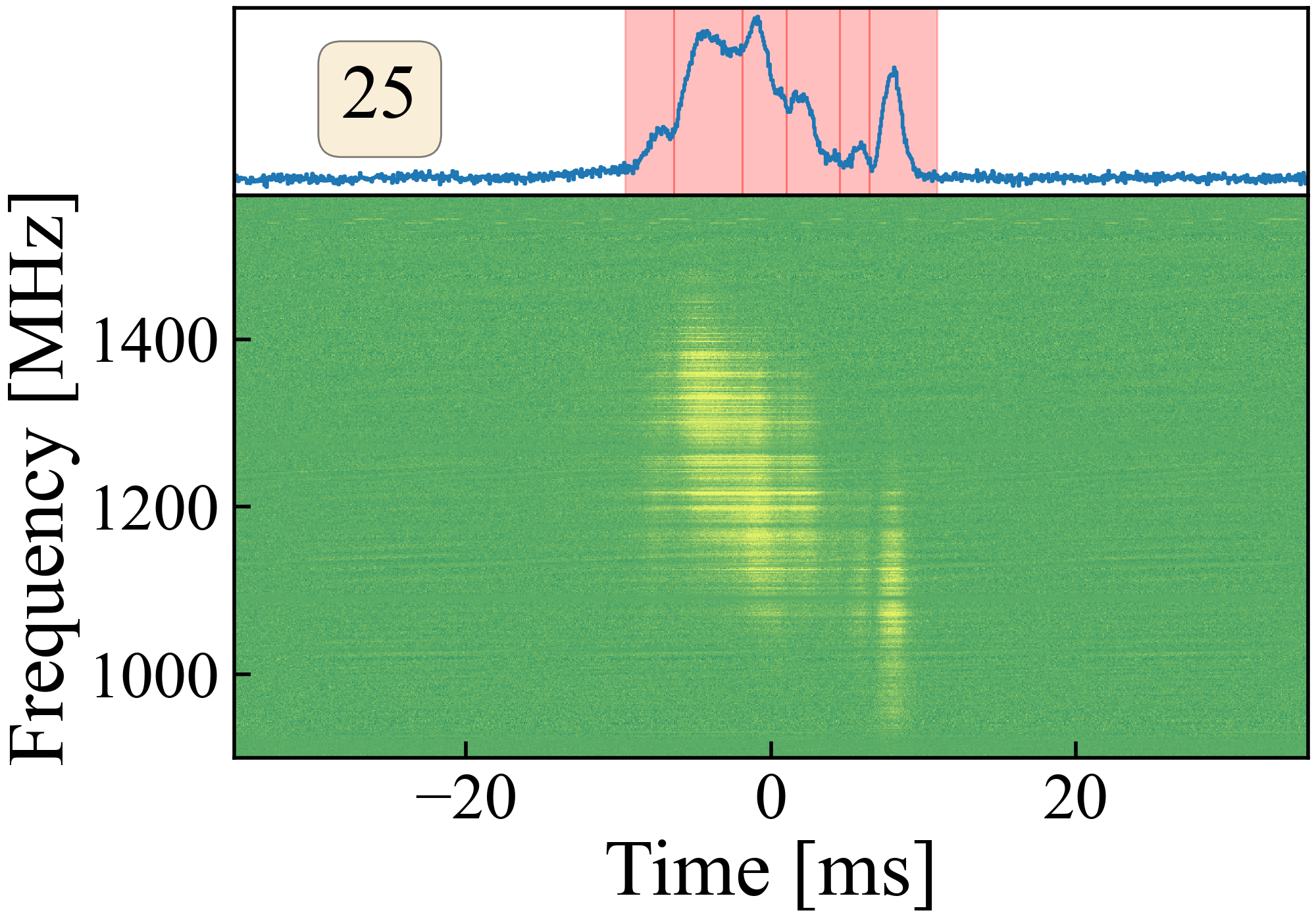}}
    \subfigure{\includegraphics[width=0.24\textwidth]{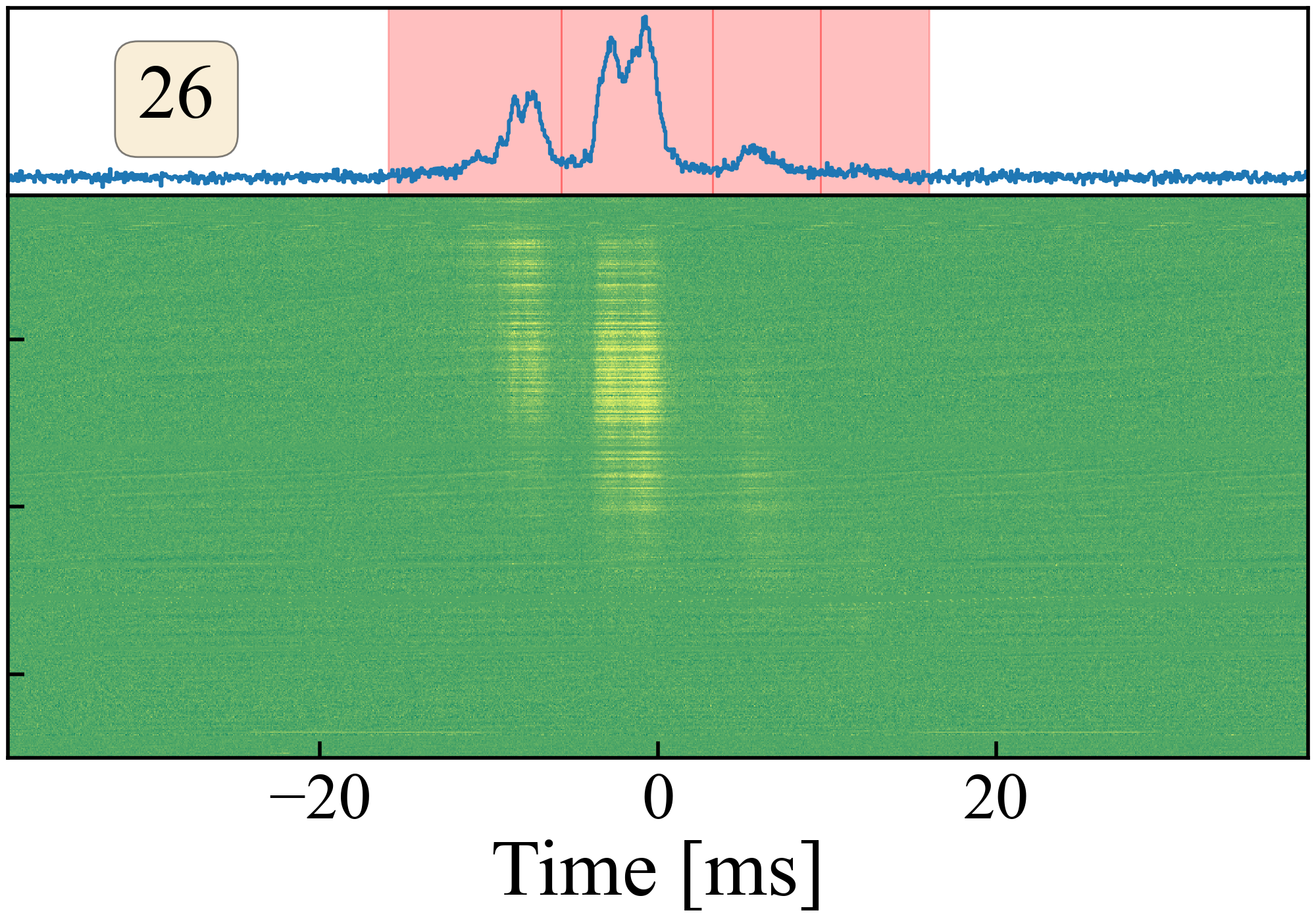}}
    \subfigure{\includegraphics[width=0.24\textwidth]{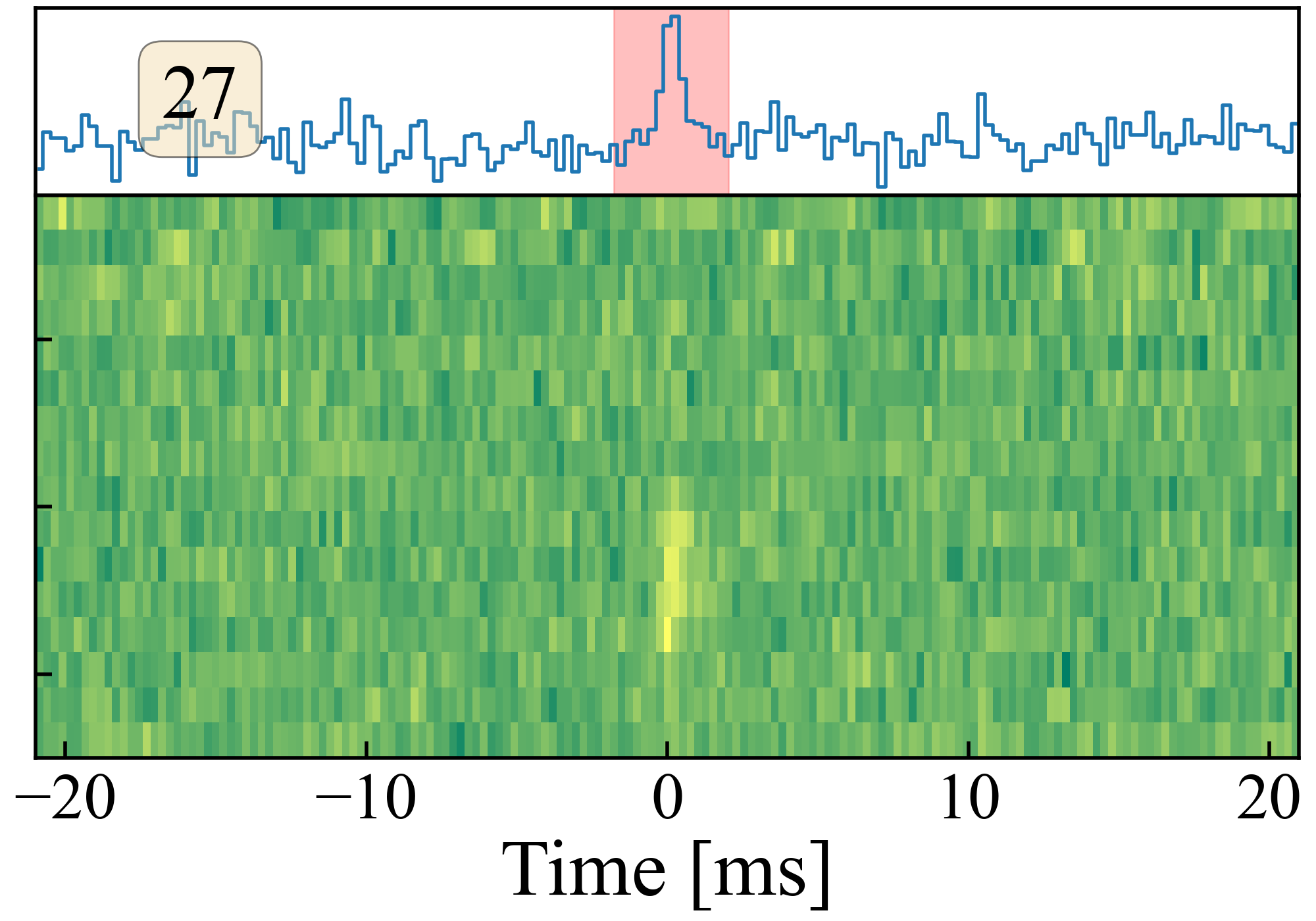}}
    \subfigure{\includegraphics[width=0.24\textwidth]{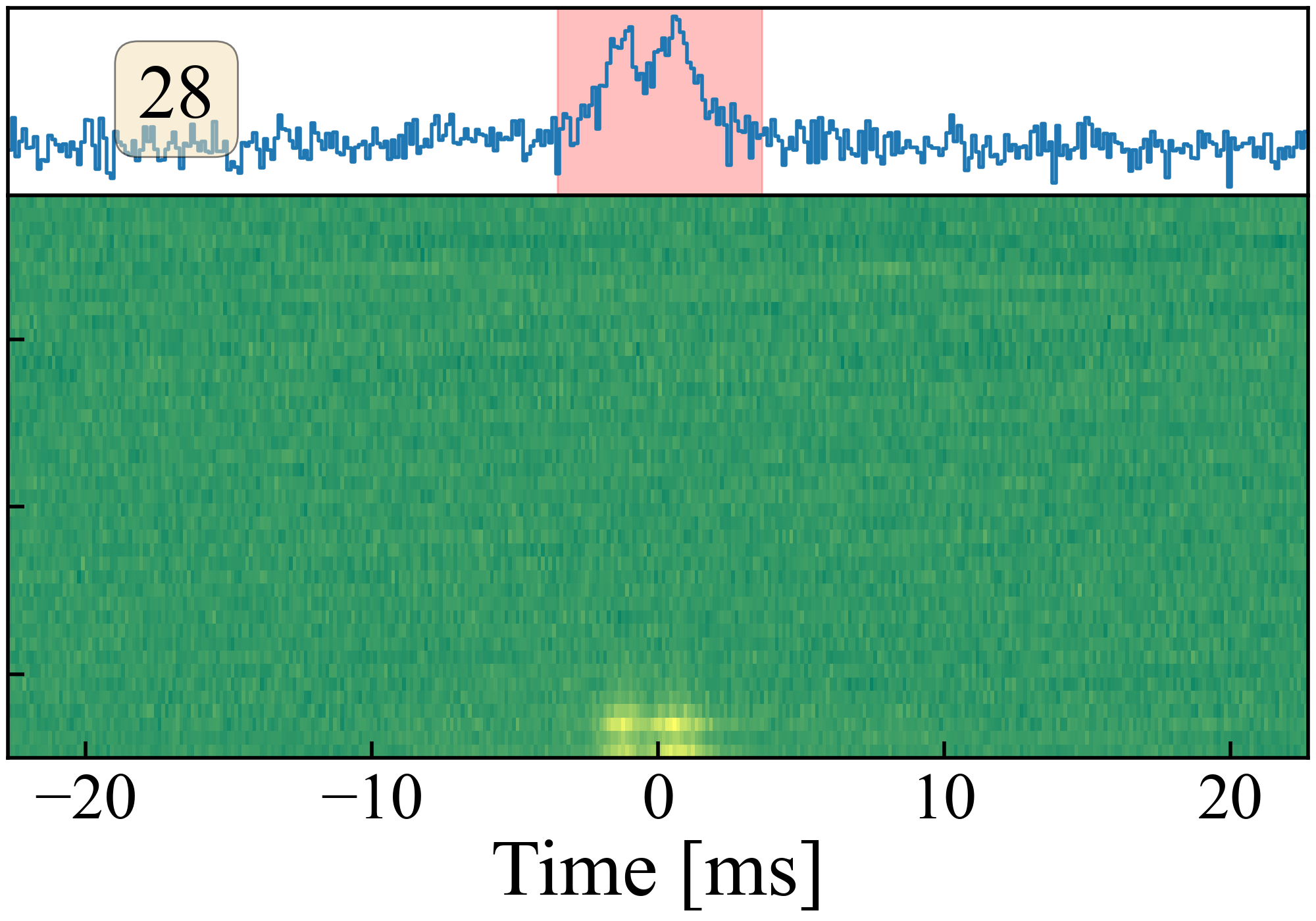}}
    \subfigure{\includegraphics[width=0.24\textwidth]{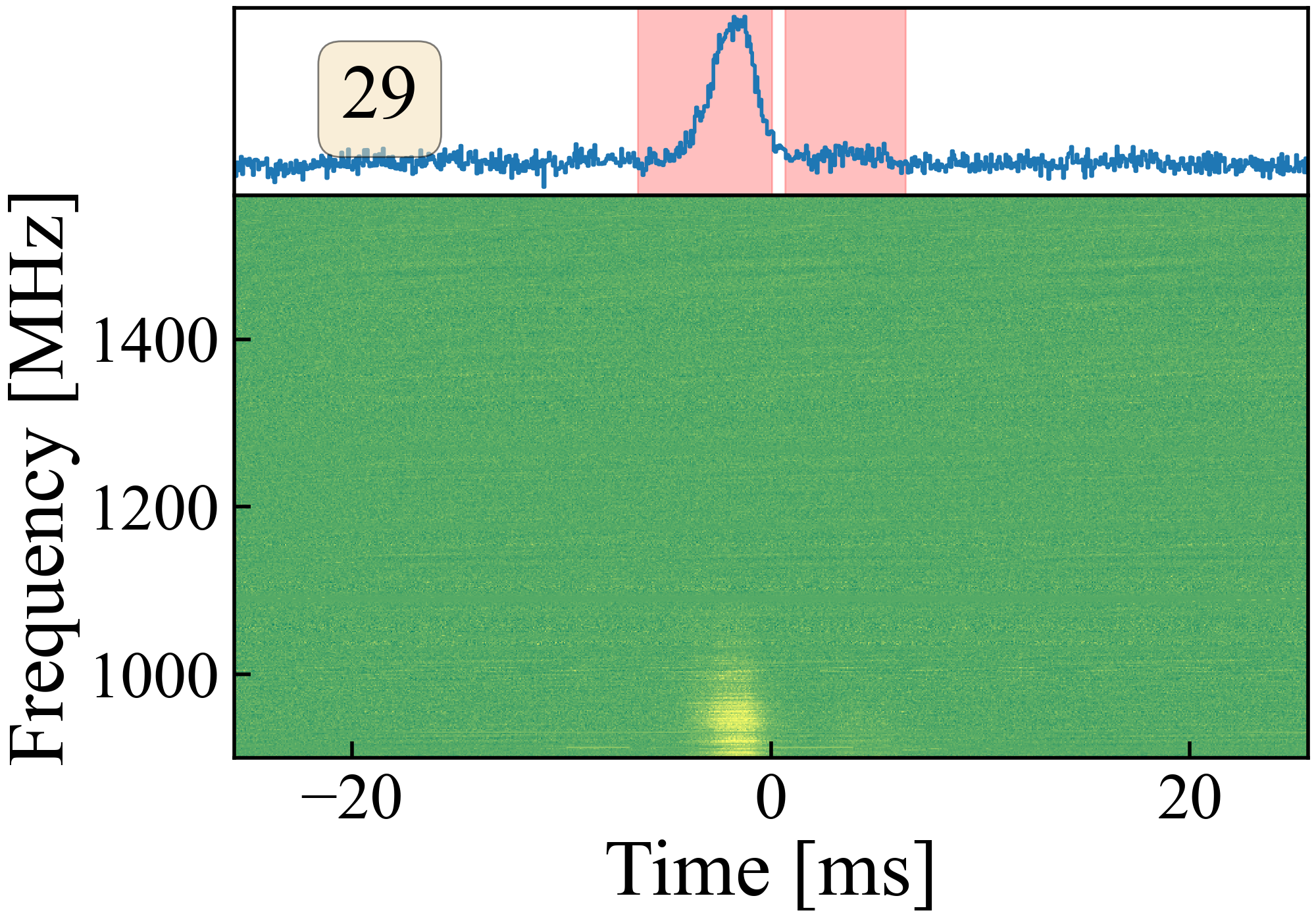}}
    \subfigure{\includegraphics[width=0.24\textwidth]{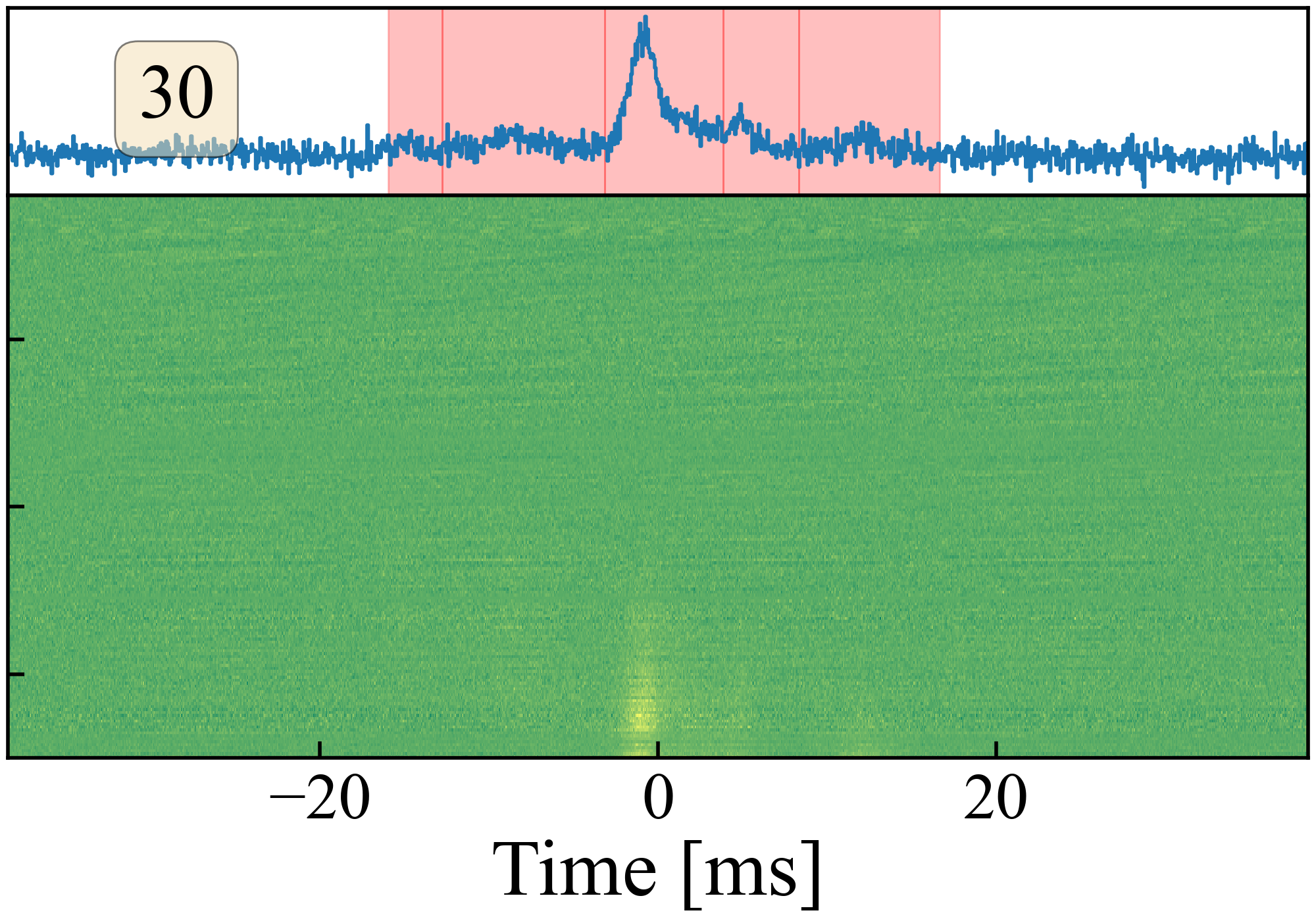}}
    \subfigure{\includegraphics[width=0.24\textwidth]{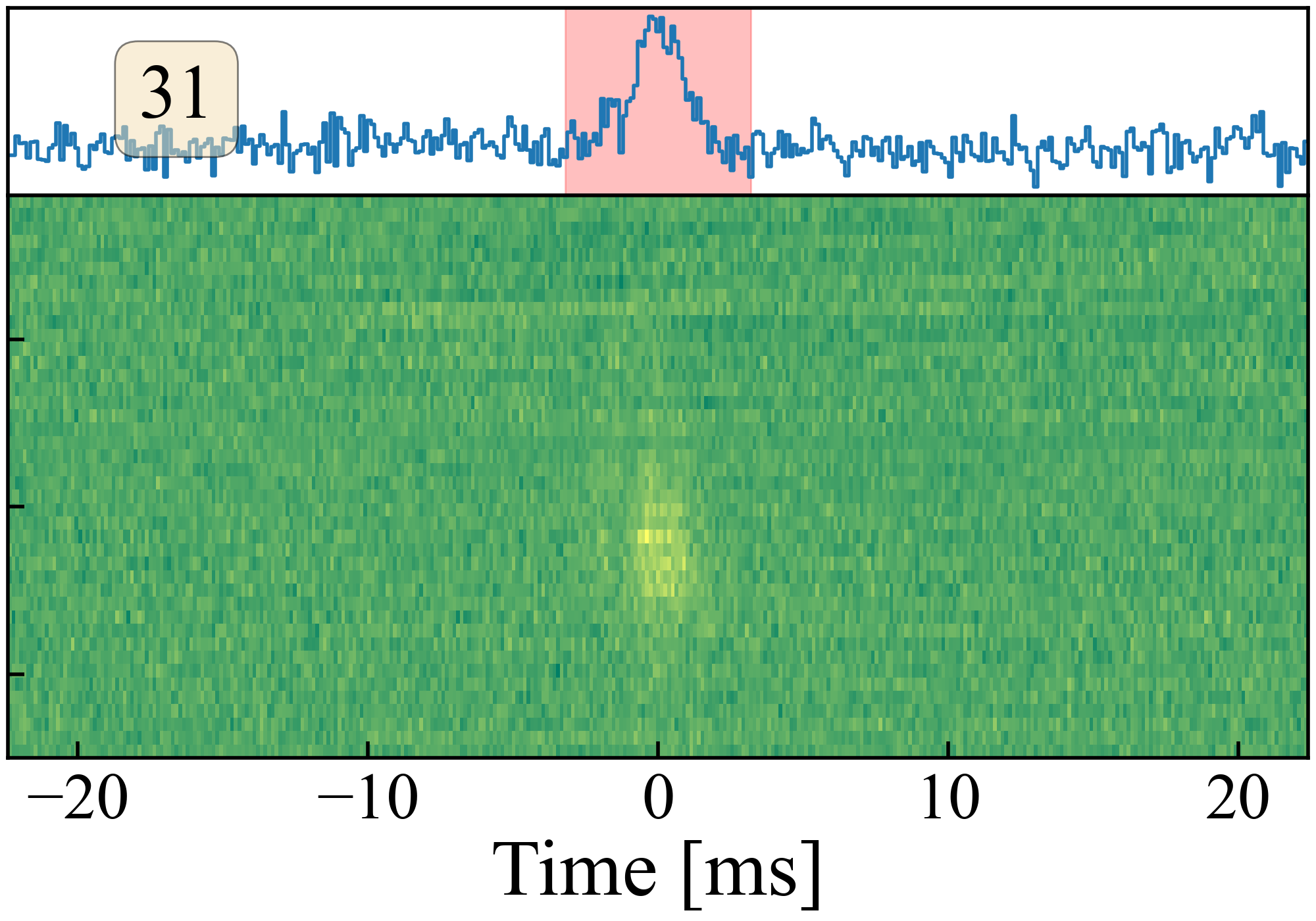}}
    \subfigure{\includegraphics[width=0.24\textwidth]{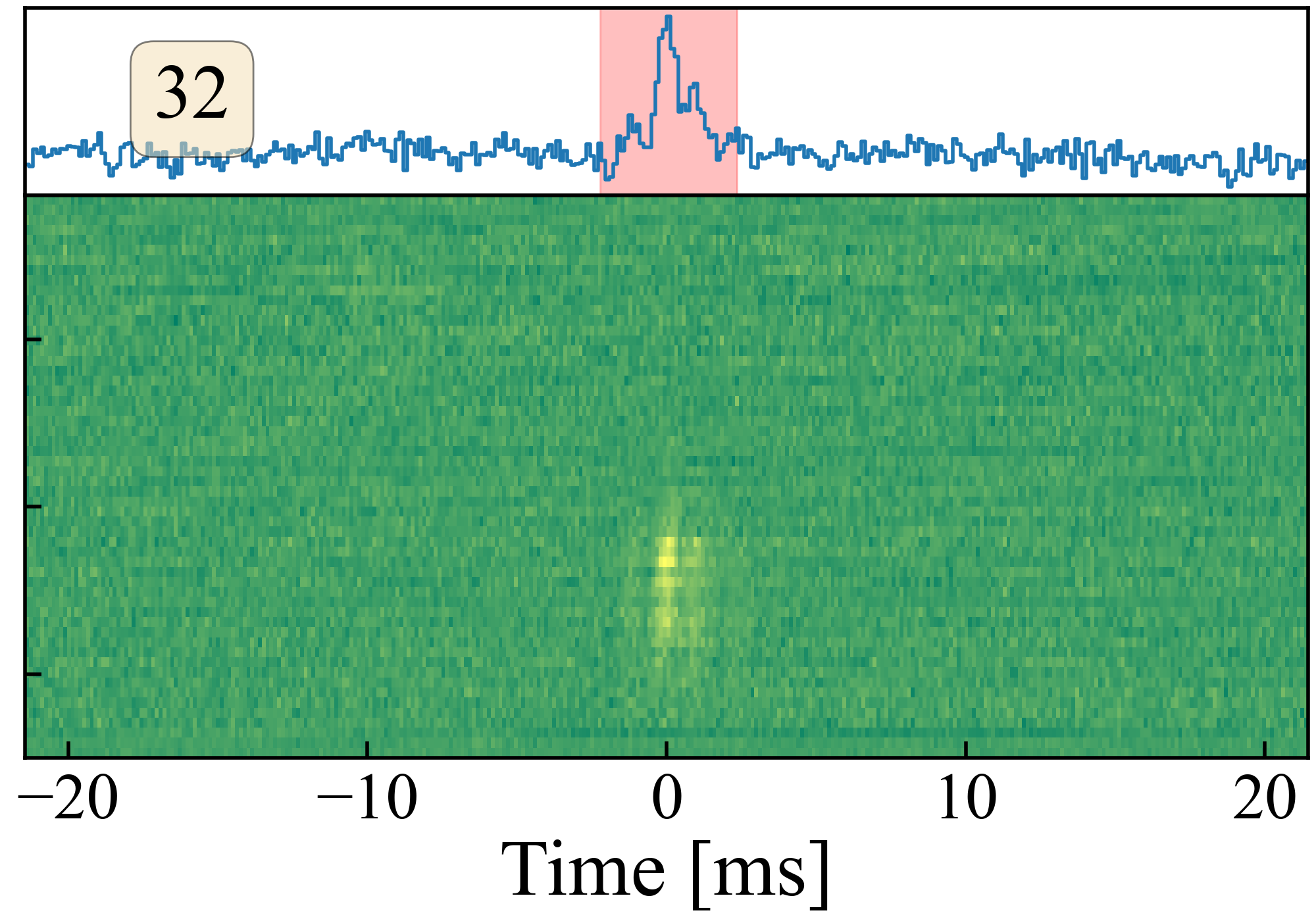}}
    \subfigure{\includegraphics[width=0.24\textwidth]{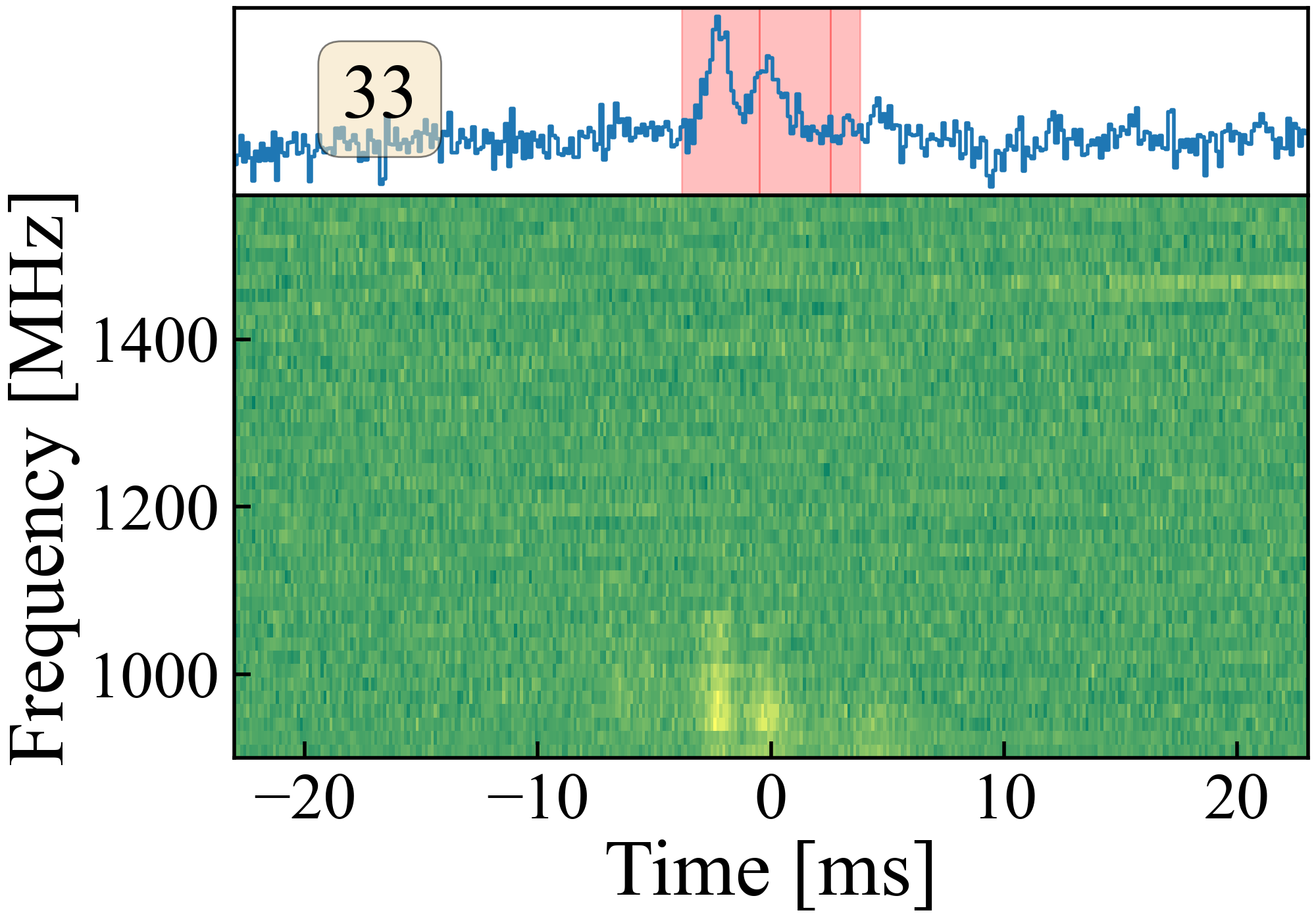}}
    \subfigure{\includegraphics[width=0.24\textwidth]{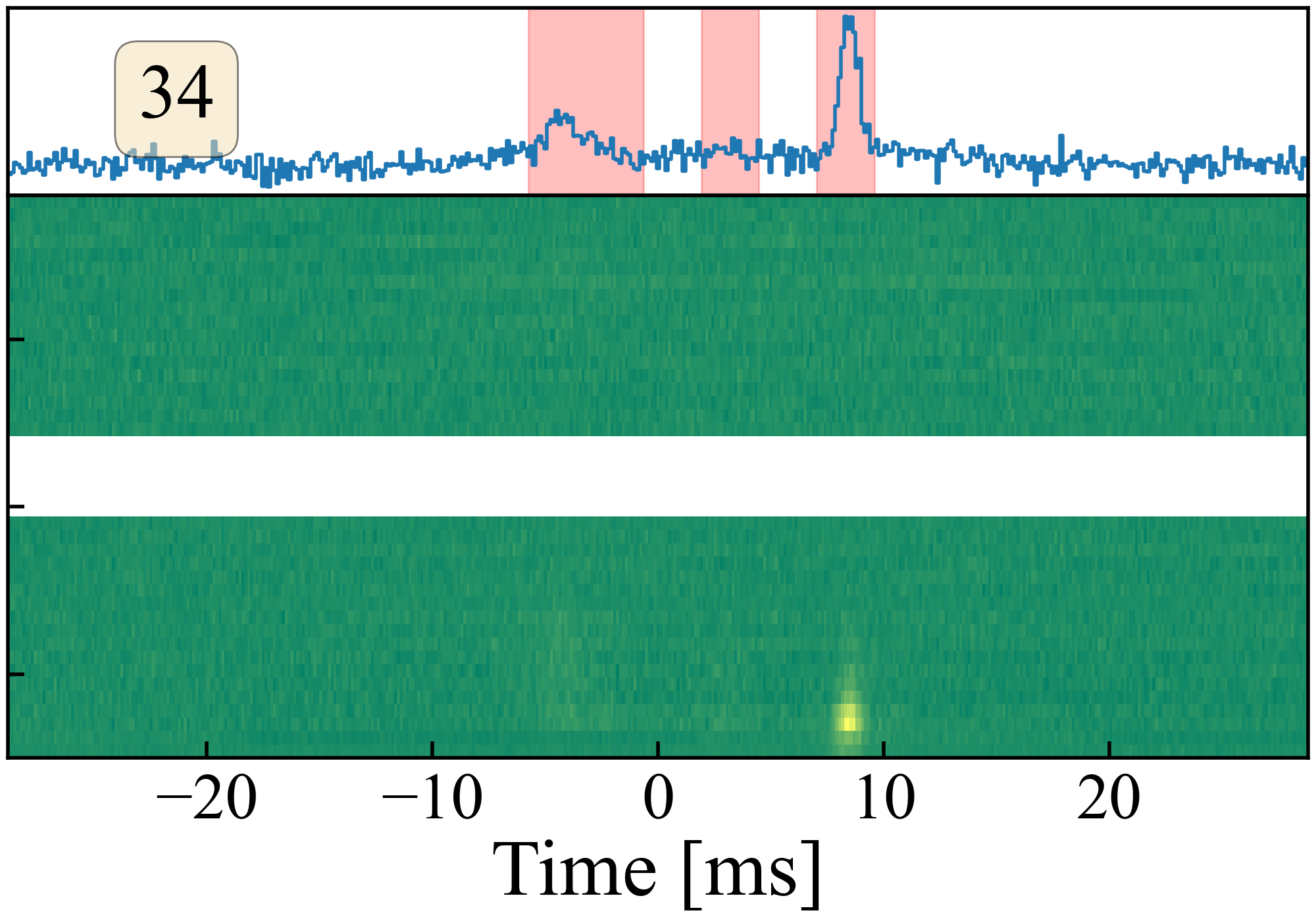}}
    \subfigure{\includegraphics[width=0.24\textwidth]{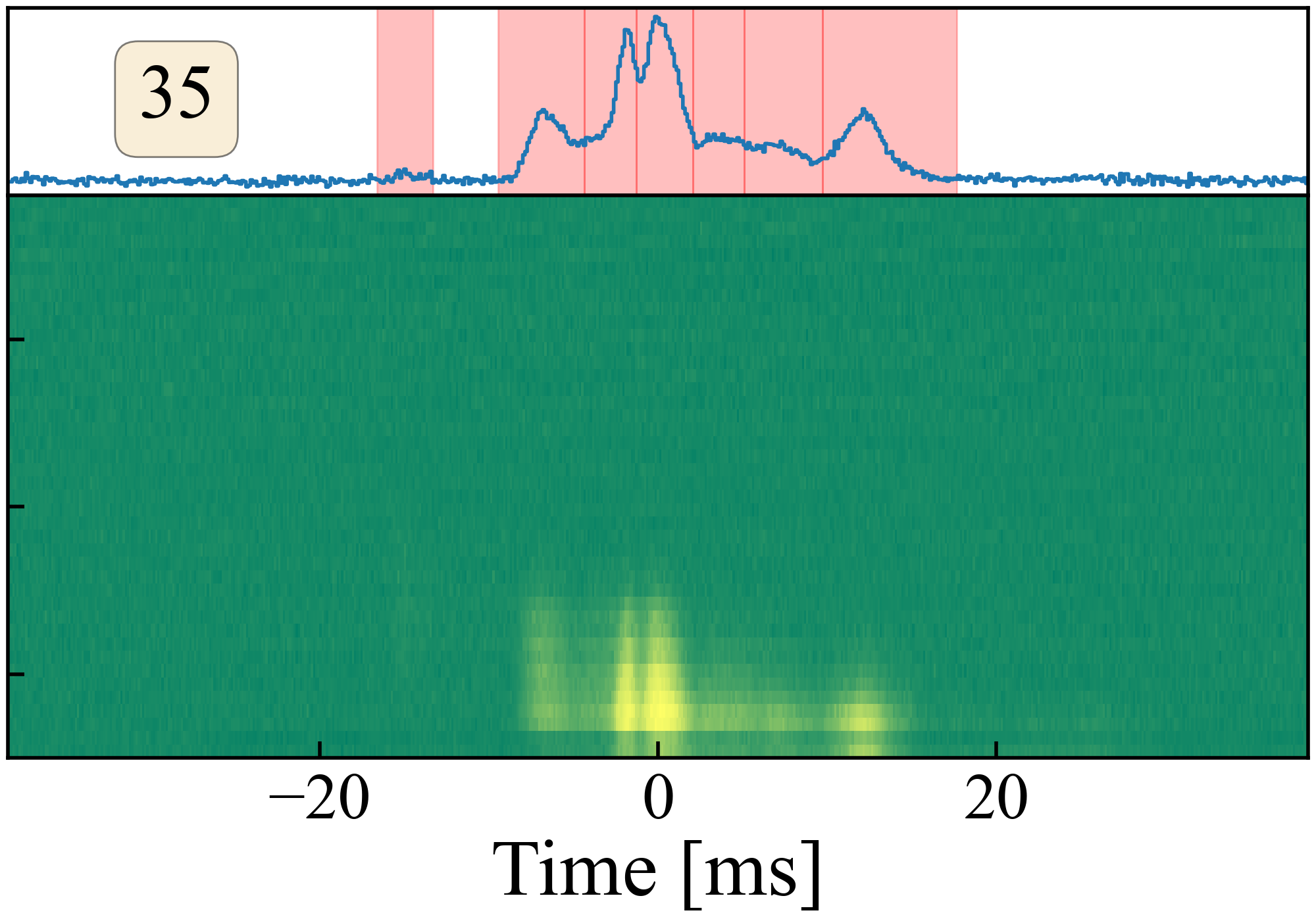}}
    \caption{Continued.}

    \label{fig:dynamic_spectra_no_sp}
\end{figure*}

\begin{figure*}
    \centering
    \includegraphics[width=\textwidth]{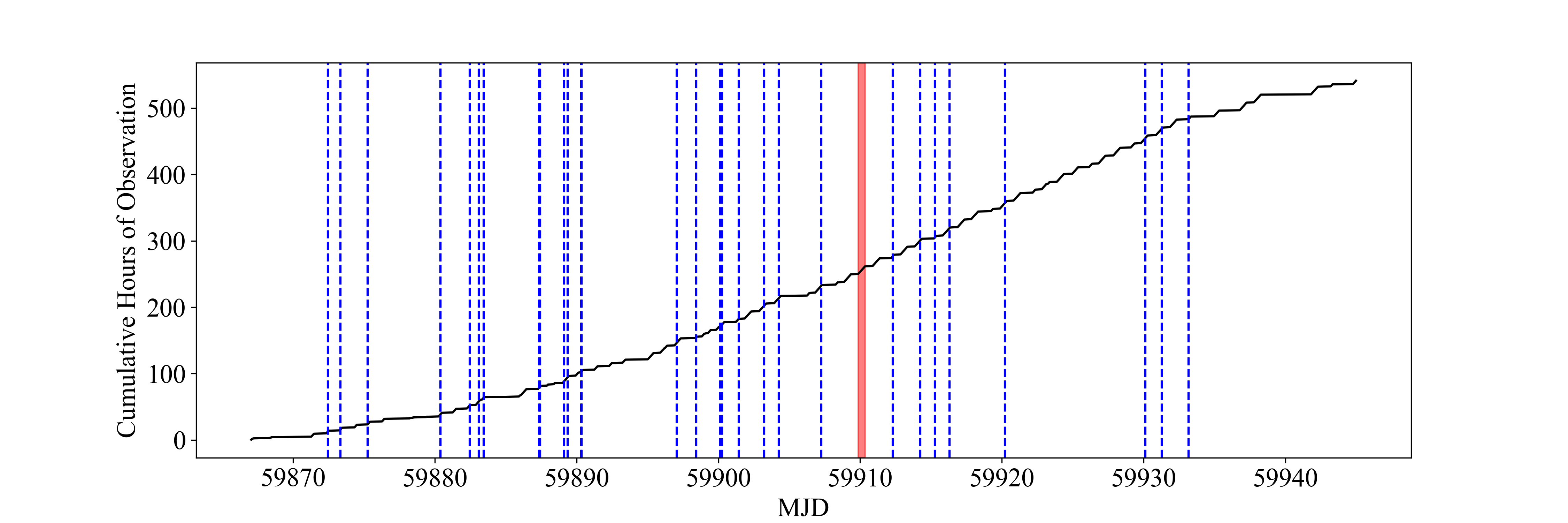}
    \caption{Cumulative observing time on FRB~20220912A over the months of the campaign (black line), with detections marked in dashed blue lines. The red shaded region indicates observations for which the higher tuning was unavailable, restricting our bandwidth by a factor of two. The average interval between observations was 1.11~days, with a maximum interval of 3.80~days and a minimum interval of a few hours.}
    \label{fig:observing_calendar}
\end{figure*}

The timeline of our detections is visualized in Figure \ref{fig:observing_calendar}.

\section{Data Reduction}
\label{sec:data_reduction}

\subsection{File Preparation}
\label{ssec:file_prep}

Once we have identified an \ac{FRB} within one of the 30-minute long \texttt{filterbank} files as described in Section \cref{ssec:search}, we crop the \texttt{filterbank} in time, centering on \texttt{SPANDAK}'s reported trigger time for the burst. We choose a 10-second cropping length, to capture the entire burst with dispersive delay and provide context for noise and \ac{RFI} properties. 
%


Next we convert the cropped \texttt{filterbanks} to archive files with \texttt{DSPSR} \citep{VanStraten2011_dspsr}, using the best-fit \ac{SNR}-maximizing \ac{DM} of the burst from \texttt{SPANDAK} contained in the metadata. This choice visually corresponded with structure-maximizing in the dedispersed waterfall plots for all but two of the bursts. For these two bursts (\#24 and \#25), we found that the \texttt{SPANDAK} over-disperses the burst, as evidenced by the morphology of the subcomponents and the anomalously large reported \ac{DM}s (224.94 and 223.24~pc~cm$^{-3}$, respectively). We thus set their archive file \ac{DM} to the average from the other 33 bursts, at 219.775~pc~cm${^{-3}}$. All archive files have the time axis of the file (10 seconds) read in as ``phase'' but are otherwise equivalent to archive files used for other radio pulsar applications.


\subsection{Calculating Flux and Fluence}
\label{ssec:flux_and_fluence}

Using the calibration measurements described in Section \ref{ssec:observations}, we calculate the flux and fluence of each of the 35 \ac{FRB}s in the sample. First, we obtain the \ac{SEFD} of each observation. All three flux calibrators in this dataset have flux models from \citet{perley2017accurate}, from which we obtain expected spectral flux densities $S_{\mathrm{exp}}$ at the central frequencies of both \ac{LO}s. We calculate the \ac{SEFD} for each antenna-polarization combination (antenna $x$, polarization $y$) as follows:

\begin{equation}
    \mathrm{SEFD}_{x,y} = (G_{\mathrm{cal}}^2 \times S_{\mathrm{exp}})^{-1}
\end{equation}

Where $G_{\mathrm{cal}}$ is the gain from the associated calibrator file, in units of $\textrm{Jy}^{-1}$, computed using the Common Astronomy Software Application (CASA) \texttt{gaincal} task \citep{bean2022casa}. We then average each of the antenna-polarization \ac{SEFD}s to get the beamformer \ac{SEFD}:

 \begin{equation}
    \mathrm{SEFD}_{\mathrm{beamformer}} = \frac{\frac{\Sigma_{x,y} \mathrm{SEFD}_{x,y}}{N_{x,y}}}{N_\textrm{el}}
\end{equation}

Where $N_{x,y}$ = 40 (the number of antenna-polarization combinations) and $N_\textrm{el}$ = 20 (the number of elements in the beamformer). This results in a polarization-averaged \ac{SEFD}$_\textrm{beamformer}$ of $387\,\pm 38$\,Jy. This number is in agreement with earlier tests conducted with the instrument, and details about recovering flux scales and beamformer efficiency for the ATA will be presented in a forthcoming paper (Farah et al., in prep.). For observations listed here, we used a uniform weighting of all antennas for the beamformer.

From this \ac{SEFD}, we can use the radiometer equation to get the minimum detectable flux density $S_{\nu, \mathrm{min}}$ of this work:

\begin{equation}
    S_{\nu, \mathrm{min}} = \frac{(\mathrm{SNR})(\mathrm{SEFD})}{\sqrt{n_{\mathrm{pol}}\tau \Delta \nu}}
\end{equation}

Where $\tau$ is the duration of an \ac{FRB}, $\Delta \nu$ is its frequency extent, and $n_{\mathrm{pol}}$ is the number of recorded polarizations (in this case, 2). Therefore, $S_{\nu, \mathrm{min}}$ for a \ac{SNR} threshold of 10, an \ac{SEFD} of 387~Jy, and a fiducial \ac{FRB} duration and bandwidth of 1~ms and 672~MHz, is 4.7~Jy.

We detected 35 bursts in 541 hours of observing, implying an average burst rate above 4.7~Jy of $6.47^{^{+1.29}} _{_{-1.09}}\times10^{-2}\,\textrm{h}^{-1}$, where the uncertainties represent 1-sigma Poisson errors \citep{Gehrels1986}.

We use the Python binding for \texttt{PSRCHIVE} \citep{Hotan2004_psrchive} to load the data as dynamic spectra, incoherently dedisperse each array to the \texttt{SPANDAK} \ac{DM} described in Section \ref{ssec:file_prep}, and remove the baseline. We then average each array in frequency to get a timeseries burst profile, and normalize it by subtracting the median and dividing by the standard deviation of a noise region consisting of $\pm 1000$ time bins (64 ms) on either side of the burst. 

At this point, we define the boundaries in the timeseries from which to extract \ac{SNR}s, fluxes, and fluences. Only 7 \ac{FRB}s consist of a single component; the rest of the sample shows complex sub-burst structure. A sub-burst is qualitatively defined here as a significant local maximum in intensity in an \ac{FRB}'s profile, especially if that structure seems distinct in frequency from the neighboring emission in the dynamic spectrum. We opt here to measure each of the sub-bursts individually, as not to downweight the flux by averaging over the periods between sub-bursts. The sub-burst delineations determined here are the same ones that are used in the process of extracting spectral properties of sub-bursts, described in Section \ref{ssec:frbgui}. 

Using the sub-burst bounds and the normalized profile defined above, we obtain the \ac{SNR} with $\mathrm{SNR} = \frac{\Sigma p}{\sqrt{N}}$, where $p$ is the normalized profile and $N$ is the number of points across the profile (i.e., the duration of the sub-burst). To transform the \ac{SNR} into the flux in Jy, we can incorporate the \ac{SEFD} as follows:

\begin{equation}
    S_\textrm{Jy} = \frac{\mathrm{SEFD} \times \mathrm{SNR}}{\sqrt{B \times N_P \times N \times t_\textrm{samp}}}
\end{equation}

where $B$ is bandwidth of the sub-burst, $N_\textrm{P}$ is the number of polarizations, and $t_\textrm{samp}$ is the sampling time. Finally, we convert $N$, the duration in bins, into a duration in ms and multiply by $S_\textrm{Jy}$ to get the fluences in Jy-ms. 

The \ac{SNR}s, fluxes, and fluences derived here are included in Table \ref{tab:frb_detections} (at the end of the paper). Uncertainties in the \ac{DM} were derived by applying a sliding boxcar with the parameters used by \texttt{SPANDAK} in the original detection and providing the width of the best-fit Gaussian to the \ac{DM}-\ac{SNR} curve. Uncertainties in the fluxes and fluences are calculated by propagating the uncertainty on the mean \ac{SEFD}.

It should be noted that the first 3 \ac{FRB}s in the sample were detected while the synthesized beam was centered on the original, offset coordinates. To address that, we modeled the shape of the synthesized beam at the topocentric coordinates for the aforementioned bursts. We then corrected the measured flux densities with the calculated attenuation values of 0.5\,dB, 0.5\,dB, and 1\,dB for bursts 1, 2, and 3 respectively.

Some of the flux values in the downsampled 8-bit \texttt{filterbank} files were clipped to the highest intensity bin, implying that we were underestimating the total flux by some amount for the brightest \ac{FRB}s. To estimate the impact of the clipping, we looked at the percentage of pixels in the highest intensity bin for the highest \ac{SNR} \ac{FRB} (\#25). For this particular \ac{FRB}, approximately 3\% of the pixels were clipped as seen when plotting the pixel intensity distribution histogram. Recovering the original values (i.e. before resizing to 8-bit depth) is almost impossible given that the original 32-bit, higher dynamic range data products were deleted. However, we still attempted to quantify the effects of pixel clipping by assuming that the intensity distribution, at the top end, follows a continuous dampened trend. We fit the intensity histogram, and ``redistribute'' the clipped pixels by extrapolating beyond the 8-bit clipping point, and then integrate the intensities and measure the flux. 
By doing so, we found that we were only underestimating the flux by approximately 3\% for this \ac{FRB}. Given that this is the most extreme example in our sample, we do not correct for this effect in the dataset.

\subsection{Extracting Spectral Properties with \texttt{frbgui}}
\label{ssec:frbgui}

Some parameters of interest for the 35 \ac{FRB} sample are calculated from the original archive files, but some spectrotemporal properties such as bandwidth and duration must be extracted with additional processing, especially for bursts with multiple sub-components. We used the spectrotemporal \ac{FRB} code \texttt{frbgui} \citep{chamma2022broad} to extract these parameters.

We load the archive files into Python, dedisperse, and remove baseline as described in the previous Section \ref{ssec:flux_and_fluence}. For each \ac{FRB}, we defined start time and end time indexes in the original archive file --- we crop the array to $\pm$ 100 bins (6.4 ms) on this range. Finally, the files are written out as dynamic spectra with accompanying metadata containing information about the time axis, frequency axis, dispersion measure, and various units.

We open these dedispersed and cropped files with \texttt{frbgui}. For each file, we define a \ac{DM} grid covering the original \texttt{SPANDAK} \ac{DM} $\pm$ 0.5 pc\,cm$^{-3}$, sampled at 0.01 pc\,cm$^{-3}$ intervals. If the average \ac{DM} of the whole sample would fall outside of this grid, we extend it by another $\pm$\,0.5 pc\,cm$^{-3}$ in the relevant direction. We then optionally perform a series of tasks to improve the \ac{SNR} of the burst:

\begin{itemize}
    \item \textit{Subtract a background sample from the entire dynamic spectrum}: This background sample is usually selected to be the first few milliseconds of the file before the burst begins. This helps remove narrowband \ac{RFI}.
    \item \textit{Add a frequency mask range}: For bursts that do not cover the entire bandwidth (most of them, as per Figure \ref{fig:dynamic_spectra_no_sp}), we can mask up to hundreds of MHz of bandpass where there is no \ac{FRB} signal in order to improve the \ac{SNR}.
    \item \textit{Remove remaining \ac{RFI}}: For bursts with significant \ac{RFI}, we implemented built-in \ac{RFI}-masking with a spectral kurtosis - Savitzky–Golay (SK-SG) filter (with a polynomial order of $\sigma=3$ and a a window-size of 15 samples), which improved the \ac{SNR}. For less affected bursts, however, this removed the brightest parts of the true \ac{FRB} signal and was not used.
    \item \textit{Downsample in time and frequency:} By downsampling, or frequency averaging and time integrating the dynamic spectrum, we can improve the \ac{SNR} --- this is especially necessary for faint \ac{FRB}s and was already part of \texttt{SPANDAK}'s routine for finding these fainter \ac{FRB}s. We select time and frequency downsampling factors that are evenly divisible into the length of the array axes, while still endeavoring to keep enough resolution to be confident of e.g., drift rates.
\end{itemize}

We then manually demarcate burst splitting with \texttt{frbgui}'s interactive interface. Given that \ac{FRB}s have been shown to have sub-bursts and structure on the order of microseconds \citep{nimmo2022burst}, and that we are downsampling to increase \ac{SNR}, there are likely fine-structure details that we cannot resolve in our dataset. Nevertheless, we do delineate the sub-bursts that are visible given our resolution, and find \ac{FRB}s with up to 7 distinct components in our 35 burst dataset.

Finally, we use a non-linear least squares algorithm\footnote{Via \texttt{scipy.optimize.curve\_fit}} to perform a 6-parameter 2D Gaussian fit to the 2D auto-correlation of the dynamic spectrum for the entire array, and then for the limited sub-arrays delineated in time by each sub-burst start and stop defined previously. Fitting to the auto-correlation increases the \ac{SNR} and mitigates the effect of e.g., scintillation and \ac{RFI} in the parameter measurements \citep[see ][ for more details about this operation]{chamma2022broad}. For each \ac{DM} in the grid, the fitter returns the bandwidth, duration, and sub-burst slope for each of the sub-bursts, as well as the bandwidth, duration, and drift rate (in MHz~ms$^{-1}$) of the overall burst, with their associated errors. In some cases, the fitter does not converge on the correct parameters for the burst. We visually reviewed each fit (for every \ac{DM}-sub-burst combination) in the \texttt{frbgui} interface and, where necessary, implemented a refit by manually inputting an initial guess based on the correct preceding fits.





The center frequency, duration, and bandwidth information for each sub-burst are shown in Table~\ref{tab:frb_detections}. Reported uncertainties in the duration are equal to the time bin size after scrunching in the dynamic spectrum used for the spectral property fitting.

\section{Data Analysis}
\label{sec:analysis}

\subsection{Spectrotemporal Analysis}
\label{ssec:spectrotemporal}

\begin{figure}
    \centering
    \includegraphics[width=0.48\textwidth]{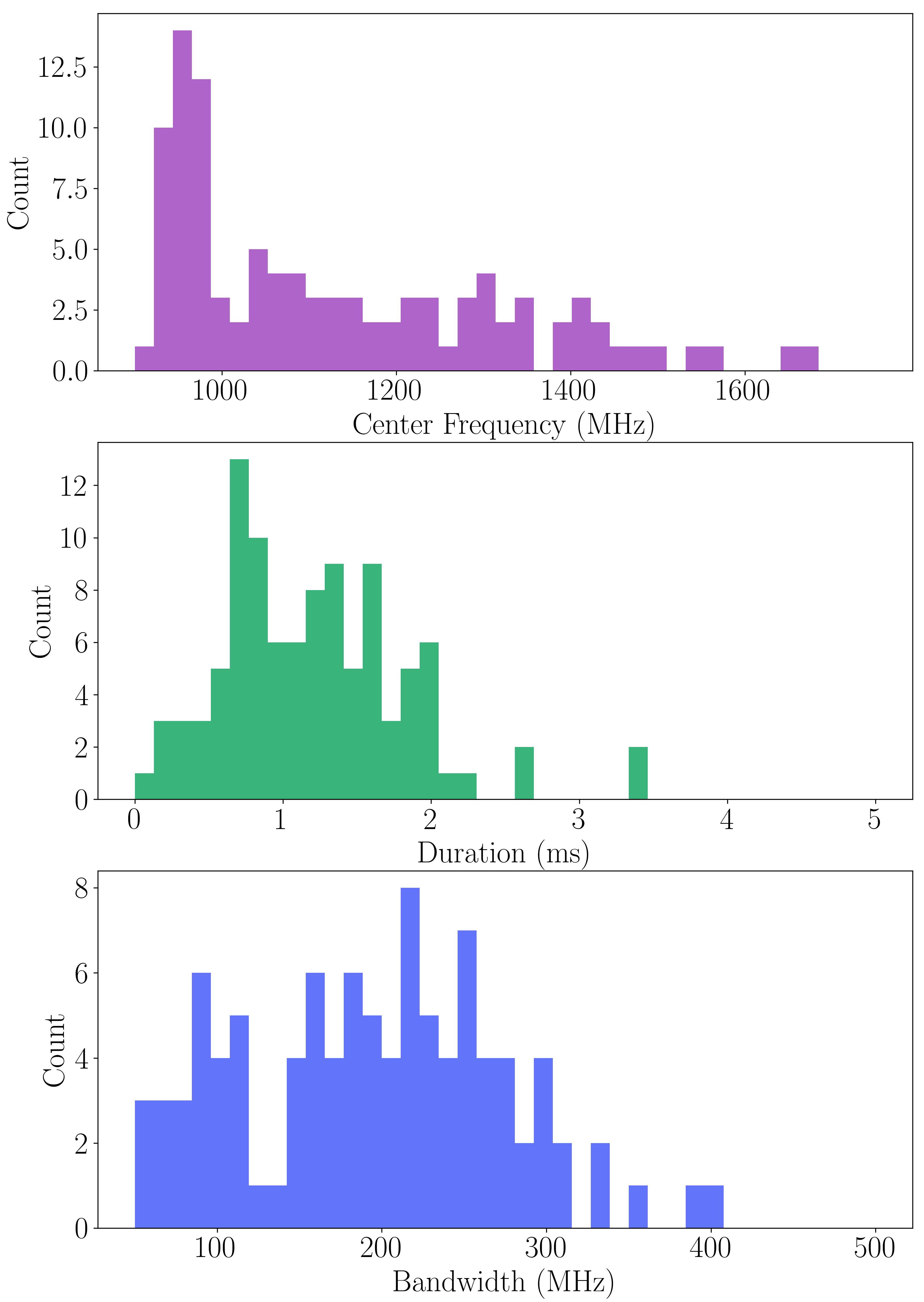}
    \caption{The distributions of center frequency in MHz, duration in ms, and bandwidths in MHz for each of the 101 sub-bursts in the dataset. We see the majority of \ac{FRB}s reside at the lower end of the bandpass, have $<2$ ms durations and show bandwidths of a few 100 MHz.}
    \label{fig:freq_duration_bw_hist}
\end{figure}

Using the \ac{FRB} and sub-burst characteristics extracted in Section \ref{sec:data_reduction}, we can investigate population-scale features of our dataset. The center frequencies, durations, and bandwidths for all 101 sub-bursts in the set are displayed in Figure \ref{fig:freq_duration_bw_hist}.

For all linear fits in Sections~\ref{sssec:freq_vs_bw}--\ref{sssec:freq_vs_dur}, we use the following methodology. In most cases, we have an independent variable $x$, a dependent variable $y$, and heteroscedastic errors in each variable $x_{\mathrm{err}}$ and $y_{\mathrm{err}}$. We account for this heteroscedasticity with a bootstrapping technique. We employ an ordinary least squares (OLS) fitting method that incorporates robust covariance estimators \citep[specifially, the HC1 estimator from ][]{mackinnon1985some}. In each of 10000 bootstrapped trials, we draw each value of $x$ or $y$ from a normal distribution with a width of $x_{\mathrm{err}}$ or $y_{\mathrm{err}}$ respectively, and then use the OLS fitting method to get a value for the slope and its associated $R^2$ value (the square of the Pearson correlation coefficient, to quantify how much variance in the dependent variable is explained by the independent variable). 

For each fit, we report a) the mean of the 10000 slopes from the trials above b) the mean of the 10000 standard errors of the slope, and c) the mean of the $R^2$ values of the 10000 trials. We calculate the significance of these correlations by running an additional 10000 trials where we permute the data, i.e., randomly shuffle the $y$ values and then re-assign them to the $x$ values. The p-value is then the percentage of permuted outcomes that meet or exceed the absolute value of the mean $R^2$ described above. Here, we use $p<0.05$ as our threshold for statistical significance.

\subsubsection{Burst Energy}
\label{sssec:energy}

To compare our sample with the energy distribution of 128 bursts seen by the \ac{GBT} \citep{feng2023extreme}, we can convert the fluences from Section \ref{ssec:flux_and_fluence} into isotropic equivalent sub-burst energies, using the redshift and luminosity distance from \citet{Planck2015, Ravi2022_frb220912a}, which assumes Planck 2016 cosmology. We find a median sub-burst energy for our sample of 1~$\times$~$10^{39}$~erg, approximately two orders-of-magnitude greater than the median from \citet{feng2023extreme}, although approximately on-par with their brightest bursts. This is consistent with the lower sensitivity of the \ac{ATA}. The brightest sub-burst in our sample, from \#25, has an energy of $1.8 \times~$~$10^{40}$~erg; the brightest burst, combining all subbursts (also \#25), has an isotropic energy of $5\times~10^{40}$~erg.


\subsubsection{Center Frequency vs. Bandwidth}
\label{sssec:freq_vs_bw}

\begin{figure}
    \centering
    \includegraphics[width=0.48\textwidth]{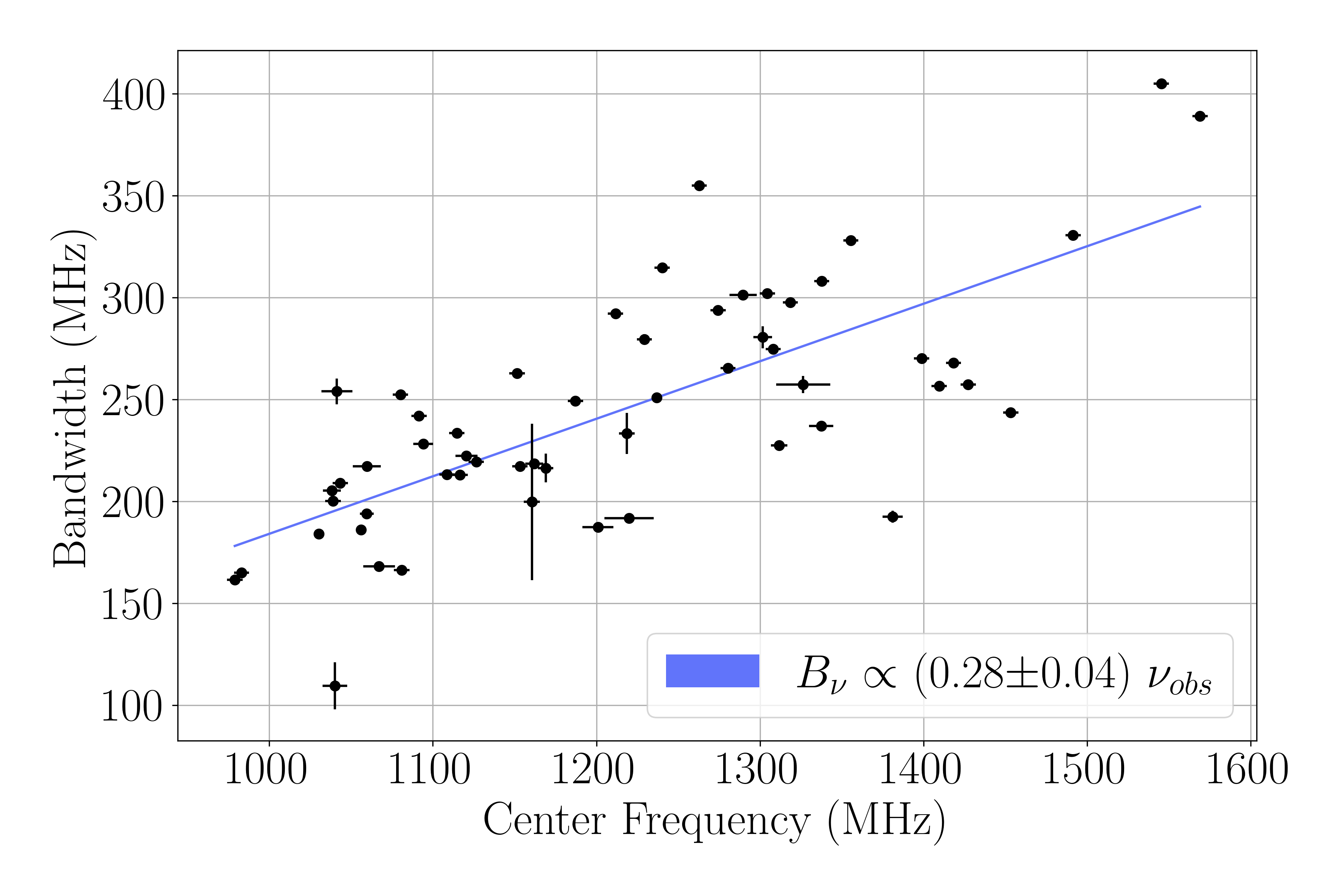}
    \caption{Center frequency in MHz versus bandwidth in MHz for 55 sub-bursts from FRB~20220912A (black) and their best-fit line (blue). 1-$\sigma$ fit errors derived by \texttt{frbgui} are shown in both center frequency and bandwidth for the points used in the linear fit, which were accounted for during the bootstrapping. We confirm the expected positive linear relationship between these two properties.}
    \label{fig:freq_vs_bw_fit}
\end{figure}

To test the correlation between the center frequency of a sub-burst and its bandwidth, we first remove any \ac{FRB} from the set whose sub-bursts significantly intersect the edge of the band by visual inspection of the de-dispersed dynamic spectra. This results in a set of 55 sub-bursts for which we evaluate a linear fit. We find a positive, linear correlation between center frequency and bandwidth with a slope of $0.28 \pm 0.04$, as shown in Figure \ref{fig:freq_vs_bw_fit}. The $R^2$ for the fit between the center frequency and bandwidth is 0.533 and the p-value is $<10^{-4}$. 

This slope is consistent within a factor of two of the slope of $0.14 \pm 0.004$ found by \citet{chamma2022broad} for FRB~20121102A, although different sources would not necessarily be expected to have the same slope. The center frequency and bandwidth relationship described here also appears consistent with the 1000-burst dataset from \citet{zhang2023fast}. 

\subsubsection{Center Frequency and Bandwidth vs. Time}
\label{sssec:freq_vs_time}

\begin{figure*}
    \centering
    \includegraphics[width=\textwidth]{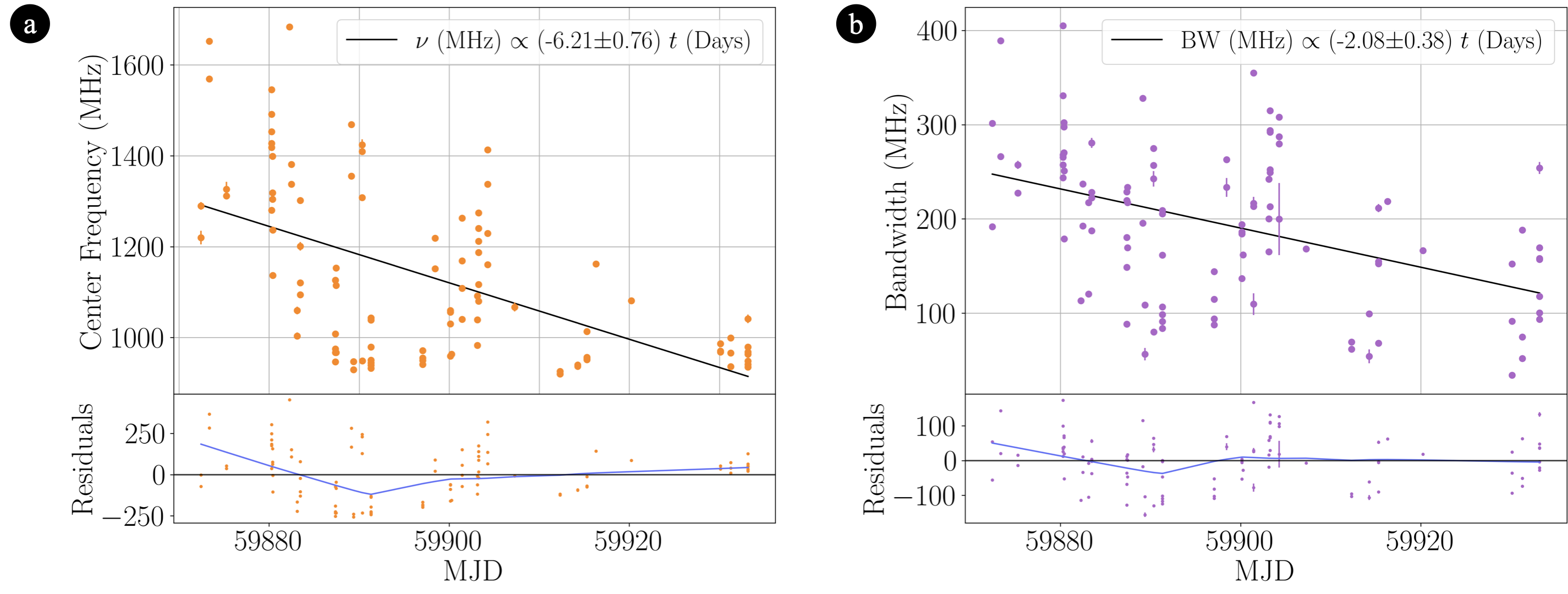}
    \caption{Two FRB~20220912A dataset parameters --- center frequency and bandwidth --- plotted over time, in MJD, from the beginning of the campaign to the end of the campaign (a time period of approximately 60 days). Panel a) indicates that the central frequency of the \ac{FRB} decreases through the campaign (with residuals from the fit and a LOWESS non-parametric guideline shown below in blue). Panel b) shows the same decrease over time for bandwidth, which is understandable given the relationship between bandwidth and center frequency shown in Figure \ref{fig:freq_vs_bw_fit}.}
    \label{fig:everything_over_time}
\end{figure*}

A unique feature of our dataset was the relatively steady rate of data acquisition over the two months of the campaign; this can be seen in the smooth slope of the cumulative observing time in Figure \ref{fig:observing_calendar}. This allows us to interrogate changes in the properties of the 101 sub-bursts from FRB~20220912A over time. We see a decrease in both the center frequency and bandwidth over the course of the campaign (Figure \ref{fig:everything_over_time}). The center frequency decreases at a rate of $6.21 \pm 0.76$ MHz per day, with a corresponding $R^2$ of 0.311 and a p-value of $<10^4$. This trend is echoed by the bandwidth, which decreases $2.08 \pm 0.38$ MHz per day, with a corresponding $R^2$ of 0.191 and a p-value of $<10^4$, due to their inherent relationship shown in Figure \ref{fig:freq_vs_bw_fit}. There were no similarly obvious changes in flux, duration, or DM over the campaign.

Other repeaters have been observed to have varying activity levels at different frequency ranges. For example, FRB~20180916B shows activity in the 1000~MHz range 3 days before activity peaks in the 100~MHz range \citep{pleunis2021lofar}. \citet{pearlman2020multiwavelength} also find that apparent \ac{FRB} activity is strongly affected by which frequencies are being recorded, with FRB~20180916B again showing this behaviour particularly strongly. However, this marks the first trend in frequency over time for a so-far non-periodic \citep[as per][]{zhang2023fast} repeating \ac{FRB}.

The residuals from the linear fit of central frequency over time in Figure \ref{fig:everything_over_time} do merit discussion. We find some remaining structure in the residuals, where the best-fit model underestimates the center frequency at the beginning of the campaign and overestimates the center frequency at MJD 59890. Without an underlying model to compare to, we do not attempt to fit a more complex function to the data, but do note this behaviour as something to be investigated in future studies. In addition, the residuals in this linear fit are not homoskedastic: the scatter around the best-fit line decreases over the campaign. These heteroskedastic errors should raise some questions about the extension of this trend past the bottom of our bandpass. Specifically, it seems possible that there is a several hundred MHz wide, downward drifting ``activity window'' in frequency, that drifts out of our bandpass over the campaign. This leads to the apparent shrinking of the scatter at the end of the linear fit, and implies that the actual slope in the central frequency may be steeper than 6.2~MHz per day.

\subsubsection{Center Frequency vs. Drift Rate}
\label{sssec:dr_vs_freq}

Other repeaters have shown a correlation between center frequency and drift rate, the linear decrease in sub-burst center frequency over time. We use the values derived by \texttt{frbgui} in Section \ref{ssec:frbgui} to assess whether we see this same correlation in the FRB~20220912A data. It should be noted that no \ac{FRB}s in our sample show upward-drifting bursts \citep[e.g., ][]{kumar2022circularly}.

To construct a subsample with reliable drift rate measurements, we must first remove all of the single-component \ac{FRB}s, which do not have a defined drift rate. In addition, the \ac{ACF}  method for sub-burst center frequency determination will underestimate the center frequencies of components that intersect the top edge of the bandpass, and correspondingly overestimate the center frequencies of components that intersect the bottom edge of the bandpass. Both effects lead to an underestimation of the drift rate: therefore, we remove any \ac{FRB}s that intersect the edges of the bandpass, bringing the subsample to 11 \ac{FRB}s. Finally, we wish to remove any \ac{FRB} where the \ac{ACF} measured the sub-burst slope of the brightest component instead of the drift rate across multiple components, which can occur in \ac{FRB}s with large \ac{SNR} variation across sub-bursts. This effect is visible in the 2D-\ac{ACF}s, and can be confirmed as an anomalously high drift rate that is consistent with a single sub-burst's slope. For this reason, we additionally remove \ac{FRB}s \#3, \#6, and \#18. 

\begin{figure}
    \centering
    \includegraphics[width=0.48\textwidth]{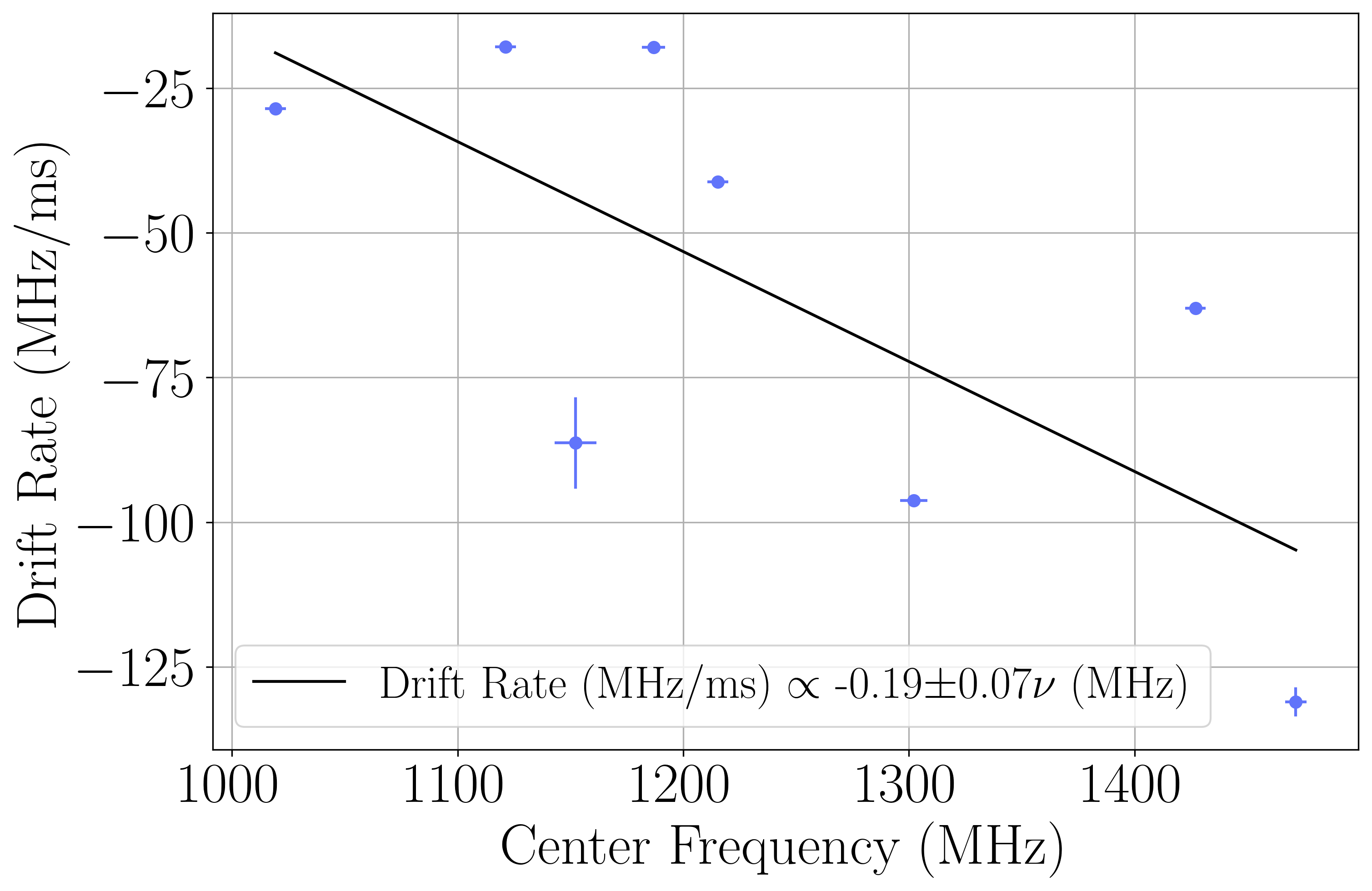}
    \caption{Drift rates in MHz~ms$^{-1}$ plotted against frequency in MHz for 8 \ac{FRB}s from FRB~20220912A that a) had more than 1 subcomponent b) did not intersect either edge of the bandpass and c) had subcomponents of similar-enough \ac{SNR}s to allow for a correct drift rate fit from the 2D \ac{ACF} of the dynamic spectrum. Visually, there is a downward trend (increasing magnitude of drift) with increasing frequency and we show the best-fit line from a linear fit (black), but we do not find that this trend is significant}.
    \label{fig:freq_vs_dr}
\end{figure}

We plot the drift rates for the 8 remaining \ac{FRB}s (\#1, \#4, \#8, \#10, \#20, \#23, \#24, and \#25) against their center frequencies in Figure \ref{fig:freq_vs_dr} and do not find a statistically-significant correlation. We used a $N=10^5$ trial bootstrap for this correlation, due to p-values that were near the threshold. The bootstrapping OLS routine finds a slope of -0.189$\pm$0.07 MHz~ms$^{-1}$MHz$^{-1}$ and an $R^2$ of 0.502, but the p-value of 0.052 is not significant. This is consistent with the data, considering the small-number statistics resulting from only 8 \ac{FRB}s in the fit. 

Other repeating \ac{FRB}s demonstrate a trend of steeper drift rates with frequency. For FRB~20180916B, compilations of studies across wide bandwidths have shown a drift rate change of $-0.02$--$-0.03$ MHz~ms$^{-1}$MHz$^{-1}$ with frequency \citep{pastor-marazuela2021chromatic, sand_FRB_2023}. Meanwhile, for FRB~20180301A (at frequencies an order-of-magnitude higher), \citet{kumar2023spectro} find a drift rate change of $-0.14$ MHz~ms$^{-1}$MHz$^{-1}$ with frequency, over approximately the same frequency range as this work. Using MeerKAT, \citet{platts2021analysis} find a similar drift rate change with frequency, -0.147 $\pm$ 0.014 MHz~ms $^{-1}$MHz$^{-1}$, for FRB~121102. It should be noted that a quadratic fit is favoured when data covers a larger range of frequencies \citep{wang2022magnetospheric, chamma2022broad}. Given that our measurement was not significant, we cannot add to this population of results with FRB~20220912A in this work; regardless, the consistency of the sign and general relationship of frequency and drift rate over this population of repeaters hints at a persistent feature in at least a subclass of \ac{FRB}s that must be explained by any proposed emission mechanism.

Sub-burst slope and drift rate can follow similar relationships \citep[e.g., ][]{rajabi2020simple,chamma2022broad}. However, we see no correlation between subburst slope and duration, or subburst slope and center frequency, in this dataset. Given the scatter in the measurements and the lack of bursts at other bands to constrain model fits, this is not unexpected.

\subsubsection{Center Frequency vs. Temporal Duration}
\label{sssec:freq_vs_dur}

\begin{figure}
    \centering
    \includegraphics[width=0.48\textwidth]{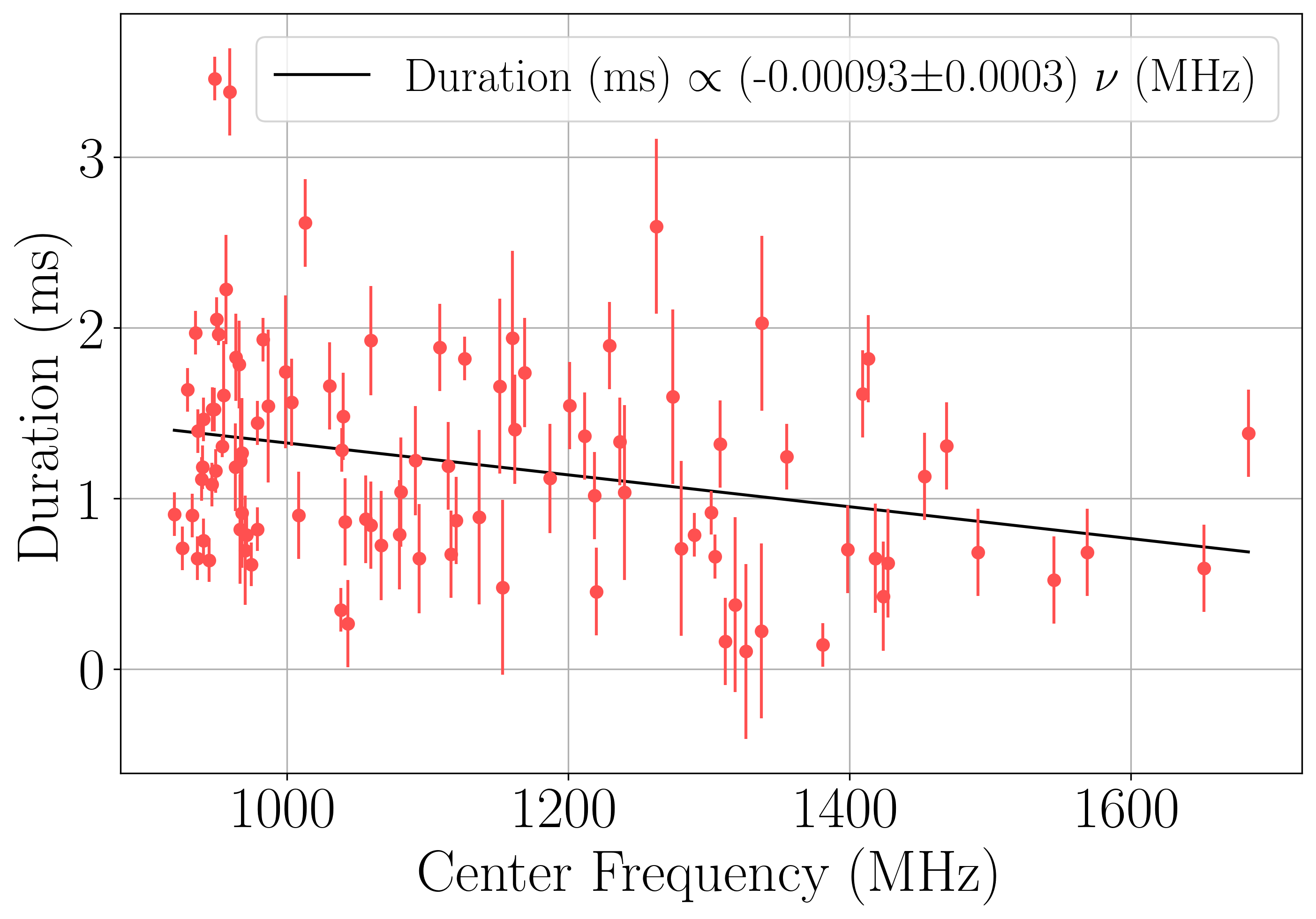}
    \caption{Duration in ms plotted against frequency in MHz for our full sample of 101 sub-bursts from FRB~20220912A. Note the slight decrease in duration over the bandpass.}
    \label{fig:freq_vs_dur}
\end{figure}

It has been noted that higher frequency sub-bursts seem to have shorter temporal durations, which is only observable across GHz of bandwidth \citep[e.g., ][]{platts2021analysis}. \citet{Gajjar2018} notes that the widths are considerably narrower (about an order of magnitude) at 8~GHz compared to 1~GHz for FRB~121102, but that the effect does not seem consistent with broadening due to scattering from turbulent plasma along the line-of-sight. Here, we do observe a slight downward trend in duration with increasing frequency, as shown in Figure \ref{fig:freq_vs_dur}. The associated linear fit has a slope of $(-9.3 \pm 3.3) \times 10^{-4}$ MHz~ms$^{-1}$, an R$^2$ value of 0.067 (indicating wide scatter over the $\sim$700~MHz available in these observations), and a p-value of 0.0098. Given the low correlation, this result should be taken only as an indication of consistency with the previously-observed trend of shorter durations at higher central frequencies. Given the channel width and the FRB DM, the expected smearing due to dispersion at the bottom of the ATA band is $\sim0.9$\,ms.

\subsubsection{Scattering}
\label{sssec:scattering}

We do not observe any significant scattering behaviour in the waterfall plots in Figure \ref{fig:dynamic_spectra_no_sp}, though we cannot constrain scattering timescales less than several hundred $\upmu$s at these frequencies, due to our sampling time of 64~$\upmu$s. This is consistent with the nominal scattering value of $\leq15$~ms at 400~MHz for this source as reported by \citet{bhusare2022uGMRT}, which would imply, given a Kolmogorov scaling with $\alpha$ = 4.4, an expected scattering timescale of $\leq$60~$\upmu$s. 


\subsection{Repeating Rate Function and All-Sky Rate}
\label{ssec:all_sky_rate}

\begin{figure}
    \centering
    \includegraphics[width=0.48\textwidth]{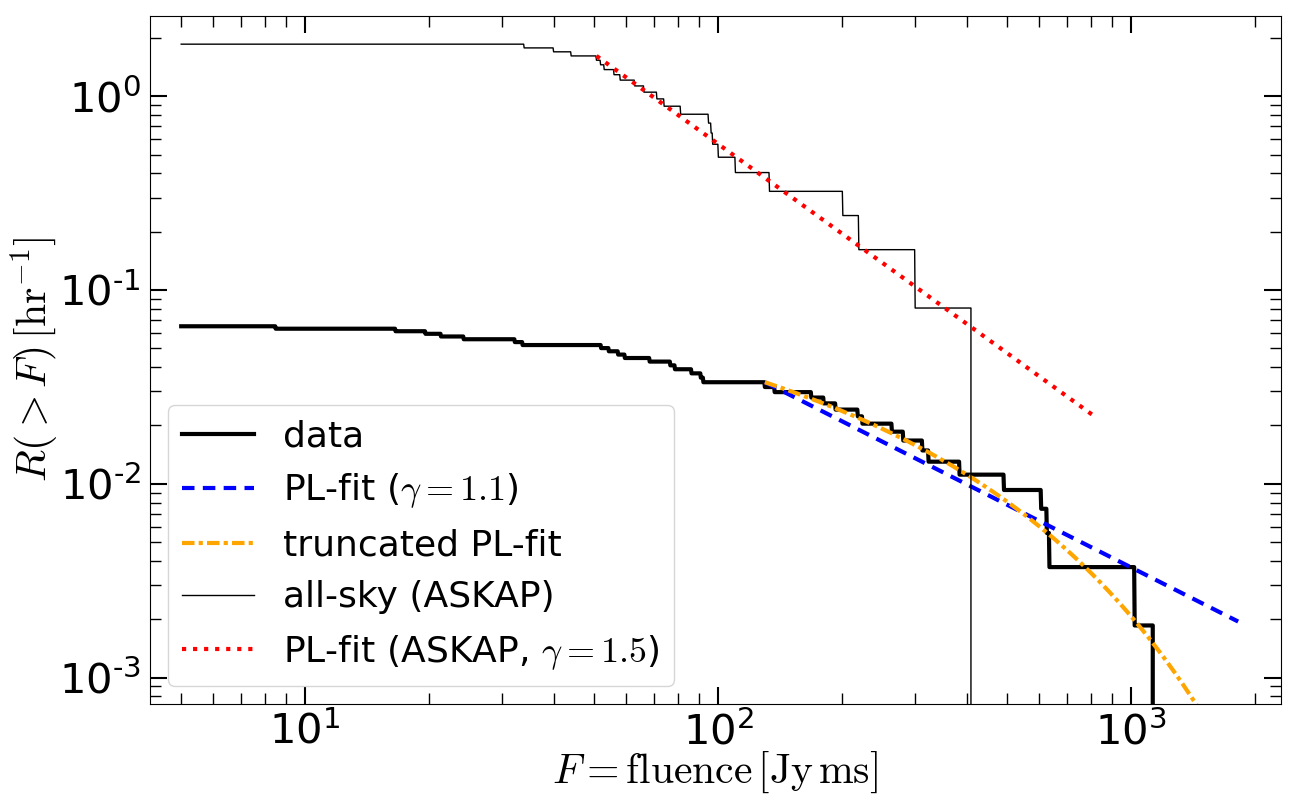}
    \caption{Fluence distributions of the ATA sample of bursts from FRB\,20220912A (thick black line) and the ASKAP Fly's Eye sample (thin black line). The vertical axis shows the cumulative rate, as computed by the raw number of detections divided by the total on-source time. The total exposure of the ASKAP Fly's Eye survey is $5.1\times10^5\rm\, deg^2 \,h$ \citep{Shannon2018}, which is converted into an all-sky equivalent exposure time of $12.4\rm\, hr$ --- the thin black line shows the all-sky rate as inferred from the ASKAP sample. The flattening of the distributions at the low fluence end ($\lesssim 100\rm\, Jy\,ms$ for the ATA sample and $\lesssim 50\rm\, Jy\,ms$ for the ASKAP sample) is likely due to incompleteness of the surveys. The steepening at the high fluence end is due to insufficient exposure times. The blue dashed and orange dash-dotted lines show the best-fit power-law and truncated power-law models to the ATA data, respectively. The red dotted line shows the best-fit power-law model for the ASKAP data.
    }
    \label{fig:fluence_distribution}
\end{figure}

We show the cumulative repeating rate function $R(>F)$ for FRB\,20220912A in Figure \ref{fig:fluence_distribution}, where $F$ is the fluence of each FRB (which may contain multiple sub-bursts). We fit the observed fluence distribution with a power-law function using the $\mathtt{powerlaw}$ package provided by \citet{alstott2014powerlaw} based on \citet{clauset2009power}. The package uses the maximum likelihood method to optimize the power-law slope $\gamma \equiv -\mathrm{d}\, \mathrm{ln}[R(>F)]/\mathrm{d}\, \mathrm{ln} F$ and further uses the Kologorov-Smirnov test to optimize the minimum fluence $F_{\rm min}$ above which a power-law provides the best description \citep[see][for the method descriptions]{clauset2009power}. 

When restricting the minimum fluence to be in the range from $40$ to $150\rm\, Jy\, ms$ (roughly corresponding to a minimum SNR between $\sim$40 and $\sim$150 for burst duration of $\sim$1 ms), we find the optimized minimum fluence to be $F_{\rm min}\simeq 130\rm\, Jy \, ms$. We also find the best power-law above this minimum fluence to be
\begin{equation}
    R(>F)=3.3\times10^{-2}\mathrm{\,h^{-1}}\, (F/F_{\rm min})^{\gamma}, \gamma=1.08,
\end{equation}
and the standard deviation of the power-law slope is $\sigma=0.25$. The slope is not well constrained by the current small sample but it is consistent with the power-law slope of the FRB luminosity (or energy) function \citep[e.g.,][]{Luo2018_luminosity_function, lu19_ASKAP_sample_study, james22_FRB_population, shin23_CHIME_luminosity_function}. We also tried fitting the ATA sample by an exponentially truncated power-law described by $R(>F)\propto F^\gamma \exp(-F/F_{\rm max})$, and the best fit result is $\gamma\simeq 0.0$ and $F_{\rm max}\simeq 570\rm\, Jy\,ms$. Although the likelihood ratio test shows that a truncated power-law fit is preferred with a $p$-value of 0.06 --- meaning that there is a 6\% chance that the improvement in the likelihood is due to random fluctuations, the truncated power-law model has an additional free parameter and hence a truncation is not required by the current data.

We then compared the repeating rate function for FRB\,20220912A with the cumulative all-sky rate function at high fluences, which is provided by the ASKAP Fly's Eye survey \citep{Shannon2018}. Using the same method as outlined above, we find that the all-sky rate function can be described by the following best-fit power-law function above an optimized minimum fluence of $F_{\rm min}=50.6\rm\, Jy\, ms$,
\begin{equation}
    R_{\rm as}(>F) = 1.6\mathrm{\,h^{-1}}\, (F/F_{\rm min})^{\gamma_{\rm as}}, \gamma_{\rm as}=1.54,
\end{equation}
and the standard deviation of the power-law slope is $\sigma_{\rm as}=0.34$. Despite that $\gamma_{\rm as}$ is statistically weakly constrained, we do theoretically expect the fluence distribution to be close to the Euclidean value of $\gamma=1.5$ because the ASKAP bursts are from the local Universe and the entire FRB population has a maximum specific energy \citep[see e.g.,][]{macquart18_FRB_event_rate_counts, lu19_ASKAP_sample_study, shin23_CHIME_luminosity_function}.

It should be noted that the specific energy of the brightest burst ($F=1128\rm\, Jy\,ms$) detected in our sample is
\begin{equation}
    E \simeq {F\times 4\pi D_{\rm L}^2(z)\over (1+z)^2} \simeq 1.5\times10^{32}\rm\, erg\,Hz^{-1},
\end{equation}
where we have used the \citet{Planck2015} cosmological parameters for the luminosity distance $D_{\rm L}$ and a source redshift of $z=0.077$ provided by \citet{Ravi2022_frb220912a}. This burst is among the most intrinsically bright ones observed to-date \citep[e.g.,][]{ryder22_most_luminous_FRB, shin23_CHIME_luminosity_function}. As can be seen in Figure \ref{fig:fluence_distribution}, the repeating rate function of  FRB\,20220912A must have a cut-off at fluence $F_{\rm max}\lesssim 10^4\rm\, Jy\,ms$ for this source to be consistent with the all-sky fluence distribution extrapolated from the ASKAP measurement assuming $\gamma_{\rm as}=1.5$ (and additionally assuming that the repetition rate of this source is constant). Such a cut-off is also consistent with the inferred maximum specific energy $E_{\rm max}\sim 10^{32}\mbox{ to }10^{33}\rm\, erg\,Hz^{-1}$ in the FRB luminosity/energy function \citep{shin23_CHIME_luminosity_function}. Further long-term monitoring of FRB\,20220912A is needed to test if the repeating rate function indeed has a cut-off.

On the other hand, we also see from Figure \ref{fig:fluence_distribution} that FRB\,20220912A contributes at least a few percent of the all-sky rate above fluence of $\sim \rm\,100\, Jy\,ms$. Although prolific repeaters like FRB\,20220912A and FRB\,20121102A represent a minority of all FRB sources, they contribute a significant fraction of the all-sky rate. On the other hand, many of the $>100 \rm\,Jy\, ms$ sources detected by the ASKAP Fly's Eye did not show repetitions even with sensitive follow-up observations. This means that the brightest FRBs in the sky are contributed by the most active repeaters as well as by the less active ones. This further restricts the maximum fluence of FRB\,20220912A to be likely significantly less than $10^4\rm\, Jy\,ms$.

\subsection{Characteristic Timescales and Sub-burst Periodicity}
\label{ssec:periodicity}

Given that so many of our \ac{FRB}s are multi-component (29/35), we investigate whether or not the sub-bursts show any consistent inter-pulse spacings. In particular, \ac{FRB} \#10 shows 4 sub-bursts that are bright and appear evenly-spaced, hinting at sub-burst periodicity. For both a general characteristic timescale analysis and a periodicity analysis, we begin with the \texttt{numpy} arrays produced in Section \ref{ssec:frbgui}, and average over the frequency axis to get a timeseries profile. We then median-subtract and apply a 5th-degree polynomial Savitzky-Golay filter with a window-size of 21 samples \citep{savitzky1964smoothing}. The smoothing filter is applied to make peak-finding possible in Section \ref{sssec:timescales}, but does attenuate potential short-period behaviour ($\lesssim 0.13$ ms) in the \ac{FRB} profile.

\subsubsection{Characteristic Timescales}
\label{sssec:timescales}

To characterize significant timescales in each FRB, we calculate the \ac{ACF} of the smoothed \ac{FRB} profile with \texttt{scipy}. 
This allows us to be more sensitive to non-sinusoidal signals with fewer repetitions in short timeseries than a power spectrum \citep[e.g., a Leahy-normalized power spectrum, see ][]{chime2022sub} while still performing a similar function. 

We then find the tallest local maximum in the \ac{ACF} with a required minimum peak width of $w_{_\textrm{ACF}} > 2$ bins (to avoid single-bin noise fluctuations).
This corresponds to the lag time $p_{_\textrm{ACF}}$ which contains the most power in the \ac{ACF} --- if the \ac{FRB} is generating the peak in the \ac{ACF}, this is likely representative of the delay in sub-bursts for a two-component \ac{FRB}. At this stage, 2 of the 35 \ac{FRB}s were removed from the sample due to not having any peaks in their \ac{ACF}s.

\ac{RFI} is often periodic and could create a false-positive periodicity in the \ac{ACF}s from the previous step. Therefore, we perform a bootstrap resampling noise-permutation test as a crude filter for ACF peaks that are not coming from the \ac{FRB} signal itself. Each file contains 100 channels of noise before and after the FRB, which constitutes 25\% or more of each file (depending on the variable \ac{FRB} duration; mean=42\%). For each \ac{FRB}, we perform 1000 trials where the noise arrays before and after the \ac{FRB} signal are randomly shuffled, and then perform the median-removal, smoothing, \ac{ACF}, and peak-finding steps as for the original \ac{FRB} profile. If the mean of the set of 1000 noise-scrambled \ac{ACF} peaks still falls within 10\% of the original, with a standard deviation that is less than the width of the peak $w_{_\textrm{ACF}}$ from the original \ac{ACF}, we treat this as a spacing that is inherent to the \ac{FRB} itself. We find that 21 of the remaining 33 FRBs pass this filter. Those 21 \ac{ACF} peak spacings were then visually checked against the timeseries profiles, to ensure that the above methodology returned results that were consistent with the visual appearance of the \ac{FRB}. It should be noted that peaks in an \ac{ACF} can be difficult to interpret or assign significances to; for example, for \ac{FRB}s with complex quasi-periodic structure, this could be representative of the distance between the two brightest sub-bursts, or an indication of evenly-spaced components.


The median and median absolute deviation of the distribution of \ac{ACF}-derived timescales is $5.82 \pm 1.16$ ms. This implies that, while individual bursts may have preferred periods or spacings, we do not see a sharp mode or otherwise tightly-clustered distribution providing evidence of a shared, strict periodicity between sub-bursts across the sample, which corroborates the report of no 1~ms--1000~s periodicity from \citet{zhang2023fast}.

In the future, this sort of technique would be improved by fitting for the \ac{FRB} shape, subtracting it from the data, and then fully permuting the remaining profile, instead of the ``bookend'' noise-permutation method described above. 

\subsubsection{Sub-Burst Periodicity}
\label{sssec:sub-burst-periodicity}

As noted by \citet{petroff2022fast}, strict periodicity even within a single sub-burst \citep[as in ][]{chime2022sub} supports a magnetospheric origin of \ac{FRB} emission, given that it is difficult to produce with an external shock model. Visually,  \ac{FRB} \# 10 is a candidate for this kind of strict periodicity. To assess the statistical significance of the periodicity in this particular burst, we employ methodology similar to \citet{chime2022sub}, as follows.

We first check for a peak in the squared Fourier transform, which we find at 6.4~ms (the \ac{ACF} peak for this \ac{FRB} was at 5.7~ms). To find actual \ac{ToAs} of each of the four subbursts, we perform a least-squares fit of a function containing four Gaussians (each with independent amplitude, width, and central time) and a constant vertical offset. Using the results of that fit, we extract the central times as the \ac{ToAs}; we then re-fit while enforcing a constant spacing between peaks, letting only period and start time of the first subburst vary while fixing the amplitudes, widths, and offset to the best-fit values from the previous fit. The resulting best-fit period is 6.1~ms.

To assess the significance of this result, we calculate the $\hat{L}[n]$ statistic from \citet{chime2022sub}, a metric of how well the \ac{ToAs} approximate the linear relationship expected for arrival time vs. integer multiples of the best-fit period (where larger scores equal better approximation). Using the \ac{ToAs} obtained in the first 4-Gaussian fit and a 6.1~ms period, we find $\hat{L}[n]_{FRB10} = 3.459$. Then, we simulate 10000 arrays of four \ac{ToAs} each, with the following conditions:

\begin{enumerate}
    \item Average spacings $\bar{d}$ that would enforce all four pulses falling within the length of the timeseries
    \item An ``exclusion factor'' (minimum spacing factor) of $\chi=0.2$
    \item \ac{ToAs} drawn from a uniform probability distribution between $\chi \bar{d} \leq d \leq (2 - \chi)\bar{d}$
\end{enumerate}

This creates series of four \ac{ToAs} that are not drawn from a periodic distribution, but will have a range of scores in $\hat{L}[n]$ which could, in cases, approach periodicity. When we compare $\hat{L}[n]_{FRB10}$ to the distribution of $\hat{L}[n]$ values from the simulations, we find that it is larger than 81\% of values in the simulation, giving a generous ``false alarm probability'' of 19\%. From this result, we cannot reject the null hypothesis of a non-periodic emission mechanism. Note that the first two sub-pulses are particularly low \ac{SNR}, so their Gaussian-fitted \ac{ToAs} may be contributing to the lower $\hat{L}[n]$ score.

We also see no obvious periodicity between bursts. However, we note that, because the \ac{ATA} can only observe the part of the \ac{FRB}'s energy distribution that is above both bimodal peaks seen by \citet{zhang2023fast} and \citet{feng2023extreme}, our sampling of periodicity or patterns in wait-time will be incomplete. 

\section{Discussion and Conclusion}
\label{sec:conclusion}

\acresetall

As described in Section \ref{ssec:search}, we detect 35 bursts from the \ac{FRB} 20220912A in 541 hours of observation with the \ac{ATA}, using the \texttt{SPANDAK} detection pipeline. The \ac{FRB}s appear throughout the 672~MHz bandpass, biased towards the lower ($\sim1$~GHz) frequency end, with an average sub-burst duration of 1.2 ms and an average sub-burst bandwidth of $\sim 200$~MHz. In Section \ref{sec:data_reduction}, we used the \texttt{frbgui} package, which leverages 2D fits to the \ac{ACF} of the dynamic spectrum of each bursts, to measure spectrotemporal features from each \ac{FRB}. In Section \ref{sec:analysis}, we described the following features in our dataset:

\begin{enumerate}
    \item A median dispersion measure of 219.775 pc~cm$^{-3}$ (Section \ref{ssec:file_prep})
    \item A median isotropic equivalent burst energy of 1~$\times$~$10^{39}$~erg (Section \ref{sssec:energy})
    \item A positive, linear correlation between center frequency and bandwidth (Section \ref{sssec:freq_vs_bw})
    \item A decrease in center frequency and bandwidth over the two months of the campaign (Section \ref{sssec:freq_vs_time})
    \item A slight decrease in duration with increasing center frequency (Section \ref{sssec:freq_vs_dur})
    \item No evidence for scattering (Section \ref{sssec:scattering})
    \item  FRB~20220912A must have a cut-off fluence $F_{\rm max}\lesssim 10^4$~Jy-ms to be consistent with the all-sky fluence distribution (Section \ref{ssec:all_sky_rate})
    \item FRB~20220912A significantly contributed to the all-sky \ac{FRB} rate at a level of a few percent for fluences $\gtrsim$100~Jy-ms (Section \ref{ssec:all_sky_rate})
    \item The majority of bursts in the observed sample were multi-component \ac{FRB}s, with median component spacings of 5.82$\pm$1.16~ms (Section \ref{sssec:timescales})
    \item No bursts showed statistically significant sub-burst periodicity (Section \ref{sssec:sub-burst-periodicity})
\end{enumerate}

Broadly speaking, there are two classes of \ac{FRB} models depending on whether the radio bursts are created within the magnetosphere of a neutron star (or black hole) or far from the magnetosphere. 

One of the close-in models is coherent curvature emission, where charged particles in macroscopic clumps of longitudinal sizes $\lesssim \lambda$ (the \ac{FRB} wavelength) radiate coherently when moving along curved magnetic field lines \citep[e.g.,][]{lu18_FRB_emission_mechanism, lu20_unified_model}. In this model, the \ac{FRB} spectrum is set by the spatial distribution of currents in the longitudinal direction. By Fourier transformation, a narrow spectrum of $\Delta \omega/\omega_{\rm center}\simeq 0.3$ can be produced by a modestly long ($N\sim \mbox{few}$) chain of current islands.

One class of the far-away models relies on synchrotron maser emission when an ultra-relativistic outflow interacts with strongly magnetized plasma forming a quasi-perpendicular shock \citep{hoshino91_syn_maser, Metzger2019, plotnikov19_syn_maser_spectrum}. The synchrotron maser model predicts $\Delta \omega/\omega_{\rm center}\gtrsim 1$ for two reasons: (1) the ring-like particle distribution function is not infinitely thin in phase space but with a fractional momentum spread of order unity --- this allows the rapid growth of modes at a rather broad range of frequencies $\Delta \omega'/\omega_{\rm center}'\sim 1$ (here, primes $'$ mean in the comoving frame of the shocked plasma) instead of a narrow spectral line \citep{plotnikov19_syn_maser_spectrum, sironi21_syn_maser_shocks}; (2) the Doppler beaming for viewing angles of $\lesssim 1/\Gamma$ across a quasi-spherical shock front creates a broad spectrum $\Delta \omega/\omega_{\rm center}\sim 1$ in the observer's frame \citep{Metzger2019, beniamini20_frb_lightcurve}. Previous observations show that many bursts, especially those from repeaters \citep{pleunis21_CHIME_burst_morphology}, have narrow bandwidths $\Delta \omega/\omega_{\rm center}\ll 1$. For the synchrotron maser model to be viable for these narrow-banded bursts, external propagation effects \citep[e.g.,][]{Cordes2017, sobacchi22_non-linear_filamentation} must be playing a role at modulating the spectral intensity across narrow frequency intervals. 

While magnetar models are currently favoured, other exotic repeater models do exist. For example, superradiance models rely on a narrowband emitter similar to a molecular maser, and triggered superradiance models invoke \ac{FRB}s initiated in the system by a more distant coherent emitter such as a pulsar \citep{dicke1954coherence, houde2019triggered}.

The conclusions in the list above do not strongly favour or disfavour any of the described classes of models, but do provide a benchmark against which to compare future observations. For example, it remains to be shown how (and if possible at all) propagation effects in the synchrotron maser model would create a linear correlation between the bandwidth and the central frequency as seen in this work. This same observation, as well as the tentative decrease in duration with center frequency, is consistent with a narrowband emission process such as that in the superradiance model \citep{rajabi2020simple}.


Regardless of the model's location of emission, our results underscore the conclusions found in \citet{feng2023extreme} and \citet{zhang2023fast}: low-efficiency models (with efficiencies of order $10^{-4}$ or lower) for the emission mechanism are not compatible with the immense energy being released and the high activity rate of this \ac{FRB} source, especially given that our sample consists of bursts with a median energy $\sim100\times$ that of the previous literature.

All of the published bursts for this source, our 35 included, are downward-drifting, potentially indicating that downward frequency drift is inherent to the emission mechanism; this is a feature consistent with both close-in and far-away magnetar models.

This work emphasizes the importance of the \ac{ATA} in \ac{FRB} science, given its wide bandwidth recording capabilities and potential to engage in unique modes of observation, for example, a ``Fly's Eye'' strategy covering up to a 389 sq. deg. field-of-view on the sky at 1~GHz \citep{siemion2011allen}. As the refurbishment continues, additional \ac{FRB}-relevant features, such as a fast imaging mode will be re-implemented on the upgraded system \citep[][]{law2009results}. More observations of the source, especially at higher frequencies with instruments like the \ac{ATA}, will help to differentiate between the various classes of \ac{FRB} progenitor models.

\section*{Acknowledgements}

The Allen Telescope Array (ATA) refurbishment program and its ongoing operations have received substantial support from Franklin Antonio. Additional contributions from Frank Levinson, Greg Papadopoulos, the Breakthrough Listen Initiative and other private donors have been instrumental in the renewal of the ATA. Breakthrough Listen is managed by the Breakthrough Initiatives, sponsored by the Breakthrough Prize Foundation.  The Paul G. Allen Family Foundation provided major support for the design and construction of the ATA, alongside contributions from Nathan Myhrvold, Xilinx Corporation, Sun Microsystems, and other private donors. The ATA has also been supported by contributions from the US Naval Observatory and the US National Science Foundation. S.Z.S. acknowledges that this material is based upon work supported by the National Science Foundation MPS-Ascend Postdoctoral Research Fellowship under Grant No. 2138147. Participation of J.K. made possible by SETI Institute REU program (NSF award \#2051007). The authors would also like to thank Ron Maddalena for his help in measuring the sensitivity of the ATA feeds.
We acknowledge use of the CHIME/FRB Public Database, provided at https://www.chime-frb.ca/ by the CHIME/FRB Collaboration. We also acknowledge use of the software packages \texttt{pandas} \citep{the_pandas_development_team_2023_7794821}, \texttt{numpy} \citep{harris2020numpy}, \texttt{astropy} \citep{astropy:2022}, and \texttt{YOUR} \citep{your_package}.

\begin{acronym}
    \acro{FRB}{Fast Radio Burst}
    \acro{CHIME}{the Canadian Hydrogen Intensity Mapping Experiment}
    \acro{ATA}{Allen Telescope Array}
    \acro{ATel}{Astronomer's Telegram}
    \acro{DSA-110}{Deep Synoptic Array}
    \acro{DM}{Dispersion Measure}
    \acro{DSP}{Digital Signal Processing}
    \acro{RF}{Radio Frequency}
    \acro{LO}{Local Oscillator}
    \acro{IF}{Intermediate Frequency}
    \acro{FPGA}{Field Programmable Gate Array}
    \acro{BLADE}{Breakthrough Listen Accelerated DSP Engine}
    \acro{RFI}{Radio Frequency Interference}
    \acro{SNR}{Signal-to-Noise Ratio}
    \acro{SEFD}{System Equivalent Flux Density}
    \acro{ACF}{Auto-Correlation Function}
    \acro{GBT}{Green Bank Telescope}
    \acro{RFSoC}{Radio Frequency System-on-Chip}
    \acro{ToAs}{times-of-arrival}
\end{acronym}

\section*{Data Availability}

The extracted FRB properties are available in the article and in its online supplementary material. The dynamic spectra in archive or filterbank formats will be shared on reasonable request to the corresponding author.


\bibliographystyle{mnras}
\bibliography{bibliography} 


\clearpage
\onecolumn

\begin{landscape}
\begin{longtable}{p{.03\textwidth} p{.13\textwidth} p{0.1\textwidth} p{.04\textwidth} p{.12\textwidth} p{.12\textwidth} p{.05\textwidth} p{.13\textwidth} p{.11\textwidth}}
    \caption{Properties of the 35 bursts from FRB\,20220912A detected by the Allen Telescope Array.}
    \\
        \hline
        \hline
        Index & MJD & DM & Width & Flux & Fluence & S/N & Center Frequency & Bandwidth\\
         & (day) & (pc\,cm${^{-3}}$) & (ms) & (Jy) & (Jy-ms) &  & (MHz) & (MHz)\\
        \hline

        1 & 59872.4423299306 & 219.81 $\pm$ 1.22 & 2.56 & 6.30 $\pm$ 0.57 & 16.13 $\pm$ 1.45 & 18.53 & 1289.66 $\pm$ 8.39 & 301.39 $\pm$ 0.87 \\
         & & & 1.92 & 3.00 $\pm$ 0.27 & 5.75 $\pm$ 0.52 & 6.08 & 1220.00 $\pm$ 15.03 & 191.81 $\pm$ 2.42 \\

        2 & 59873.3504993634 & 219.63 $\pm$ 1.61 & 2.56 & 65.74 $\pm$ 5.77 & 168.30 $\pm$ 14.78 & 177.38 & 1652.05 $\pm$ 4.62 & 266.21 $\pm$ 0.01 \\
         & & & 3.20 & 14.71 $\pm$ 1.29 & 47.04 $\pm$ 4.13 & 53.61 & 1568.95 $\pm$ 4.64 & 389.06 $\pm$ 0.19 \\

        3 & 59875.2672106481 & 219.46 $\pm$ 0.90 & 0.77 & 25.60 $\pm$ 1.88 & 19.66 $\pm$ 1.45 & 29.29 & 1311.63 $\pm$ 5.00 & 227.51 $\pm$ 0.35 \\
         & & & 0.64 & 11.26 $\pm$ 0.83 & 7.21 $\pm$ 0.53 & 12.52 & 1326.39 $\pm$ 16.55 & 257.39 $\pm$ 4.28 \\

        4 & 59880.2720233796 & 219.98 $\pm$ 1.95 & 3.20 & 5.90 $\pm$ 0.59 & 18.87 $\pm$ 1.90 & 25.10 & 1545.38 $\pm$ 4.69 & 405.03 $\pm$ 0.92 \\
         & & & 1.92 & 25.14 $\pm$ 2.53 & 48.27 $\pm$ 4.85 & 74.93 & 1491.42 $\pm$ 4.65 & 330.73 $\pm$ 0.07 \\
         & & & 3.20 & 44.74 $\pm$ 4.50 & 143.16 $\pm$ 14.39 & 147.75 & 1453.30 $\pm$ 4.62 & 243.71 $\pm$ 0.01 \\
         & & & 1.60 & 108.09 $\pm$ 10.86 & 172.95 $\pm$ 17.38 & 259.44 & 1427.25 $\pm$ 4.62 & 257.44 $\pm$ 0.01 \\
         & & & 2.88 & 59.11 $\pm$ 5.94 & 170.22 $\pm$ 17.11 & 194.19 & 1418.31 $\pm$ 4.62 & 267.98 $\pm$ 0.01 \\
         & & & 3.84 & 21.01 $\pm$ 2.11 & 80.69 $\pm$ 8.11 & 79.35 & 1280.43 $\pm$ 4.62 & 265.50 $\pm$ 0.03 \\

        5 & 59880.3955575231 & 219.81 $\pm$ 1.12 & 3.20 & 8.48 $\pm$ 0.85 & 27.12 $\pm$ 2.73 & 29.47 & 1398.85 $\pm$ 4.69 & 270.24 $\pm$ 0.29 \\
         & & & 1.60 & 24.64 $\pm$ 2.48 & 39.42 $\pm$ 3.96 & 63.59 & 1318.60 $\pm$ 4.63 & 297.62 $\pm$ 0.08 \\
         & & & 2.88 & 35.13 $\pm$ 3.53 & 101.18 $\pm$ 10.17 & 122.56 & 1304.51 $\pm$ 4.62 & 302.13 $\pm$ 0.02 \\

        6 & 59880.4029308449 & 219.78 $\pm$ 1.82 & 7.68 & 12.60 $\pm$ 1.27 & 96.78 $\pm$ 9.73 & 65.43 & 1236.88 $\pm$ 2.32 & 250.98 $\pm$ 0.02 \\
         & & & 6.40 & 19.80 $\pm$ 1.99 & 126.70 $\pm$ 12.73 & 79.22 & 1136.83 $\pm$ 2.31 & 178.88 $\pm$ 0.01 \\

        7 & 59882.2707171412 & 219.72 $\pm$ 1.43 & 2.56 & 9.43 $\pm$ 0.94 & 24.15 $\pm$ 2.40 & 18.76 & 1683.63 $\pm$ 4.62 & 113.21 $\pm$ 0.12 \\

        8 & 59882.4726475694 & 219.23 $\pm$ 0.55 & 1.60 & 2.79 $\pm$ 0.28 & 4.47 $\pm$ 0.44 & 5.73 & 1381.16 $\pm$ 6.14 & 192.61 $\pm$ 2.94 \\
         & & & 1.60 & 7.53 $\pm$ 0.75 & 12.05 $\pm$ 1.20 & 17.13 & 1337.40 $\pm$ 7.50 & 237.15 $\pm$ 0.92 \\

        9 & 59883.1097671759 & 221.16 $\pm$ 1.39 & 6.40 & 3.95 $\pm$ 0.36 & 25.31 $\pm$ 2.32 & 15.90 & 1059.68 $\pm$ 8.55 & 217.32 $\pm$ 1.40 \\
         & & & 6.40 & 4.18 $\pm$ 0.38 & 26.75 $\pm$ 2.45 & 12.51 & 1003.38 $\pm$ 7.99 & 120.38 $\pm$ 1.10 \\

        10 & 59883.4502390046 & 219.60 $\pm$ 0.68 & 6.40 & 2.79 $\pm$ 0.26 & 17.86 $\pm$ 1.64 & 12.75 & 1301.65 $\pm$ 5.71 & 280.61 $\pm$ 5.38 \\
         & & & 6.40 & 2.87 $\pm$ 0.26 & 18.38 $\pm$ 1.68 & 10.73 & 1200.97 $\pm$ 9.44 & 187.49 $\pm$ 1.11 \\
         & & & 6.40 & 4.24 $\pm$ 0.39 & 27.14 $\pm$ 2.49 & 17.26 & 1120.52 $\pm$ 6.76 & 222.49 $\pm$ 0.78 \\
         & & & 5.12 & 5.57 $\pm$ 0.51 & 28.51 $\pm$ 2.61 & 20.53 & 1094.15 $\pm$ 6.03 & 228.34 $\pm$ 1.10 \\

        11 & 59887.3512349537 & 219.60 $\pm$ 2.23 & 9.60 & 15.66 $\pm$ 1.57 & 150.34 $\pm$ 15.05 & 84.69 & 1126.54 $\pm$ 4.62 & 219.55 $\pm$ 0.04 \\
         & & & 3.84 & 10.03 $\pm$ 1.00 & 38.53 $\pm$ 3.86 & 35.02 & 1008.34 $\pm$ 4.62 & 228.68 $\pm$ 0.45 \\
         & & & 2.56 & 14.88 $\pm$ 1.49 & 38.08 $\pm$ 3.81 & 37.65 & 974.84 $\pm$ 4.62 & 180.38 $\pm$ 0.39 \\
         & & & 3.84 & 6.78 $\pm$ 0.68 & 26.04 $\pm$ 2.61 & 19.09 & 966.87 $\pm$ 4.91 & 148.70 $\pm$ 1.85 \\
         & & & 5.76 & 10.29 $\pm$ 1.03 & 59.28 $\pm$ 5.93 & 27.37 & 946.64 $\pm$ 4.62 & 88.50 $\pm$ 0.20 \\

        12 & 59887.4423248843 & 219.86 $\pm$ 1.35 & 2.88 & 11.63 $\pm$ 1.16 & 33.48 $\pm$ 3.35 & 34.26 & 1153.26 $\pm$ 4.62 & 217.24 $\pm$ 0.09 \\
         & & & 6.08 & 27.24 $\pm$ 2.73 & 165.64 $\pm$ 16.58 & 120.94 & 1114.70 $\pm$ 4.62 & 233.57 $\pm$ 0.01 \\
         & & & 5.12 & 12.46 $\pm$ 1.25 & 63.81 $\pm$ 6.39 & 43.26 & 967.08 $\pm$ 4.62 & 169.55 $\pm$ 0.04 \\

        13 & 59889.1244077662 & 220.03 $\pm$ 0.98 & 4.48 & 5.92 $\pm$ 0.60 & 26.53 $\pm$ 2.70 & 20.98 & 1469.10 $\pm$ 4.64 & 195.59 $\pm$ 0.28 \\
         & & & 6.40 & 4.80 $\pm$ 0.49 & 30.71 $\pm$ 3.12 & 26.32 & 1355.54 $\pm$ 4.66 & 328.12 $\pm$ 0.31 \\

        14 & 59889.3744443403 & 219.77 $\pm$ 1.35 & 7.68 & 13.49 $\pm$ 1.37 & 103.60 $\pm$ 10.54 & 46.63 & 947.01 $\pm$ 4.64 & 108.60 $\pm$ 0.76 \\
         & & & 6.40 & 5.17 $\pm$ 0.53 & 33.11 $\pm$ 3.37 & 11.77 & 929.49 $\pm$ 4.66 & 56.48 $\pm$ 6.54 \\

        15 & 59890.3176200116 & 219.57 $\pm$ 1.64 & 2.56 & 2.58 $\pm$ 0.26 & 6.60 $\pm$ 0.67 & 7.63 & 1423.93 $\pm$ 12.58 & 242.58 $\pm$ 8.22 \\
         & & & 7.68 & 16.86 $\pm$ 1.70 & 129.51 $\pm$ 13.06 & 88.89 & 1409.52 $\pm$ 4.62 & 256.73 $\pm$ 0.03 \\
         & & & 5.12 & 28.24 $\pm$ 2.85 & 144.57 $\pm$ 14.58 & 125.74 & 1308.04 $\pm$ 4.63 & 274.81 $\pm$ 0.02 \\

        16 & 59890.3637615394 & 220.34 $\pm$ 1.43 & 16.64 & 4.60 $\pm$ 0.46 & 76.52 $\pm$ 7.72 & 19.92 & 948.73 $\pm$ 4.79 & 80.03 $\pm$ 0.05 \\

        17 & 59891.3109625000 & 219.49 $\pm$ 1.00 & 3.84 & 9.85 $\pm$ 1.05 & 37.82 $\pm$ 4.01 & 30.63 & 978.96 $\pm$ 4.82 & 161.54 $\pm$ 0.51 \\
         & & & 7.04 & 16.30 $\pm$ 1.73 & 114.76 $\pm$ 12.18 & 55.80 & 950.16 $\pm$ 4.62 & 106.73 $\pm$ 0.06 \\
         & & & 1.92 & 20.17 $\pm$ 2.14 & 38.72 $\pm$ 4.11 & 33.33 & 944.70 $\pm$ 4.65 & 91.18 $\pm$ 0.27 \\
         & & & 3.84 & 38.79 $\pm$ 4.12 & 148.96 $\pm$ 15.81 & 94.28 & 939.47 $\pm$ 4.62 & 98.62 $\pm$ 0.02 \\
         & & & 3.84 & 11.40 $\pm$ 1.21 & 43.79 $\pm$ 4.65 & 25.54 & 932.68 $\pm$ 4.89 & 83.77 $\pm$ 0.43 \\

        18 & 59891.3164585648 & 219.29 $\pm$ 0.61 & 1.92 & 7.66 $\pm$ 0.81 & 14.71 $\pm$ 1.56 & 19.17 & 1043.39 $\pm$ 4.62 & 209.05 $\pm$ 0.32 \\
         & & & 1.92 & 9.09 $\pm$ 0.96 & 17.46 $\pm$ 1.85 & 22.55 & 1038.36 $\pm$ 5.53 & 205.43 $\pm$ 0.26 \\

        19 & 59897.0712164352 & 219.40 $\pm$ 1.11 & 3.20 & 16.15 $\pm$ 1.61 & 51.69 $\pm$ 5.16 & 40.72 & 971.51 $\pm$ 2.37 & 144.12 $\pm$ 1.01 \\
         & & & 4.48 & 55.00 $\pm$ 5.49 & 246.42 $\pm$ 24.58 & 146.33 & 955.11 $\pm$ 2.31 & 114.66 $\pm$ 0.05 \\
         & & & 5.12 & 24.72 $\pm$ 2.47 & 126.58 $\pm$ 12.62 & 63.69 & 949.56 $\pm$ 2.31 & 94.08 $\pm$ 0.22 \\
         & & & 3.84 & 17.51 $\pm$ 1.75 & 67.22 $\pm$ 6.70 & 37.72 & 940.87 $\pm$ 2.32 & 87.74 $\pm$ 0.49 \\

        20 & 59898.4399226852 & 219.91 $\pm$ 1.65 & 4.16 & 2.48 $\pm$ 0.25 & 10.32 $\pm$ 1.03 & 9.12 & 1218.58 $\pm$ 4.79 & 233.42 $\pm$ 10.08 \\
         & & & 5.44 & 10.64 $\pm$ 1.07 & 57.86 $\pm$ 5.80 & 47.43 & 1151.51 $\pm$ 4.77 & 262.89 $\pm$ 0.23 \\

        21 & 59900.1470432060 & 220.03 $\pm$ 1.45 & 2.56 & 9.37 $\pm$ 0.93 & 24.00 $\pm$ 2.38 & 24.36 & 1059.62 $\pm$ 4.17 & 194.06 $\pm$ 1.14 \\
         & & & 2.56 & 22.82 $\pm$ 2.26 & 58.42 $\pm$ 5.79 & 58.10 & 1056.02 $\pm$ 2.32 & 186.19 $\pm$ 0.12 \\
         & & & 3.84 & 30.42 $\pm$ 3.01 & 116.82 $\pm$ 11.58 & 94.34 & 1030.33 $\pm$ 2.31 & 184.12 $\pm$ 0.03 \\
         & & & 14.08 & 28.78 $\pm$ 2.85 & 405.15 $\pm$ 40.15 & 147.22 & 959.33 $\pm$ 2.31 & 136.69 $\pm$ 0.00 \\

        22 & 59900.2719472685 & 220.00 $\pm$ 1.47 & 5.76 & 13.64 $\pm$ 1.35 & 78.57 $\pm$ 7.79 & 48.58 & 963.68 $\pm$ 4.63 & 161.91 $\pm$ 0.01 \\

        23 & 59901.4354073148 & 219.66 $\pm$ 1.76 & 10.24 & 6.69 $\pm$ 0.69 & 68.46 $\pm$ 7.09 & 49.16 & 1262.87 $\pm$ 4.62 & 354.94 $\pm$ 1.52 \\
         & & & 6.40 & 4.11 $\pm$ 0.43 & 26.28 $\pm$ 2.72 & 18.64 & 1169.06 $\pm$ 4.68 & 216.43 $\pm$ 7.03 \\
         & & & 8.32 & 15.62 $\pm$ 1.62 & 129.93 $\pm$ 13.46 & 80.24 & 1108.69 $\pm$ 4.62 & 213.32 $\pm$ 0.27 \\
         & & & 5.76 & 16.99 $\pm$ 1.76 & 97.86 $\pm$ 10.14 & 52.06 & 1040.13 $\pm$ 7.50 & 109.59 $\pm$ 11.58 \\

        24 & 59903.1278839120 & 224.94 $\pm$ 1.76 & 5.76 & 5.13 $\pm$ 0.51 & 29.53 $\pm$ 2.93 & 22.38 & 1091.56 $\pm$ 4.64 & 242.03 $\pm$ 1.30 \\
         & & & 4.48 & 11.86 $\pm$ 1.18 & 53.15 $\pm$ 5.28 & 41.54 & 1039.08 $\pm$ 4.77 & 200.17 $\pm$ 0.29 \\
         & & & 4.16 & 11.31 $\pm$ 1.12 & 47.07 $\pm$ 4.68 & 34.67 & 983.03 $\pm$ 4.62 & 165.10 $\pm$ 0.21 \\

        25 & 59903.2531478009 & 223.24 $\pm$ 1.62 & 3.20 & 23.33 $\pm$ 2.32 & 74.64 $\pm$ 7.42 & 86.56 & 1240.18 $\pm$ 4.62 & 314.81 $\pm$ 0.12 \\
         & & & 4.48 & 96.10 $\pm$ 9.55 & 430.55 $\pm$ 42.77 & 407.73 & 1274.37 $\pm$ 4.62 & 293.91 $\pm$ 0.00 \\
         & & & 2.88 & 97.12 $\pm$ 9.65 & 279.71 $\pm$ 27.79 & 329.37 & 1211.68 $\pm$ 4.62 & 292.13 $\pm$ 0.00 \\
         & & & 3.52 & 44.98 $\pm$ 4.47 & 158.33 $\pm$ 15.73 & 155.80 & 1187.19 $\pm$ 4.62 & 249.33 $\pm$ 0.02 \\
         & & & 1.92 & 20.92 $\pm$ 2.08 & 40.17 $\pm$ 3.99 & 49.47 & 1116.64 $\pm$ 4.64 & 213.09 $\pm$ 0.22 \\
         & & & 4.48 & 32.35 $\pm$ 3.21 & 144.93 $\pm$ 14.40 & 127.18 & 1080.22 $\pm$ 4.62 & 252.39 $\pm$ 0.02 \\

        26 & 59904.2669536921 & 219.46 $\pm$ 1.15 & 10.24 & 17.17 $\pm$ 1.71 & 175.87 $\pm$ 17.48 & 108.96 & 1413.36 $\pm$ 4.62 & 287.34 $\pm$ 0.34 \\
         & & & 8.96 & 41.23 $\pm$ 4.10 & 369.43 $\pm$ 36.71 & 253.34 & 1337.74 $\pm$ 4.62 & 308.05 $\pm$ 0.08 \\
         & & & 6.40 & 9.60 $\pm$ 0.95 & 61.44 $\pm$ 6.11 & 47.50 & 1229.29 $\pm$ 4.63 & 279.59 $\pm$ 2.61 \\
         & & & 6.40 & 2.82 $\pm$ 0.28 & 18.05 $\pm$ 1.79 & 11.79 & 1160.55 $\pm$ 4.84 & 199.84 $\pm$ 38.37 \\

        27 & 59907.2594524074 & 219.77 $\pm$ 0.43 & 3.84 & 2.21 $\pm$ 0.21 & 8.48 $\pm$ 0.82 & 6.39 & 1067.12 $\pm$ 9.71 & 168.12 $\pm$ 1.44 \\

        28 & 59912.2962470023 & 219.83 $\pm$ 1.76 & 3.20 & 12.44 $\pm$ 1.26 & 39.82 $\pm$ 4.03 & 22.08 & 925.77 $\pm$ 4.64 & 69.26 $\pm$ 0.02 \\
         & & & 3.84 & 13.22 $\pm$ 1.34 & 50.77 $\pm$ 5.14 & 24.26 & 920.02 $\pm$ 4.63 & 61.70 $\pm$ 0.01 \\

        29 & 59914.2529407407 & 220.31 $\pm$ 2.56 & 6.40 & 30.83 $\pm$ 3.15 & 197.31 $\pm$ 20.15 & 93.40 & 936.72 $\pm$ 4.62 & 99.31 $\pm$ 0.06 \\
         & & & 5.76 & 3.53 $\pm$ 0.36 & 20.31 $\pm$ 2.07 & 7.49 & 940.03 $\pm$ 4.64 & 54.19 $\pm$ 7.29 \\

        30 & 59915.2896052546 & 219.60 $\pm$ 1.11 & 9.60 & 2.44 $\pm$ 0.25 & 23.47 $\pm$ 2.39 & 13.22 & 1013.09 $\pm$ 7.50 & 211.43 $\pm$ 4.31 \\
         & & & 7.04 & 15.45 $\pm$ 1.58 & 108.78 $\pm$ 11.09 & 61.20 & 951.36 $\pm$ 4.62 & 154.67 $\pm$ 0.15 \\
         & & & 4.48 & 5.26 $\pm$ 0.54 & 23.58 $\pm$ 2.40 & 16.52 & 954.24 $\pm$ 4.67 & 152.58 $\pm$ 2.56 \\
         & & & 8.32 & 2.92 $\pm$ 0.30 & 24.30 $\pm$ 2.48 & 8.34 & 956.78 $\pm$ 6.55 & 68.03 $\pm$ 1.25 \\

        31 & 59916.3118642361 & 220.11 $\pm$ 1.20 & 6.40 & 5.25 $\pm$ 0.52 & 33.59 $\pm$ 3.32 & 22.81 & 1162.06 $\pm$ 5.44 & 218.62 $\pm$ 0.20 \\

        32 & 59920.2334853009 & 219.83 $\pm$ 1.25 & 4.48 & 12.14 $\pm$ 0.79 & 54.38 $\pm$ 3.55 & 25.46 & 1080.98 $\pm$ 4.71 & 166.28 $\pm$ 0.11 \\

        33 & 59930.1239017014 & 219.55 $\pm$ 0.98 & 3.20 & 7.75 $\pm$ 0.79 & 24.81 $\pm$ 2.52 & 20.48 & 970.46 $\pm$ 4.74 & 152.34 $\pm$ 1.21 \\
         & & & 3.20 & 8.53 $\pm$ 0.87 & 27.31 $\pm$ 2.78 & 17.46 & 968.14 $\pm$ 4.67 & 91.45 $\pm$ 1.13 \\
         & & & 1.28 & 5.64 $\pm$ 0.57 & 7.21 $\pm$ 0.73 & 4.47 & 986.71 $\pm$ 4.89 & 34.28 $\pm$ 1.41 \\

        34 & 59931.2705442477 & 219.77 $\pm$ 1.43 & 5.12 & 4.27 $\pm$ 0.43 & 21.87 $\pm$ 2.22 & 15.82 & 999.10 $\pm$ 4.73 & 188.25 $\pm$ 1.60 \\
         & & & 2.56 & 3.74 $\pm$ 0.38 & 9.58 $\pm$ 0.97 & 5.15 & 966.17 $\pm$ 5.40 & 51.96 $\pm$ 2.06 \\
         & & & 2.56 & 21.32 $\pm$ 2.16 & 54.58 $\pm$ 5.53 & 35.14 & 936.37 $\pm$ 4.91 & 74.58 $\pm$ 0.11 \\

        35 & 59933.1711922801 & 219.86 $\pm$ 1.12 & 3.20 & 2.07 $\pm$ 0.22 & 6.62 $\pm$ 0.69 & 7.28 & 1041.30 $\pm$ 9.47 & 254.05 $\pm$ 6.32 \\
         & & & 5.12 & 24.94 $\pm$ 2.62 & 127.72 $\pm$ 13.40 & 90.72 & 979.10 $\pm$ 4.62 & 169.45 $\pm$ 0.03 \\
         & & & 3.20 & 58.65 $\pm$ 6.15 & 187.69 $\pm$ 19.69 & 162.29 & 968.05 $\pm$ 4.62 & 156.93 $\pm$ 0.01 \\
         & & & 3.20 & 84.50 $\pm$ 8.86 & 270.39 $\pm$ 28.37 & 234.56 & 963.53 $\pm$ 4.62 & 157.97 $\pm$ 0.01 \\
         & & & 3.20 & 33.11 $\pm$ 3.47 & 105.94 $\pm$ 11.11 & 79.37 & 948.45 $\pm$ 4.62 & 117.81 $\pm$ 0.03 \\
         & & & 4.48 & 26.32 $\pm$ 2.76 & 117.92 $\pm$ 12.37 & 68.89 & 940.82 $\pm$ 4.62 & 100.29 $\pm$ 0.05 \\
         & & & 8.00 & 25.42 $\pm$ 2.67 & 203.35 $\pm$ 21.33 & 85.77 & 935.08 $\pm$ 4.62 & 93.35 $\pm$ 0.01 \\
        \hline
        \label{tab:frb_detections}
\end{longtable}
\end{landscape}

\bsp	
\label{lastpage}
\end{document}